\documentclass[journal,draftcls,onecolumn,12pt,twoside]{IEEEtran}
\usepackage{color,enumitem}
\usepackage{graphicx}
\usepackage{amsmath,amssymb,amsthm}
\usepackage{latexsym}
\usepackage{mathrsfs}
\usepackage{cite}
\usepackage{verbatim}

\newtheorem{lemma}{Lemma}
\newtheorem{theorem}{Theorem}

\usepackage{algorithm}
\usepackage{algorithmic}

\newcommand{\bP}{\mathbf{P}}

\newcommand{\bZ}{\mathbf{Z}}
\newcommand{\bR}{\mathbf{R}}
\newcommand{\bV}{\mathbf{V}}
\newcommand{\interior}[1]{%
  {\kern0pt#1}^{\mathrm{o}}%
}
\usepackage{bm}
\usepackage{stfloats}
\usepackage[font=footnotesize, justification=centering]{caption}

\ifCLASSINFOpdf
   \graphicspath{{../jpg/}{../pdf/}{../jpeg/}{../eps/}}
   \DeclareGraphicsExtensions{.jpg,.pdf,.jpeg,.png,.eps}
\else
  \DeclareGraphicsExtensions{.eps}
\fi

\ifCLASSOPTIONcompsoc
  \usepackage[caption=false,font=normalsize,labelfont=sf,textfont=sf]{subfig}
\else
  \usepackage[caption=false,font=footnotesize]{subfig}
\fi
\ifCLASSOPTIONcompsoc
\else
\fi

\usepackage{setspace}

\begin{document}
%
\title{Movement-efficient Sensor Deployment in Wireless Sensor Networks\\ with Limited Communication Range\vspace{-10pt}}


\author{Jun~Guo,~\IEEEmembership{Student Member,~IEEE},
        and~Hamid~Jafarkhani,~\IEEEmembership{Fellow,~IEEE}\vspace{-30pt}
\thanks{
The authors are with Center for Pervasive Communications and Computing,
University of California, Irvine (e-mail: guoj4@uci.edu; hamidj@uci.edu).
This work was supported in part by the NSF Award CCF-1815339.
The paper was presented in part at ICC-18 \cite{GJICC}.
}}

\maketitle
\vspace{-15pt}
\begin{abstract}
We study a mobile wireless sensor network (MWSN) consisting of multiple mobile sensors or robots.
Three key factors in MWSNs, sensing quality, energy consumption, and connectivity, have attracted plenty of attention, but the interaction of these factors is not well studied.
To take all the three factors into consideration, we model the sensor deployment problem as a constrained source coding problem.
Our goal is to find an optimal sensor deployment (or relocation) to optimize the sensing quality with a limited communication range and a specific network lifetime constraint.
We derive necessary conditions for the optimal sensor deployment in both homogeneous and heterogeneous MWSNs.
According to our derivation, some sensors are idle in the optimal deployment of heterogeneous MWSNs.
Using these necessary conditions, we design both centralized and distributed algorithms to provide a flexible and explicit trade-off between sensing uncertainty and network lifetime.
The proposed algorithms are successfully extended to more applications, such as area coverage and target coverage, via properly selected density functions.
Simulation results show that our algorithms outperform the existing relocation algorithms.
\end{abstract}

\begin{IEEEkeywords}
\vspace{-10pt}
Sensor deployment, coverage, heterogeneous, mobile wireless sensor networks, source coding.
\end{IEEEkeywords}
\vspace{-10pt}

%
\IEEEpeerreviewmaketitle

\section{Introduction}
\label{secIntro}
Deploying multiple nodes, sensors or robots, to monitor the environment is the primary objective of the mobile wireless sensor networks (MWSNs).
To evaluate the sensing quality, the binary disk coverage model, in which each sensor can only cover a disk with the radius $R_s$, is widely used in MWSNs \cite{GJICC,CCLG,BW,MAKF,JS,SD,YBZ,GJ,VD,FJSY,MLCS,KTMN,ICC,DCC,yousefizadeh1}.
There have been many coverage measurement and deployment algorithms, for different sensing tasks, in the literature; look at \cite{CCLG,BW,MAKF} and the references therein.

Four popular coverage categories are (i) area coverage, (ii) target coverage, (iii) barrier coverage, and (iv) even deployment of the sensors.
A natural sensing task is to maximize the area coverage, which is formulated by the total area covered by sensors.
In another popular coverage task, target coverage, the specific target locations are detected and reported by static sensors.
In this case, sensors or robots are required to collect detailed information from the discrete targets.
A full-target-coverage is achieved if and only if every discrete target in the 2-dimensional region is covered by at least one sensor.
In another popular coverage task, barrier coverage, sensors are moving along the boundary to detect intruders as they cross the border of a region or domain.
To obtain full-barrier-coverage, one should place sensors to cover the whole barrier or boundary.
Finally, an even deployment of the sensors requires them to form a Centroidal Voronoi Tessellation (CVT).
It is mainly used when there is no specific target.
The widely used CVT model (see more details in Section \ref{sec:model}) can be presented as a quantizer with the sensing uncertainty  as its distortion \cite{JS,SD,YBZ,GJ,VD,GJICC,FJSY,MLCS,KTMN,Erdem1,Erdem2,MAKF,XKYLM,VNH,ML,BW,CCLG}.

Connectivity is also an important requirement in MWSNs.
In MWSNs, mobile sensor nodes are relocated to collect physical information, such as magnetism, temperature, and voice, and then forward the collected data to the outside world through access points (APs).
Therefore, the collected data is useless if it cannot be forwarded to the AP via single-hop or multiple-hop communications.
When sensors are connected by wirelines, the connectivity is guaranteed automatically.
But, the connectivity is still a challenge in MWSNs where sensors are communicating with each other through wireless channels.
A common communication model \cite{BW,DCC,ICC,yousefizadeh1} assumes that each sensor node is able to communicate with sensors in a limited communication range $R_c$.

Energy efficiency is another key issue in MWSNs, as most sensors have limited battery energy, and it is inconvenient or even infeasible to replenish the batteries of numerous densely deployed sensors \cite{GJcon}.
In general, the energy consumption of a device includes communication energy, data processing energy \cite{Moshfeghi}, sensing energy, and movement energy.
In fact, sensor movement has a much higher energy consumption compared to other types of energy \cite{KD,JW}, and then dominates the energy consumption.
Guiling et al. \cite{GW} study the optimal angular velocity and the optimal acceleration to minimize the energy consumption for motion.
Simulation results in \cite{GW} show that the energy consumption for one-step motion with the optimal angular velocity setting is approximately linear to the movement distance.
In fact, the linear movement energy consumption is a popular assumption and widely adopted in the literature \cite{YM,WL,KS,SP,YK,SMJ,ZW,AJ}.
Particularly, the movement energy consumption in some specific sensors is 5.976J/m \cite{YM}.
Total energy consumption and network lifetime are two common energy-related measures.
But, compared with total energy minimization, network lifetime maximization, which balances the energy consumption among sensors, is a more worthy and challenging problem in MWSNs.
\vspace{-15pt}
\subsection{Related Work}\label{sec:relatedWorks}
\vspace{-5pt}
A huge body of literature exists on energy-efficient sensor relocation.
Reducing the energy consumption with a full-coverage guarantee is well studied in \cite{WL,KS,SP,YK,SMJ,ZW,AJ}.
Hungarian Algorithm is applied to minimize the total energy consumption after the full-area-coverage is achieved by Genetic Algorithm \cite{WL}.
Similarly, the grid-based algorithms are proposed in \cite{SP} to reduce the total energy consumption while keeping the full-area-coverage and full-connectivity.
Kuei-Ping et al. \cite{KS} propose a distributed partition avoidance lazy movement (PALM) protocol, which avoids unnecessary movement, to ensure both full-area-coverage and connectivity.
Shuhui et al. \cite{SMJ} provide a scan-based relocation algorithm, SAMRT, which is supposed to be energy-effective with densely deployed sensors.
Note that the above methods put sensing quality as the first priority, and total energy consumption is minimized among solutions that provide full-area-coverage.
To provide a flexible trade-off between area coverage and energy consumption, virtual force based algorithms, DSSA \cite{ZW}, HEAL \cite{AJ}, and VFA \cite{YK,YY}, are proposed.
In \cite{ZW}, the authors take into account the local sensor density, and thus avoid unnecessary movements in the region with densely deployed sensors.
In \cite{AJ}, HEAL is designed to mend area coverage holes while minimizing the moving distance. However, the main assumption that there are enough sensors to achieve full-area-coverage, limits it's usage.
Instead of saving the total energy consumption, another virtual force based algorithm, VFA, is proposed in \cite{YK} to prolong the network lifetime during the area coverage maximization.
A variant of VFA is designed in \cite{YY} to maximize the area coverage in a heterogeneous MWSN with both mobile and static sensors.
However, connectivity is not considered in \cite{ZW,AJ,YK,YY}.

The sensor relocations for target coverage and barrier coverage are also well studied by researchers.
Rout et al. \cite{MRRR} design a virtual-force based algorithm, OATIDA, to obtain both full-target-coverage and full-connectivity on a region with obstacles, while energy consumption is ignored.
Chen et al. \cite{ZXFG} propose a two-phase algorithm to achieve full-target-coverage with minimum total energy consumption.
In the first phase, the target area is divided into some subareas according to the target set.
And then mobile sensors in the second phase are scheduled efficiently to cover all subareas.
Unfortunately, connectivity is not considered in either phase.
Liao et al. \cite{ZJSJG} investigate how to deploy mobile sensors with minimum total energy consumption to form a MWSN that provides both full-target-coverage and full-connectivity.
Although all three factors are considered in \cite{ZJSJG}, full-target-coverage and full-connectivity are implemented sequentially, which requires redundant sensors.
Different from the above relocation schemes which seek the perfect sensing quality, i.e., full-target-coverage, Njoya et al. design an evolutionary-based framework to make the trade-off between target coverage and network lifetime.
Nonetheless, the connectivity requirement is missing in \cite{AWAE}.
Similar to the above studies of target coverage, the existing literature on barrier coverage also seeks the perfect sensing quality, i.e., full-barrier-coverage.
Chen et al. \cite{DYJH} focus on 1-dimensional barriers, and then provide an energy-efficient relocation plan to obtain full-barrier-coverage.
In \cite{SLH}, a greedy algorithm with binary search is applied to achieve maximum network lifetime and 2-dimensional full-barrier-coverage simultaneously.
A faster algorithm which achieves the same purpose as \cite{SLH} is provided in \cite{ZHB}.
Still, above sensor relocation algorithms designed for barrier coverage ignore the connectivity requirement.

Furthermore, the sensor relocation for even deployment (or sensing uncertainty) has been investigated in recent years.
Li et al. \cite{FJSY} explore directional sensors whose sensing uncertainty varies among different directions, and then design two iterative algorithms to optimize the sensor deployment.
The authors also claim that full-connectivity is ensured when sensor density is high.
But, energy consumption is not taken into their objective function.
Taking both connectivity and sensing uncertainty into account, we analyze the necessary conditions for the optimal sensor relocation in heterogeneous MWSNs in our previous work \cite{GJ}.
Unfortunately, another important factor, energy consumption, is not taken into consideration.
A natural approach to save energy is to add an energy-related penalty term into the objective function.
In \cite{YBZ}, the authors propose two algorithms, Lloyd-$\alpha$ and DEED, to minimize sensing uncertainty with a movement related penalty function.
For Lloyd-$\alpha$, the movement in each iteration is scaled by a parameter $\alpha\in[0,1]$.
In DEED, the penalty function is properly selected with a positive definite matrix depending on a parameter $\delta$, and then the movement is optimized with the help of the gradient and Hessian matrix of the distortion.
Note that one has to manually adjust the parameter $\alpha$ or $\delta$ to satisfy a specific total energy constraint.
To overcome this weakness, two Lloyd-like algorithms without any intermediate parameter are proposed in \cite{GJICC}.
These two algorithms can be employed to minimize sensing uncertainty with a total energy constraint or a network lifetime constraint.
However, the above two papers do not consider the connectivity requirement.

\vspace{-15pt}
\subsection{Our Contributions}\label{sec:Contributions}
\vspace{-5pt}
In summary, sensing quality, connectivity, and energy consumption (or network lifetime) are three major factors in a successful MWSN.
Although there exist great achievements in the sensor relocation problems with one or two of the above factors, the optimal sensor relocation with all three factors is significantly challenging and has not been well studied.
In particular, to the best of our knowledge, a sensor deployment which aims to improve area-coverage (or sensing uncertainty) with (i) a limited communication range and (ii) a required network lifetime has not yet been considered in the literature.
Moreover, the existing relocation methods can only be applied to deal with one of the above-mentioned sensing tasks.
For example, the algorithms designed for area coverage cannot be applied to target coverage.
However, it is possible that multiple sensing tasks, e.g., target and barrier coverage, are simultaneously required in practice.
As a result, finding a general solution for different kinds of sensing tasks is needed.

In this paper, we study the sensor relocation problem in MWSNs and make the following contributions:
(1) Taking sensing quality, connectivity, and energy consumption into consideration, we propose a constrained optimization problem for sensor deployment.
(2) In a centralized scenario, we provide the necessary conditions for the optimal deployment.
(3) With the help of the necessary conditions, we design centralized Lloyd-like algorithms to optimize the sensor deployment with (i) the  network lifetime constraint and (ii) limited communication range.
(4) In a distributed, self-organized, scenario, we propose a method to keep full connectivity.
(5) A distributed realization of node deployment is provided to overcome both limited energy and limited communication range.
(6) We extend the above optimization problems to maximize the target coverage via properly selecting the density functions.

The rest of the paper is organized as follows.
We first review the related works on the sensor deployment or relocation in MWSNs in Section \ref{sec:relatedWorks}.
Then, we introduce the system model and formulate the problems of sensing, energy consumption, and connectivity in Section \ref{sec:model}.
In Section \ref{sec:ProblemB}, we discuss centralized sensor deployments for the MWSNs in which the network lifetime and communication range are considered, and then propose a centralized algorithm.
In Section \ref{sec:ProblemC}, we propose a distributed algorithm to relocate sensors such that the required network lifetime and full-connectivity are fulfilled during the relocation.
After that, we extend the proposed algorithms to other self-deployment scenarios in Section \ref{sec:extension}.
Finally, we present numerical simulations in Section \ref{sec:simulation} and conclude our work in Section \ref{sec:conclusion}.

\section{System model}\label{sec:model}
Let $\Omega$ be a simple convex polygon in $\Re^2$ including its interior.
Given $N$ sensors in the target area $\Omega$, sensor deployment before and after the relocation
are, respectively, defined by $\bP^0 = (p^0_1, \dots, p^0_N)\subset\Omega^{N}$ and $\bP = (p_1, \dots, p_N)\subset\Omega^{N}$, where $p^0_n$ is Sensor $n$'s initial location and $p_n$ is Sensor $n$'s final location.
Let $\mathcal{I}_{\Omega}=\{1,\dots,N\}$ be the set of sensors in the MWSN.
For any point $w\in\Omega$, the density function $f(\omega)$ reflects the density of an event at point $w$.
A cell partition $\bR(\bP)$ of $\Omega$ is a collection of disjoint subsets of $\{R_n(\bP)\}_{n\in\mathcal{I}_{\Omega}}$ whose union is $\Omega$.
We assume that Sensor $n$ only monitors the events that occurred in its cell partition $R_n(\bP)$, $\forall n\in\mathcal{I}_{\Omega}$.
Let $\|\cdot\|$ denote the Euclidean distance, $\partial{A}$ be the boundary of a set $A\subset\Omega$, and $\mathbb{B}(c,r)=\{\omega|\ \|\omega-c\|\le r\}$ be a disk centered at $c$ with radius $r$.

We define the access point (AP) as the sensor node that can communicate with the outside information world.
Without loss of generality, we assume that Sensor $1$ acts as the AP.
Let $\mathcal{S}(\bP)$ be the set of sensor nodes that can communicate with the AP when the sensor deployment is $\bP$.
The sensors in $\mathcal{S}(\bP)$ are referred to as active sensors while sensors out of $\mathcal{S}(\bP)$ are referred to as inactive sensors.
Note that in general not all nodes can communicate with the AP and $card(\mathcal{S}(\bP))\le n$, where $card(\mathcal{A})$ is the number of elements in set $\mathcal{A}$.
We define the active sensor deployment, which is a subset of all sensor locations, $\mathcal{H}(\bP)$ as the vector of sensor
locations for the $card(\mathcal{S}(\bP))$ sensor nodes connected to the AP.
When $\mathcal{S}(\bP)$ includes all sensor nodes, we have $\bP = \mathcal{H}(\bP)$ and $card(\mathcal{S}(\bP))=n$.
Let $\mathcal{T}$ be the set of sensor deployments that provide full connectivity, i.e., $\mathcal{T} = \{P|card(\mathcal{S}(\bP)) = n\}$.
In binary disk communication model \cite{BW,DCC,ICC}, two sensor nodes can communicate with each other within one hop if and only if the distance between
the two is smaller than $R_c$, where $R_c$ is referred to as the communication range.
A sensor node can transfer data outside if and only if there exists a path from the sensor to the AP.
The path consists of a sequence of sensor nodes where each hop distance is smaller than the communication range, $R_c$.
Sensor nodes that are connected to the AP construct the backbone network.
If all sensors are included in the backbone network we call the network fully connected.
Otherwise, the network is divided into several disconnected sub-graphs.
For convenience, we assume that the initial sensor deployment constructs a fully connected network, i.e., $\bP^0 = \mathcal{H}(\bP^0)$.

To evaluate the sensing uncertainty in heterogeneous MWSNs, we consider the Centroidal Vonoroi Tessellation function \cite{SD,GJ,YBZ,VD,GJICC} defined as
\vspace{-10pt}
\begin{equation}
D(\bP)=\sum_{n=1}^{N}\int_{R_n(\bP)}\eta_n\|p_n-\omega\|^2f(\omega)d\omega,
\label{distortion0}
\end{equation}
\vspace{-2pt}
where the sensing cost parameters ${\eta_n\in(0,1]}$ are constants
that depend on Sensor $n$'s characteristics and $f(\omega)$ is a density function that reflects the frequency of random events taking place over the target region.
In homogeneous MWSNs, sensors have identical parameters, i.e., $\!\eta_{\!n}\!\!=\!\!1,\forall\!n\!\in\!I_{\!\Omega}$.
Note that the sensing uncertainty is only determined by the final deployment $\bP$.

However, as explained previously, when the communication range $R_c$ is limited, some sensor nodes cannot transfer their data back to the AP.
As a result, only the sensor nodes in the backbone network can contribute to the sensing and therefore the performance should be revised as
\vspace{-5pt}
\begin{equation}
D(\bP)=\sum_{n\in\mathcal{S}\left(\bP\right)}\int_{R_n(\mathcal{H}(\bP))}\eta_n\|p_n-\omega\|^2f(\omega)d\omega,
\label{distortion1}
\vspace{-3pt}
\end{equation}
\vspace{-2pt}
The optimal partition for the performance function (\ref{distortion1}) is Multiplicatively Weighted Voronoi Diagram (MWVD) \cite{GJ}, which can be applied to both homogeneous and heterogeneous MWSNs.
The MWVD of $\Omega$ generated by $\bP$ is the collection of sets $\{V_n(\bP)\}_{n\in I_{\Omega}}$ defined by
\vspace{-2pt}
\begin{equation}
V_n\!(\bP)\!=\!\{\omega\!\in\!\Omega|\eta_n\|\omega\!-\!p_n\|^2\!\le\!\eta_m\|\omega\!-\!p_m\|^2,\forall m\!\in\!I_{\Omega}\}.
\label{MWVD}
\end{equation}
\vspace{-2pt}
In particular, the MWVD for homogeneous MWSNs degenerates to the Voronoi Diagram \cite{VD}.
From now on, we use
$\bV(\bP)=\{V_n(\bP)\}_{n\in\mathcal{I}_{\Omega}}$ to replace partition $\bR(\bP)=\{R_n(\bP)\}_{n\in\mathcal{I}_{\Omega}}$.
Placing (\ref{MWVD}) back to (\ref{distortion1}), we get distortion
\vspace{-2pt}
\begin{equation}
D(\bP)=\sum_{n\in\mathcal{S}\left(\bP\right)}\int_{V_n(\mathcal{H}(\bP))}\eta_n\|p_n-\omega\|^2f(\omega)d\omega.
\label{distortion}
\vspace{-3pt}
\end{equation}
\vspace{-2pt}
The same distortion can also be applied to formulate the communication energy consumption among densely deployed sensors where the $f(\cdot)$ presents the sensor density function \cite{GJcon}.

Next, we review a classic energy consumption model for the mobile sensor networks.
Since the sensor movement dominates the power consumption, we only consider the power consumption for sensor movement.
As we mentioned in Section \ref{secIntro}, the energy consumption for one-step movement is linearly related to the moving distance.
Therefore, the energy consumption for Sensor $n$ moving from $a$ to $b$ can be defined as
\vspace{-2pt}
\begin{equation}
\mathscr{E}_n(a,b) = \xi_n\|b-a\|,
\label{Eab}
\vspace{-2pt}
\end{equation}
\vspace{-2pt}
where the moving cost parameter $\xi_n$ is a predetermined constant that depends on Sensor $n$'s energy efficiency.

\section{Centralized sensor deployment with a network lifetime constraint}\label{sec:ProblemB}
In a centralized sensor deployment scenario, a fusion center or base station collects global information (all sensor locations and parameters) and then computes and determines the final destinations for the sensors.
After receiving the decisions from the fusion center, sensors move to their final destinations directly.
It is self-evident that this point-to-point relocation is the most efficient strategy in terms of energy consumption.
\vspace{-15pt}
\subsection{Problem formulation}
\vspace{-5pt}
Since sensors move to their final destinations directly, the energy consumption for Sensor $n$ is formulated as
\begin{equation}
E_n(\bP) = \mathscr{E}_n(p^0_n,p_n) = \xi_n\|p_n-p^0_n\|,
\label{individualPower}
\vspace{-2pt}
\end{equation}
where $p_n$ is Sensor $n$'s final destination.
Our main goal is minimizing the sensing uncertainty defined by (\ref{distortion}) given a constraint on the network lifetime $T$.
To guarantee a required network lifetime, each sensor should be assigned an energy threshold (or maximum movement distance) for relocation \cite{DYJH,SLH,ZHB,ASDB}.
Therefore, the corresponding constrained optimization problem, which is referred to as Problem $\mathcal{A}$, is
\vspace{-5pt}
\begin{align}
\vspace{-10pt}
& \underset{\bP}{\text{minimize}} \;\;\;\;\;\;\;\;\;\;\;\; D(\bP) \\
& \text{~~~~s.t.} \;\;\;\;\; E_n(\bP)\leq\gamma_n, n\in I_{\Omega},
\label{individualConstraint}
\vspace{-25pt}
\end{align}
where $\gamma_n$ is the maximum energy consumption on Sensor $n$.
Let ${e}_n$ be the battery energy of Sensor $n$ at the initial time and $\alpha$ (watt) be Sensor $n$'s power consumption (which is dominated by communication, sensing, and computation) after the relocation.
To ensure the network lifetime, $T$, we have $\min_n\left(e_n-E_n(\bP)\right)\ge \alpha T$, and thus $\gamma_n=e_n-\alpha T, n\in I_{\Omega}$.
\vspace{-15pt}
\subsection{The Optimal Sensor Deployment}\label{sec:opt2}
\vspace{-5pt}
\begin{lemma}
Given a fully connected initial deployment, i.e., $\bP^0 = \mathcal{H}(\bP^0)$, the optimal deployment $\bP^*$ for Problem $\mathcal{A}$ in a homogeneous MWSN 
is also fully connected, i.e., $\bP^* = \mathcal{H}(\bP^*)$.
\label{connectivity}
\end{lemma}
\vspace{-8pt}
The proof is provided in Appendix \ref{appendixL1}.

According to Lemma \ref{connectivity}, homogeneous networks will keep connectivity after optimal sensor movements.
To analyze the network connectivity, we introduce two important concepts: desired region (DR) and feasible region (FR).
Let $\mathcal{I}\subseteq\mathcal{I}_{\Omega}$ be an arbitrary sensor set.
For convenience, the sensors in $\mathcal{I}$ and $\mathcal{I}_{\Omega}-\mathcal{I}$ are referred to as internal and external sensors.
For each sensor, $n$, the set of all locations of $n$ that result in a connected $\mathcal{I}$ is called the DR of $n$.
An internal sensor's DR for sensor set $\mathcal{I}$ is defined as the region in which if the sensor is placed, the sensors in $\mathcal{I}$ are connected.
As a special case, if an internal sensor's DR is empty, the sensors in $\mathcal{I}$ cannot construct a connected network.
Without the internal sensor, $n\in\mathcal{I}$, the rest of the internal sensors, $\mathcal{I}-\{n\}$, consists of $K_n$ disjoint components: $U_{n1}(\bP, \mathcal{I}),U_{n2}(\bP, \mathcal{I}),\cdots,U_{nK_n}(\bP, \mathcal{I})$, where the sensors in each component are connected and $\bigcup_{k=1}^{K_n}U_{nk}(\bP, \mathcal{I})=\mathcal{I}-\{n\}$.
The internal sensors are connected if and only if Sensor $n$ connects to all $\{U_{nk}(\bP, \mathcal{I})\}$s.
Thus, internal sensors' DRs for set $\mathcal{I}$ are formulated as
\vspace{-5pt}
\begin{equation}
\mathbb{D}_n(\bP, \mathcal{I}) = \bigcap_{k=1}^{K_n}\left[\bigcup_{j\in U_{nk}(\bP,\mathcal{I})}\mathbb{B}\left(p_j,R_c\right)\right], \forall n\in\mathcal{I}.
\end{equation}
\vspace{-5pt}
\setlength{\footnotesep}{-2pc}
Although we represent DRs as functions of $\bP$ for convenience, Sensor $n$'s DR is in fact determined by all sensors except itself.
For an internal sensor $n\in\mathcal{I}$, the condition $p_n\in\mathbb{D}_n(\bP, \mathcal{I})$\footnote{Remark: When Sensor $n$ is placed in its DR for $\mathcal{I}$, some external sensors, $m\in\mathcal{I}_{\Omega}-\mathcal{I}$, may also connect to internal sensors.}
guarantees that all internal sensors can communicate with each other.
In particular, if the AP is included in $\mathcal{I}$, we have $\mathcal{I}\subseteq\mathcal{S}(\bP)$.
In addition, it is trivial to show that for two sensors $m, n\in\mathcal{I}$, $p_m\in\mathbb{D}_m(\bP,\mathcal{I})$ is equivalent to $p_n\in\mathbb{D}_n(\bP,\mathcal{I})$.

\begin{figure}[!t]
\centering
\subfloat[]{\includegraphics[width=2.1in]{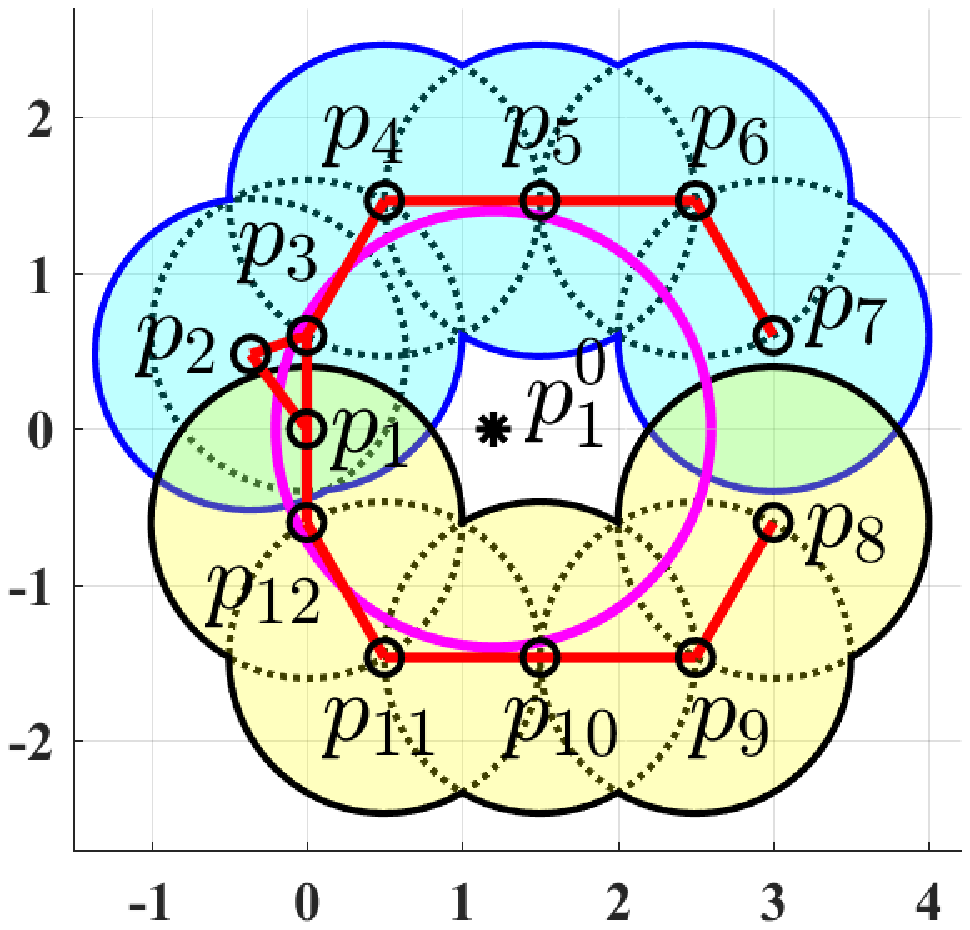}
\label{exDR}}
\hfil
\subfloat[]{\includegraphics[width=2.1in]{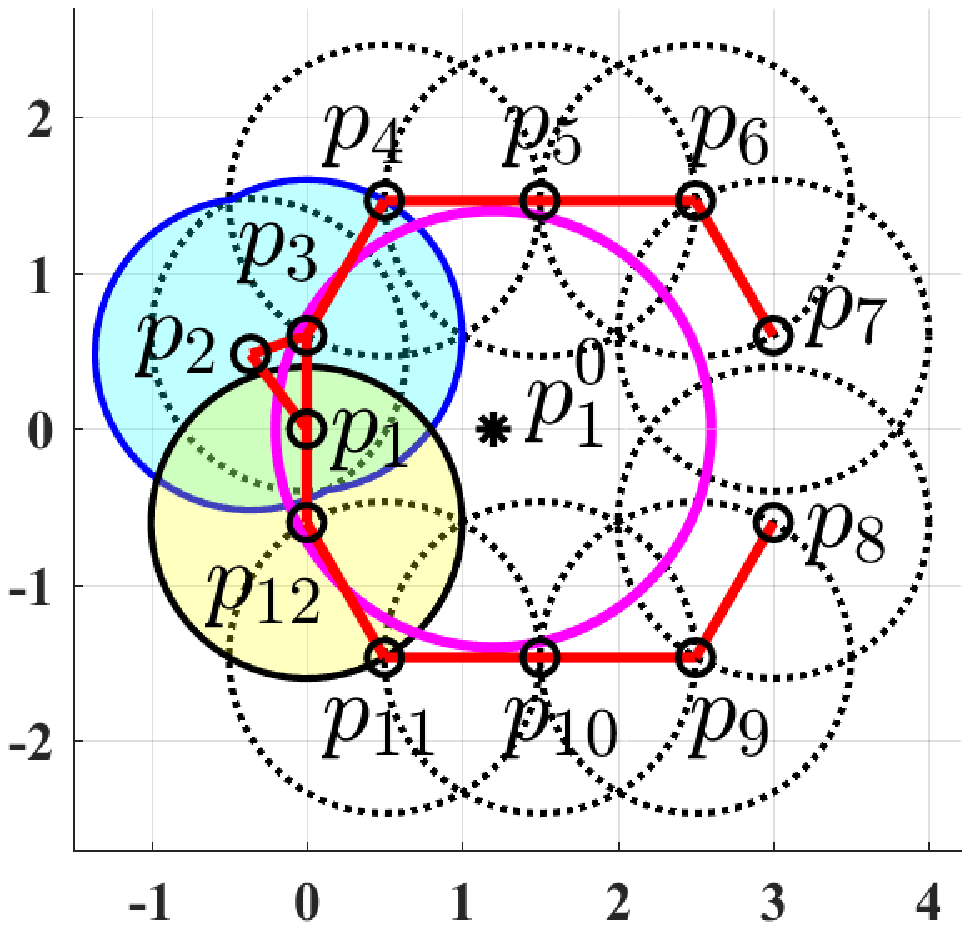}
\label{exADR}}
\hfil
\subfloat[]{\includegraphics[width=2.1in]{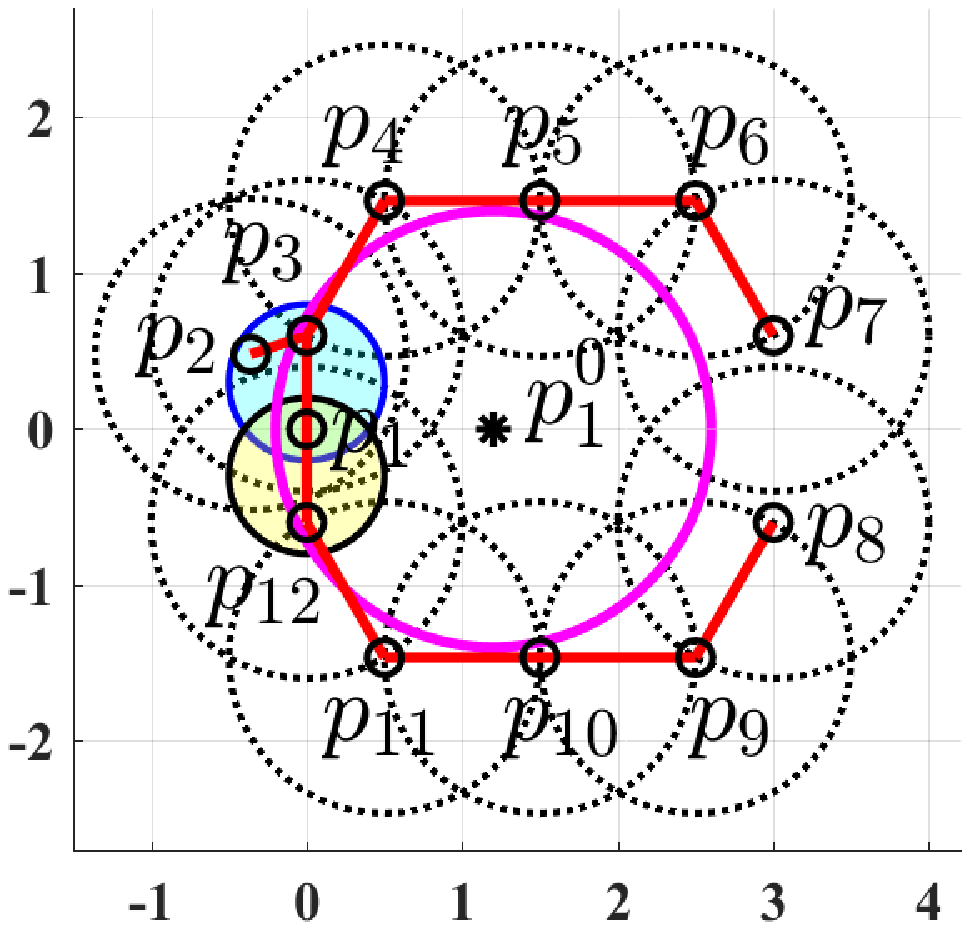}
\label{exSDR}}
\hfil
\captionsetup{justification=justified}
\caption{Example1: (a) DR and FR for Sensor 1; (b) ADR and AFR for Sensor 1; (c) SDR and SFR for Sensor 1;
DR, ADR, and SDR are shown by green. FR, AFR, and SFR are shown by the intersections of green regions and the magenta circles.
Communication ranges, movement range, and Connections are, respectively, denoted by black doted curves, magenta solid curve, and red lines.
}
\label{3Examples}
\end{figure}

An example for 12 sensors with $R_c=1$ is illustrated in Fig. \ref{exDR}.
The internal sensor set $\mathcal{I}$ is defined as all sensors, i.e., $\mathcal{I}=\mathcal{I}_{\Omega}=\{1,\dots,12\}$.
Consider $n=1$, to calculate $\mathbb{D}_1\left(\bP,\mathcal{I}_{\Omega}\right)$, the rest of the sensors are divided into $K_1=2$ components $U_{11}=\{2,3,4,5,6,7\}$ and $U_{12}=\{8,9,10,11,12\}$.
According to the definition of DR, the green overlap between the cyan region $\left[\bigcup_{j=2}^{7}\mathbb{B}\left(p_j,R_c\right)\right]$ and the yellow region $\left[\bigcup_{j=8}^{12}\mathbb{B}\left(p_j,R_c\right)\right]$ in Fig. \ref{exDR} constructs Sensor $1$'s DR, $\mathbb{D}_1\left(\bP,\mathcal{I}_{\Omega}\right)$.
Obviously, if $p_1$ is placed within $\mathbb{D}_1\left(\bP,\mathcal{I}_{\Omega}\right)$, all 12 sensors can communicate with each other.
However, if the internal sensor set $\mathcal{I}$ is defined as $\{1,4,5,6,9,10,11\}$, the corresponding DR for Sensor 1 will be empty, indicating that the sensors $\{1,4,5,6,9,10,11\}$ cannot construct a connected network.

Next, we define $\mathcal{W}\left(\bP,\mathcal{I}\right)\triangleq\bigcup_{n\in\mathcal{I}_{\Omega}-\mathcal{I}}\mathbb{B}\left(p_n,R_c\right)$.
It is self-evident that the internal sensors placed in $\mathcal{W}\left(\bP,\mathcal{I}\right)$ connect to at least one external sensor.
As a result, the sensor set $\mathcal{I}$ is the exact backbone network if and only if $1\in\mathcal{I}$ and $p_n\in \mathbb{D}_n(\bP,\mathcal{I})\bigcap \mathcal{W}^c\left(\bP,\mathcal{I}\right)$, where $\mathcal{W}^c\left(\bP,\mathcal{I}\right)=\Omega-\mathcal{W}\left(\bP,\mathcal{I}\right)$ is the complement of $\mathcal{W}\left(\bP,\mathcal{I}\right)$.
In what follows, we take the energy constraints (\ref{individualConstraint}) into account, and propose the concept of FR defined by
\vspace{-5pt}
\begin{equation}
\mathbb{F}_n(\bP, \mathcal{I})\triangleq\mathbb{D}_n(\bP, \mathcal{I})\bigcap\mathbb{B}\!\left(p^0_n, \frac{\gamma_n}{\xi_n}\right), n\in\mathcal{I},
\vspace{-12pt}
\end{equation}
where the energy constraint, $\xi_n\|p_n-p^0_n\|\le\gamma_n$, is satisfied by the condition $p_n\in\mathbb{B}\!\left(p^0_n, \frac{\gamma_n}{\xi_n}\right)$.

\setlength{\textfloatsep} {0pt plus 2pt minus 4pt}

The example of FR for Sensor 1 is illustrated in Fig. \ref{exDR}.
The magenta circle demonstrates Sensor $1$'s movement range $\mathbb{B}\left(p^0_1, \frac{\gamma_n}{\xi_n}\right)$.
Then, the intersection of green regions and the magenta circle in Fig. \ref{exDR} is Sensor $1$'s FR, $\mathbb{F}_1\left(\bP,\mathcal{I}_{\Omega}\right)$.
Obviously, if $p_1$ is placed within $\mathbb{F}_1\left(\bP,\mathcal{I}_{\Omega}\right)$, we have (a) all 12 sensors can communicate with each other and (b) Sensor 1's energy constraint is also satisfied.

Note that $p_n\in\mathbb{F}_n(\bP,\mathcal{I})$ implicitly implies that $\mathbb{F}_n(\bP,\mathcal{I})\ne\emptyset$.
Accordingly, the set of deployments that not only constructs the backbone network $\mathcal{S}\left(\bP\right)=\mathcal{I}$ but also satisfies the energy constraints can be formulated as
\vspace{-5pt}
\begin{equation}
\Gamma\left(\mathcal{I}\right)=\{\mathcal{\bP}|1\in\mathcal{I}, p_n\in \mathbb{F}_n(\bP,\mathcal{I})\bigcap\mathcal{W}^c\left(\bP,\mathcal{I}\right), \forall n\in\mathcal{I}\}.
\end{equation}
\vspace{-5pt}
Based on the aforementioned concepts, we propose the following necessary condition for the optimal deployment.
\begin{theorem}
Let $\bP^*\!\!=\!\!(p^*_1,\!\dots\!,p^*_N)$ be the optimal deployment for Problem $\mathcal{A}$.
The necessary conditions for the optimal deployment are\\
(i) $c_n(\bP^*)\notin\left[\mathbb{F}_n(\bP^*,\mathcal{S}(\bP^*))\bigcap\mathcal{W}\!\left(\bP^*\!,\mathcal{S}(\bP^*\!)\right)\right], \forall n\in\mathcal{S}(\bP^*)$\\
(ii) $\!p^*_n\!=\!
    \begin{cases}
    \!c_n(\bP^*), &\mbox{if $c_n(\bP^*\!)\!\in\mathbb{F}_{\!n}\!\left(\bP^*\!,\mathcal{S}(\bP^*\!)\right)\bigcap\mathcal{W}^c\!\left(\bP^*\!,\mathcal{S}(\bP^*\!)\right)$}\\
    \!\arg\!\!\min\limits_{q\in\partial{\mathbb{F}_n\!(\bP^*,\mathcal{S}(\bP^*))}}\!\!\|q\!-\!c_n(\bP^*)\|, &\mbox{if $c_n(\bP^*\!)\!\in\Omega-\mathbb{F}_{\!n}(\bP^*\!,\mathcal{S}(\bP^*\!))$}
    \end{cases},
    \forall n\in\mathcal{S}(\bP^*)$

\label{PropB}
\end{theorem}
The proof is provided in Appendix \ref{appendixP1}.

According to Theorem \ref{PropB}, if $n$ is a sensor in the backbone network $\mathcal{S}(\bP^*)$, its optimal location, $p^*_n$, is either at the centroid $c_n(\bP^*)$ or on the boundary of $\mathbb{F}_n(\bP^*,\mathcal{S}(\bP^*))$.
Note that Theorem \ref{PropB} only provides the necessary conditions for the sensors in the backbone network $\mathcal{S}(\bP^*)$ because sensors that are not in backbone network make no contribution to the distortion (\ref{distortion}).
In particular, by Lemma \ref{connectivity}, all sensors in homogeneous MWSNs should be included in the backbone network, i.e., $\mathcal{S}\left(\bP^*\right)=\mathcal{I}_{\Omega}$, and therefore the necessary conditions in Theorem \ref{PropB} can be extended to all sensors in homogeneous MWSNs.
Since there are no inactive sensors in homogeneous MWSNs, we have $\mathcal{W}\!\left(\bP^*\!,\mathcal{S}(\bP^*\!)\right)=\emptyset$.
Then, the necessary conditions for the optimal deployment in homogeneous MWSNs can be refined as
\begin{equation}
\!p^*_n\!=\!
    \begin{cases}
    \!c_n(\bP^*), &\mbox{if $c_n(\bP^*\!)\!\in\mathbb{F}_{\!n}\!\left(\bP^*\!,\mathcal{I}_{\Omega}\right)$}\\
    \!\arg\!\!\min\limits_{q\in\partial{\mathbb{F}_n\!(\bP^*,\mathcal{I}_{\Omega})}}\!\!\|q\!-\!c_n(\bP^*)\|, &\mbox{if $c_n(\bP^*\!)\!\notin\mathbb{F}_{\!n}(\bP^*\!,\mathcal{I}_{\Omega})$}
    \end{cases},
    \forall n\in\mathcal{I}_{\Omega}
\end{equation}

With the help of the necessary conditions in Theorem \ref{PropB}, we design centralized Lloyd-like algorithms to find the optimal sensor deployment with a network lifetime constraint in the next subsection.
\vspace{-10pt}
\subsection{Centralized Lloyd-like Algorithms}\label{sec:CCML}
To optimize the sensor deployments in homogeneous and heterogeneous MWSNs, we propose two centralized Lloyd-like algorithms: Centralized Constrained Movement Lloyd (CCML) Algorithm and Backward-stepwise Centralized Constrained Movement Lloyd (BCCML) Algorithm.
CCML Algorithm, which is designed for homogeneous MWSNs, keeps all sensors in the backbone network.
Based on CCML Algorithm, BCCML Algorithm recursively selects the optimal sensor set to construct the backbone network for heterogeneous MWSNs.

\subsubsection{CCML Algorithm}
According to our analysis in Section \ref{sec:opt2}, Sensor $n$'s movement should be restrained within its desired region, $\mathbb{D}_n(\bP, \mathcal{I}_{\Omega})$, in order to keep full-connectivity, i.e., $\mathcal{I}=\mathcal{I}_{\Omega}$.
Since the desired region is primarily influenced by the neighboring sensor nodes, we can approximate it by
\begin{equation}
\tilde{\mathbb{D}}_n(\bP, \mathcal{I}_{\Omega})=\bigcap_{k=1}^{K_n}\left[\bigcup_{j\in U_{nk}(\bP,\mathcal{I}_{\Omega})\bigcap\mathcal{N}_n(\bP)}\mathbb{B}\left(p_j,R_c\right)\right],
\label{ADR}
\end{equation}
where $\mathcal{N}_n(\bP)$ is the set of Sensor $n$'s neighbors.
Then, FR is approximated by
\begin{equation}
\tilde{\mathbb{F}}_n(\bP, \mathcal{I}_{\Omega})=\tilde{\mathbb{D}}_n(\bP, \mathcal{I}_{\Omega})\bigcap\mathbb{B}\left(p^0_n, \frac{\gamma_n}{\xi_n}\right)
\label{AFR}
\end{equation}
Note that the approximation in (\ref{AFR}) can be calculated locally, but to calculate the exact feasible region, one needs global information.
The above two approximations are referred to as approximated desired region (ADR) and approximated feasible region (AFR).
The examples of ADR, and AFR for Sensor 1 are illustrated in Fig. \ref{exADR}.
Different from the DR shown in Fig. \ref{exDR}, the ADR only considers Sensor 1's neighbors, $\mathcal{N}_1=\{2,3,12\}$.
Thus, the green overlap between cyan region $\left[\bigcup_{j=2}^{3}\mathbb{B}\left(p_j,R_c\right)\right]$ and yellow region $\mathbb{B}\left(p_{12},R_c\right)$ in Fig. \ref{exADR} construct Sensor $1$'s ADR, $\tilde{\mathbb{D}}_1\left(\bP,\mathcal{I}_{\Omega}\right)$.
Then, the intersection of the green region and the magenta circle in Fig. \ref{exADR} is Sensor $1$'s AFR, $\tilde{\mathbb{F}}_1\left(\bP,\mathcal{I}_{\Omega}\right)$.
Note that Sensor $1$'s FR in Fig. \ref{exDR} consists of two disconnected regions while Sensor $1$'s AFR in Fig. \ref{exADR} is a connected region.

Now, we provide the details of CCML Algorithm.
Like Lloyd Algorithm, the proposed algorithm iterates between two steps:
(1) Partition optimization: Partitioning is done by MWVDs;
(2) Location optimization: each sensor moves to the closest point to its centroid $c_n(\bP)$ within $\tilde{\mathbb{F}}_n(\bP, \mathcal{I}_{\Omega})$.
\begin{theorem}
CCML Algorithm is an iterative improvement algorithm, i.e., the distortion decreases at each iteration and converges.
\label{T1}
\end{theorem}
\vspace{-10pt}
\begin{IEEEproof}
CCML Algorithm is an iterative improvement algorithm only if both steps in CCML Algorithm do not increase the distortion (\ref{distortion}) subject to the constraints (\ref{individualConstraint}).
In Section \ref{sec:model}, we have proved that MWVD is the optimal cell partition for a given deployment.
Therefore, Step (1) of CCML will not increase the distortion.
During Step (2) of CCML, the cell partition is fixed as MWVD.
In Appendix A, we show that Sensor $n$'s optimal location should minimize its distance to the centroid $c_n(\bP)$ when the cell partition is fixed.
In addition, by the analysis in Section \ref{sec:opt2}, Sensor $n$'s movement should be restricted in $\tilde{\mathbb{F}}_n(\bP, \mathcal{I}_{\Omega})$ in order to guarantee both (i) energy constraints (\ref{individualConstraint}) and (ii) full-connectivity which has been proved (in Lemma \ref{connectivity}) as a necessary condition for the optimum solution.
Accordingly, Step (2) of CCML will not increase the distortion.
Therefore, CCML Algorithm is an iterative improvement algorithm.
Furthermoer, the distortion has a lower bound 0.
As a result, the distortion of CCML Algorithm is non-increasing with a lower bound, indicating that the distortion converges.
\end{IEEEproof}
\subsubsection{BCCML Algorithm}
Sensors with low-battery energy have small energy to spend on motion, which results in small movement ranges.
To keep the connection with a low-battery node, e.g., Sensor $n$, the neighboring sensors' movements will be restricted by the limited communication range of Sensor $n$, even if Sensor $n$'s neighbors have access to large battery energy.
In this case, if Sensor $n$ is not used in the MWSN, the neighboring sensors will have more freedom to move and probably further decrease the overall distortion.
Given the current sensor set $\mathcal{I}$, sensors that decrease the distortion when removed from $\mathcal{I}$ are referred to as bottleneck sensors.
To select the optimal sensor set as our backbone network, BCCML Algorithm starts with all sensors and repeatedly eliminates the least significant bottleneck sensor in terms of reducing the distortion until no bottleneck sensor is left.
Intuitively, a bottleneck sensor $n$ should satisfy the following conditions:
(i) After eliminating $n$, the rest of sensors in $\mathcal{I}$ should be connected, i.e., $K_n=1$ and $U_{n1}(\bP,\mathcal{I})=\mathcal{I}-\{n\}$.
In other words, Sensor $n$ is a leaf node in the network.
Otherwise, the network will be divided into multiple sub-graphs after eliminating $n$, and then fewer sensors will be used in the sensing task.
(ii) Sensor $n$ should use its entire energy, i.e., $\xi_n\|p_n-p^0_n\|=\gamma_n$.
(iii) At least one of its neighbors has redundant energy, i.e., $\exists m\in\mathcal{N}_n, \xi_m\|p_m-p^0_m\|<\gamma_m$.
The above three conditions are referred to as the bottleneck criterion.
To speed up the computation, BCCML Algorithm merely eliminates bottleneck sensors satisfying the bottleneck criterion.
\vspace{-10pt}
\section{Distributed sensor deployment with a network lifetime constraint}\label{sec:ProblemC}
\subsection{Problem formulation}
In the distributed scenario, there is no fusion center, and sensors determine their own destinations.
In general, sensors are supposed to only collect neighboring information (the locations and parameters of its neighbors and itself).
As we discussed in Section \ref{sec:ProblemB}, the most energy-efficient relocation strategy is moving a sensor from its initial location to the final destination in one step.
Unfortunately, this one-step relocation strategy requires global information and cannot be implemented in a distributed scenario.
In fact, the distributed sensor relocation methods in the literature are categorized into continuous and discrete time systems \cite{BW, MAKF, JS, SD, KTMN, MLCS, FJSY, GJ,YBZ, ML, XKYLM, VNH, Erdem2}.
In the continuous time systems\footnote{Remark: The energy formulation (\ref{Eab}) works for a one-step movement where the optimal velocity and acceleration is determined by the distance \cite{GW}.
However, the movement in continuous time systems is not step-wise and the corresponding motion energy is a function of velocity \cite{ML}.
Thus, the energy model in this paper cannot be applied to continuous time systems.
The continuous sensor relocation in MWSNs is an interesting future work.}, sensors keep communicating with their neighbors during continuous movements.
Then, dynamic systems are widely used to control sensors's first order dynamics, velocity, and/or second order dynamics, acceleration \cite{ML,KTMN,MLCS}.
However, in the discrete time system, sensors only communicate with their neighbors at some discrete time instances, and their relocation is divided into multiple steps \cite{YBZ}.
Regarding the discrete nature of the relocation, the sensors should be synchronized with each other in some fashion.
Some iterative algorithms, such as Lloyd-like algorithms and virtual-force based algorithms, have been applied to this scenario \cite{Erdem1,Erdem2,GJ,YBZ,XKYLM,VNH,FJSY}.
To reduce the communication costs during the relocation, we use a discrete time system to control sensors' movements.



In what follows, we concentrate on the sensor relocation with multiple stops.
The sensor deployment at the $k$-th stop is defined by $\bP^k = (p^k_1, \dots, p^k_N)\subset\Omega^{N}$, where $p^k_n$ is Sensor $n$'s location at the $k$-th stop.
Let $K$ be the maximum number of stops (iterations) for each sensor.
For convenience, each sensor is extended to have $K$ stops.
For a sensor with $J$ physical stops, e.g. Sensor $m$, its redundant stops are extended as $p_m^k=p_m^J, \forall k\in\{J+1,\dots,K\}$.
In particular, $\bP^0 = (p^0_1, \dots, p^0_N)\subset\Omega^{N}$ and $\bP^K = (p^K_1, \dots, p^K_N)\subset\Omega^{N}$ are the initial and final deployments, respectively.
Sensor $n$'s total movement distance is $\sum_{k=1}^{K}\|p^k_n-p^{k-1}_n\|$, and therefore Sensor $n$'s individual energy consumption is formulated as
\begin{equation}
\sum_{k=1}^{K}\mathscr{E}(p^{k-1}_n,p^k) = \xi_n\sum_{k=1}^{K}\|p^k_n-p^{k-1}_n\|.
\end{equation}
Now, we discuss the distributed realization for the node deployment with (i) network lifetime constraint and (ii) limited communication range.
According to the analysis in \cite{YBZ}, the movement distance should be constrained at each iteration in order to avoid zigzag movements.
Therefore, we limit Sensor $n$'s movement distance in the $k$-th iteration by an upper bound $d^k_n$.
Note that $d^k_n$ can be a constant or a function of the previous and current deployments.
For instance, $d^k_n=\alpha\|p^{k-1}-c^{k-1}\|$ in Lloyd-$\alpha$ \cite{YBZ}, where $\alpha\in(0,1]$.
Moreover, to guarantee the required network lifetime, another constraint $\xi_n\sum_{i=1}^k\|p^{i}_n-p^{i-1}_n\|\le\gamma_n$ should be taken into account.
Furthermore, full-connectivity, which is ignored in most distributed sensor relocation algorithms, is definitely required to obtain neighboring information.
Then, another constraint $\mathcal{H}\left(\bP^k\right)=\bP^k$ should also be considered.
With the above constraints, each sensor in the distributed scenario optimizes its next stop, $\bP^k$, in terms of the previous and current neighboring information, $\bP^i, \forall i<k$.
In particular, the cell partition in Lloyd-like algorithms is generated by the current deployment \cite{Erdem2,Erdem1,GJ,YBZ,XKYLM,VNH,FJSY}, $R_n=V_n(\bP^{k-1})$.
The corresponding optimization problem, which is referred to as Problem $\mathcal{B}$, is thus represented as
\begin{align}
& \underset{\bP^k}{\text{minimize}} \;\;\;\;\;\;\;\; \sum_{n=1}^{N}\int_{V_n(\bP^{k-1})}\eta_n\|p^k_n-\omega\|^2f(\omega)d\omega\label{localdistortion} \\
& \text{~~~~s.t.}\quad\quad\quad\quad\quad\quad\quad      \mathcal{H}\left(\bP^k\right)=\bP^k  \label{fullconnect}\\
& \quad\quad\quad\quad\quad\quad \|p^k_n-p^{k-1}_n\|\leq\min\left(\frac{\tilde{e}^k_n}{\xi_n}, d^k_n\right), n\in I_{\Omega} \label{DistributedIndividualConstraint2}
\end{align}
where $\tilde{e}^k_n=\gamma_n-\xi_n\sum_{i=1}^{k-1}\|p^{i}_n-p^{i-1}_n\|$ is the residual energy at the $k$-th iteration.
Since full-connectivity is guaranteed by constraint (\ref{fullconnect}), all sensors contribute to the distortion (\ref{localdistortion}).
\vspace{-5pt}
\subsection{Semi-desired Region and Semi-feasible Region}\label{sec:opt3}
\vspace{-5pt}
Before studying the optimal solution for (\ref{localdistortion}), we analyze the constraints (\ref{fullconnect}) and (\ref{DistributedIndividualConstraint2}).
In the distributed scenario, sensors are supposed to relocate simultaneously.
However, the full-connectivity strategy used in CCML and BCCML requires a one-by-one relocation scheme, which is not possible in large-scale distributed networks.
To follow full-connectivity constraint (\ref{fullconnect}) in a large-scale distributed network, we introduce another important concept, semi-desired region (SDR), which is a shrunk version of the approximated desired region $\tilde{\mathbb{D}}_n\left(\bP, \mathcal{I}_{\Omega}\right)$.
Let $\mathcal{G}(\bP)=\left(\mathcal{V(\bP)}, \mathcal{E}(\bP)\right)$ be the undirected connectivity graph comprising a set of vertices $\mathcal{V}=\{p_1,\dots,p_N\}$ and a set of edges $\mathcal{E}=\{e_{ij}\}$.
The edge cost, $w_{ij}=\|p_i-p_j\|$, is defined as the Euclidean distance between the end vertices of the edge, $e_{ij}$, and there exists an edge between $p_i$ and $p_j$ in $\mathcal{G}(\bP)$ if and only if $w_{ij}\le R_c$.
For a fully connected graph, $\mathcal{G}(\bP)$, the corresponding minimum spanning tree (MST) is defined as $\tilde{\mathcal{G}}(\bP)=\left(\mathcal{V}(\bP), {\tilde{\mathcal{E}}}(\bP)\right)$, where $\tilde{\mathcal{E}}(\bP)\subset\mathcal{E}(\bP)$ is a subset of size $|\mathcal{V(\bP)}|-1$.
Let $\mathcal{N}^s_n(\bP)=\{m|\tilde{e}_{nm}\in\tilde{\mathcal{E}}(\bP) \text{ or } \tilde{e}_{mn}\in\tilde{\mathcal{E}}(\bP)\}$ be the set of Sensor $n$'s neighbors in MST.
Then, the SDRs are defined as
\vspace{-2pt}
\begin{equation}
\mathbb{D}^{s}_n(\bP)=\bigcap_{m\in\mathcal{N}^s_n(\bP)}\mathbb{B}\left(\frac{p_m+p_n}{2},\frac{R_c}{2}\right), \forall n\in\mathcal{I}_{\Omega}.
\label{SDR}
\end{equation}
\vspace{-2pt}
An example of SDR is illustrated in Fig. \ref{exSDR}.
In this example, 12 sensors with communication range $R_c=1$ are deployed on the plane, indicating that $\mathcal{I}_{\Omega}=\{1,\dots,12\}$.
The edges in MST are denoted by red lines in Fig. \ref{exSDR}.
Also, Sensor $1$'s movement range, $\mathbb{B}\left(p^0_1,d_n\right)$, is demonstrated by a magenta circle.
According to the definition of semi-desired region, the green overlap between cyan region $\mathbb{B}\left(\frac{p_1+p_2}{2},\frac{R_c}{2}\right)$ and yellow region $\mathbb{B}\left(\frac{p_1+p_{12}}{2},\frac{R_c}{2}\right)$ in Fig. \ref{exSDR} constructs Sensor $1$'s semi-desired region, $\mathbb{D}^s_1\left(\bP\right)$.
From Figs. \ref{exDR}, \ref{exADR}, and \ref{exSDR}, it is also clear that the semi-desired region is a subset of the approximated desired region and the desired region, i.e., $\mathbb{D}^s_1(\bP)\subseteq\tilde{\mathbb{D}}_1(\bP,\mathcal{I}_{\Omega})\subseteq\mathbb{D}_1(\bP,\mathcal{I}_{\Omega})$.

\vspace{-5pt}
\begin{theorem}
Starting with a fully connected network ($\mathcal{S}(\bP^{k})=\mathcal{I}_{\Omega}$), the network is still fully connected ($\mathcal{S}(\bP^{k+1})=\mathcal{I}_{\Omega}$) if sensors simultaneously move within their respective semi-desired regions i.e., $p^{k+1}_n\in\mathbb{D}^s_n\left(\bP^{k}\right), \forall n\in\mathcal{I}_{\Omega}$.
\label{semi-connectivity}
\end{theorem}
\vspace{-15pt}
The proof is provided in Appendix \ref{appendixP3}.

\vspace{-5pt}
Then, we define the semi-feasible regions as
\begin{equation}
\mathbb{F}^{s}_n(\bP, \tilde{e}^k_n, d^k_n)=\mathbb{D}^{s}_n(\bP)\bigcap\mathbb{B}\left(p^0_n,\min\left(\frac{\tilde{e}^k_n}{\xi_n}, d^k_n\right)\right), \forall n\in\mathcal{I}_{\Omega},
\label{SFR}
\end{equation}
It is trivial to show that both  (\ref{fullconnect}) and (\ref{DistributedIndividualConstraint2}) are satisfied if sensors move within their semi-feasible regions.
Let $\mathcal{P}^k=\{\bP^k|\mathcal{H}\left(\bP^k\right)=\bP^k,\|p^k_n-p^{k-1}_n\|\leq\min\left(\frac{\tilde{e}^k_n}{\xi_n}, d^k_n\right), \forall n\in\mathcal{I}_{\Omega}\}$ be the set of deployments that follow constraints (\ref{fullconnect}) and (\ref{DistributedIndividualConstraint2}), and $\widehat{\mathcal{P}}^k=\{\bP^k|p^k_n\in\mathbb{F}^{s}_n(\bP, \tilde{e}^k_n, d^k_n), \forall n\in\mathcal{I}_{\Omega}\}$ be the set of deployments that are placed within semi-feasible regions.
Then, we have $\widehat{\mathcal{P}}^k\subseteq\mathcal{P}^k$.
To simplify the problem, we replace $\mathcal{P}^k$ by $\widehat{\mathcal{P}}^k$, and the optimization problem is represented as $N$ independent problems:
\begin{align}
& \underset{\bP^k}{\text{minimize}} \;\;\;\;\;\;\;\; \int_{V_n(\bP^{k-1})}\eta_n\|p^k_n-\omega\|^2f(\omega)d\omega\label{localdistortion2} \\
& \text{~~~~s.t.} \quad\quad\quad\quad p^k_n\in\mathbb{F}^{s}_n(\bP, \tilde{e}^k_n, d^k_n) \label{sdr2}
\end{align}
where $n\in\mathcal{I}_{\Omega}$. By parallel axis theorem, (\ref{localdistortion2}) can be rewritten as
\begin{equation}
\int_{V_n(\bP^{k-1})}\eta_n\|c^{k-1}_n-\omega\|^2f(\omega)d\omega + \eta_n\|p^k_n-c^{k-1}_n\|^2v^{k-1}_n,
\label{localdistortion3}
\end{equation}
where $v^{k-1}_n=\int_{V_n(\bP^{k-1})}f(\omega)d\omega$ and $c^{k-1}_n=\frac{\int_{V_n(\bP^{k-1})}\omega f(\omega)d\omega}{\int_{V_n(\bP^{k-1})}f(\omega)d\omega}$.
Since the first term in (\ref{localdistortion3}) is a constant, Sensor $n$'s distortion is an increasing function of the distance from $p^k_n$ to $c^{k-1}_n$.
Accordingly, the optimal solution for (\ref{localdistortion2}) with constraint (\ref{sdr2}) is the point closet to $c^{k-1}_n$ within $\mathbb{F}^{s}_n(\bP, \tilde{e}^k_n, d^k_n)$, i.e., $p^k_n=\arg\min_{q\in\mathbb{F}^{s}_n(\bP, \tilde{e}^k_n, d^k_n)}\|q-c^{k-1}_n\|$.
By moving sensors to $\arg\min_{q\in\mathbb{F}^{s}_n(\bP, \tilde{e}^k_n, d^k_n)}\|q-c^{k-1}_n\|$ at each iteration, we get a distributed realization, Distributed Constrained Movement Lloyd (DCML) Algorithm.
According to the above analysis, DCML Algorithm will result in a deployment that guarantees both connectivity and the required network lifetime.
Like CCML Algorithm, DCML Algorithm is also an iterative improvement algorithm in which the distortion is non-increasing and converges.
The proof is similar to that of Theorem \ref{T1} and therefore omitted here.
\vspace{-5pt}
\section{Extension}\label{sec:extension}
\vspace{-5pt}
In this section, we extend the proposed algorithms, CCML and DCML, to other kinds of sensing tasks: area coverage and target coverage.
We employ the binary coverage model \cite{GJICC,CCLG,BW,MAKF,JS,SD,YBZ,GJ,VD,FJSY,MLCS,KTMN,DCC,ICC,yousefizadeh1} in which Sensor $n$ can only detect the points within its sensing range ${r_n}$.
Intuitively, in order to decrease the sensing uncertainty, CCML and DCML deploy sensors into high-density regions, and thus the points with high density are more likely to be covered.
To cover the objects in different tasks, the density function $f(\omega)$ in (\ref{distortion1}) should be predetermined to highlight the points around the objects of interest.
In the following three subsections, we introduce three kinds of coverage and propose the corresponding density functions.
\vspace{-5pt}
\subsection{Area Coverage}
Without any prior information about the target region, the density function is chosen to be uniform, i.e., $f(\omega)=\frac{1}{\int_{\Omega}d\omega}, \forall \omega\in\Omega$.
Under such circumstances, maximizing the area covered by sensors is a primary task.
To evaluate the corresponding sensing performance, we employ area coverage \cite{SD,GJ,YK} (the proportion of covered area) defined by
\begin{equation}
C^{\mathcal A}(\bP)=\frac{\int_{\bigcup_{n=1}^N\mathbb{B}\left(p_n,r_n\right)}d\omega}{\int_{\Omega}d\omega}=\frac{\sum_{n=1}^{N}\int_{V_n(\bP)\bigcap \mathbb{B}\left(p_n, r_n\right)}d\omega}{\sum_{n=1}^{N}\int_{V_n(\bP)}d\omega}.
\label{CA}
\end{equation}
The experimental results in Section \ref{sec:simulation} show that CCML and DCML algorithms with uniform density function provide a large area coverage in addition to a small distortion.

\subsection{Target Coverage}
In another popular scenario, sensors are deployed to collect detailed information from the targets with known locations \cite{MRRR,ZXFG,ZJSJG,AWAE}.
Let ${\mathcal T}=\{t_1, t_2,\dots,t_M\}$ be the set of known targets, $\widehat{\mathcal T}=\{t|\min\limits_{n}\left(\frac{\|t-p_n\|}{r_n}\right)\le 1 ,t\in\mathcal T\}$ be the set of targets that covered by at least one sensor.
Then, area coverage $C^{\mathcal T}(\bP)$ - the proportion of covered target points - can be written as
\begin{equation}
C^{\mathcal T}(\bP)=\frac{\mathbf{card}(\widehat{\mathcal T})}{\mathbf{card}(\mathcal T)}=\frac{\sum_{n\in\mathcal{S}(\bP)}\int_{V_n(\bP)\bigcap \mathbb{B}\left(p_n, r_n\right)}\left[\sum_{m=1}^{M}\phi\left(\|\omega-t_m\|\right)\right]d\omega}{\sum_{n\in\mathcal{S}(\bP)}\int_{V_n(\bP)}\left[\sum_{m=1}^{M}\phi\left(\|\omega-t_m\|\right)\right]d\omega},
\label{CT}
\end{equation}
where $\mathcal\phi(\cdot)$ is the unit impulse response, $\mathbf{card}(S)$ is the cardinality of the set $S$.


To emphasize the importance of discrete targets, we model the density function as a Gaussian mixture centered at discrete targets.
The corresponding density function can be written as
\vspace{-5pt}
\begin{equation}
f(w) = \sum_{m=1}^{M}A_me^{-\frac{\|w-t_m\|^2}{r_n^2}}
\label{GM}
\vspace{-5pt}
\end{equation}
where $A_m$ reflects the comparative importance of the target $q_m$.
Similarly, CCML and DCML can also be extended to maximize barrier coverage \cite{DYJH,SLH,ZHB}.
The details are omitted to save space.
\section{Performance Evaluation}\label{sec:simulation}
We provide the simulation results for three different MWSNs: (1) MWSN1: A homogeneous MWSN in which all sensors have the same characteristics.
(2) MWSN2: A heterogeneous MWSN including sensors with different sensing and moving cost parameters.
(3) MWSN3: A heterogeneous MWSN in which four sensors are associated with low residual energies, $e_n$s.
In addition, we employ uniform density function, $f(\omega)=1$, for MWSN1 and MWSN2.
The non-uniform density function in \cite{SD,GJ} is employed for MWSN3.
The non-uniform density function is the sum of five Gaussian functions of the form $5exp(6(-(x-x_{center})^2-(y-y_{center})^2))$. The centers $(x_{center},y_{center})$ are (2,0.25), (1,2.25), (1.9,1.9), (2.35,1.25) and (0.1,0.1).
Moreover, the target region, $\Omega$, which is also the same as in \cite{SD,GJ}, is determined by the polygon vertices
{(0,0), (2.125,0), (2.9325,1.5), (2.975,1.6), (2.9325,1.7), (2.295,2.1), (0.85,2.3), (0.17,1.2)}.
Also, we set the power consumption after sensor relocation as $\alpha=1$.
As a result, the network lifetime can be calculated by $T=\min\left(e_n-E_n(\bP)\right)$, where $e_n$ is Sensor $n$'s battery energy and $E_n(\bP)$ is Sensor $n$'s energy consumed by relocation.
Other parameters for the above three MWSNs are provided in Table \ref{SPT}.
Moreover, we generate initial sensor deployments randomly, i.e., every node location is generated with uniform distribution on $\Omega$.
To guarantee the initial full-connectivity, we sequentially generate random node locations, and only keep a node if it connects with at least one previous node.
The maximum number of iterations is set to 100.
\setlength{\intextsep}{0pt plus 0pt minus 6pt}
\setlength{\textfloatsep} {0pt plus 2pt minus 6pt}
\begin{table}[!bth]
\setlength\abovecaptionskip{0pt}
\centering
\caption{Simulation Parameters}
\begin{tabular}{|c|c|c|c|c|c|c|c|c|c|}
\hline
Parameters & $\!N\!$ & $\!\!\eta_{1}\!-\!\eta_{8}\!\!$ & $\!\!\eta_{9}\!-\!\eta_{32}\!\!$ & $\!\!\xi_1\!-\!\xi_8\!\!$ & $\!\!\xi_9\!-\!\xi_{32}\!\!$ & $\!\!R_{s,1}\!-\!R_{s,8}\!\!$ & $\!\!R_{s,9}\!-\!R_{s,32}\!\!$ & $\!\!e_{1}\!-\!e_{28}\!\!$ & $\!\!e_{29}\!-\!e_{32}\!\!$\\
\hline
MWSN1 & 32 & 1 & 1 & 1 & 1 & 0.2 & 0.2 & 2 & 2\\
\hline
MSWN2 & 32 & 1 & 4 & 2 & 1 & 0.3 & 0.15 & 2 & 2\\
\hline
MSWN3 & 32 & 1 & 4 & 2 & 1 & 0.3 & 0.15 & 2 & 0.8\\
\hline
\end{tabular}
\label{SPT}
\end{table}
\vspace{-5pt}

To evaluate the performance, we compare the distortion (\ref{distortion}) and area coverage ({\ref{CA}}) CCML, BCCML, and DCML with those of VFA \cite{YY}, Lloyd-$\alpha$ \cite{YBZ}, and DEED \cite{YBZ}.
We run the algorithms for: (i) the centralized scheme where each sensor's energy consumption for relocation is determined by the distance from the initial location and the final location; and (ii) the distributed scheme where each sensor's energy consumption for relocation is determined by the total distance of its specific (100-stop) movement path.
Several important simulation details are provided as follows.
Since network lifetime is not considered in VFA \cite{YY}, it is impossible to apply the original VFA to satisfy the required network lifetime.
As we mentioned before, Sensor $n$'s energy consumption for relocation is limited by $\gamma_n=e_n-\alpha T$ to achieve the network lifetime $T$.
Thus, we propose a variant of VFA in which each sensor stops moving after the predetermined energy, $\gamma_n$, is consumed.
Furthermore, when the communication range $R_c$ is limited, the lack of full-connectivity prevents VFA, Lloyd-$\alpha$, and DEED from operating in a distributed scheme.
To compare them with our DCML Algorithm, sensors need to have global information in VFA, Lloyd-$\alpha$, and DEED.
Another issue is that sensors may be divided into multiple disconnected sub-graphs after running VFA, Lloyd-$\alpha$, and DEED because of the limited communication range.
Under such circumstances, we compute the distortions associated with different sub-graphs and report the minimum one.
In other words, we focus on the best performances that VFA, Lloyd-$\alpha$, and DEED can reach in MWSNs when communication range is limited.
Nonetheless, when we compute the distortion for our proposed algorithms, only sensors in the actual backbone network (the sub-graph including AP, i.e., Sensor 1), are taken into account, which gives our algorithm more advantage over the existing algorithms.

\begin{figure}[!t]
\centering
\subfloat[]{\includegraphics[width=2.1in]{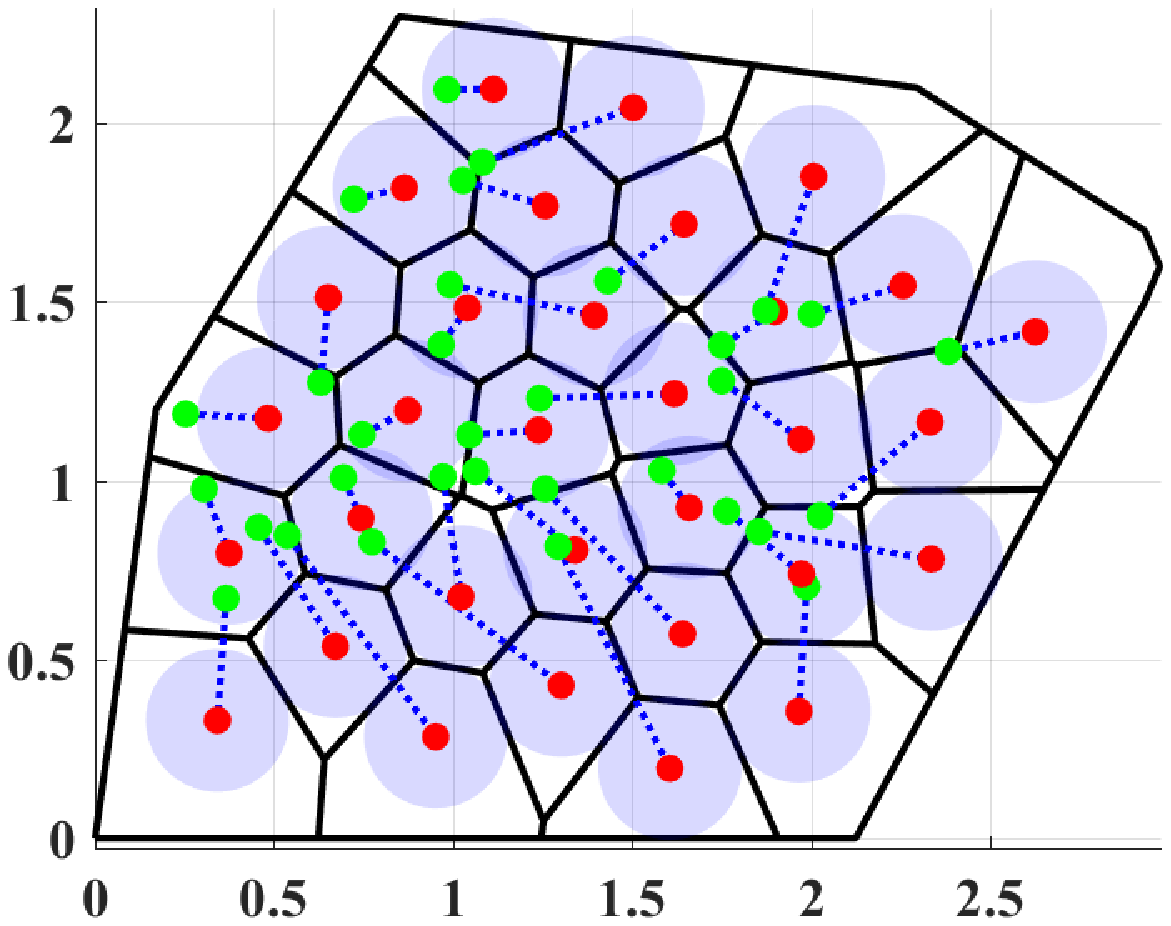}
\label{CVFADeploymentWSN1}}
\hfil
\subfloat[]{\includegraphics[width=2.1in]{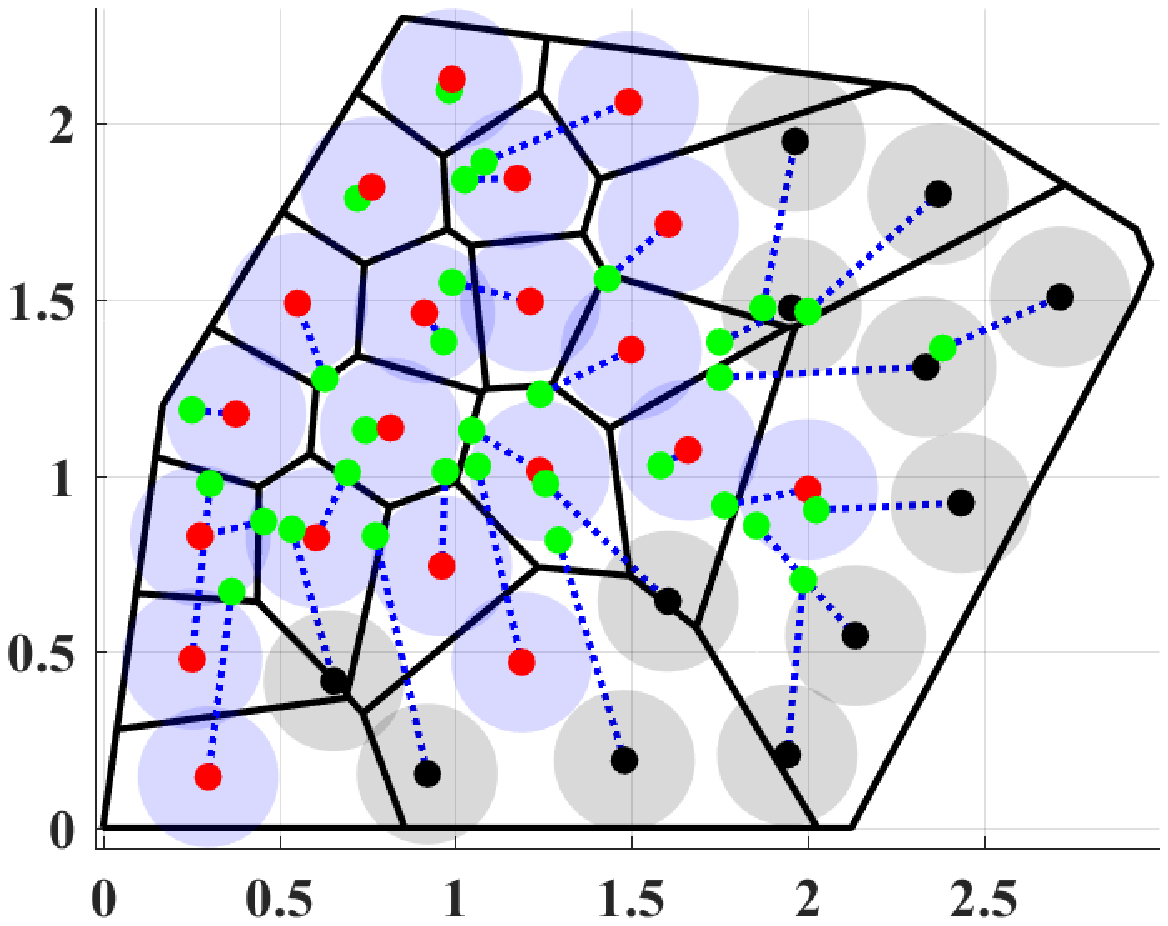}
\label{CLloydAlphaDeploymentWSN1}}
\hfil
\subfloat[]{\includegraphics[width=2.1in]{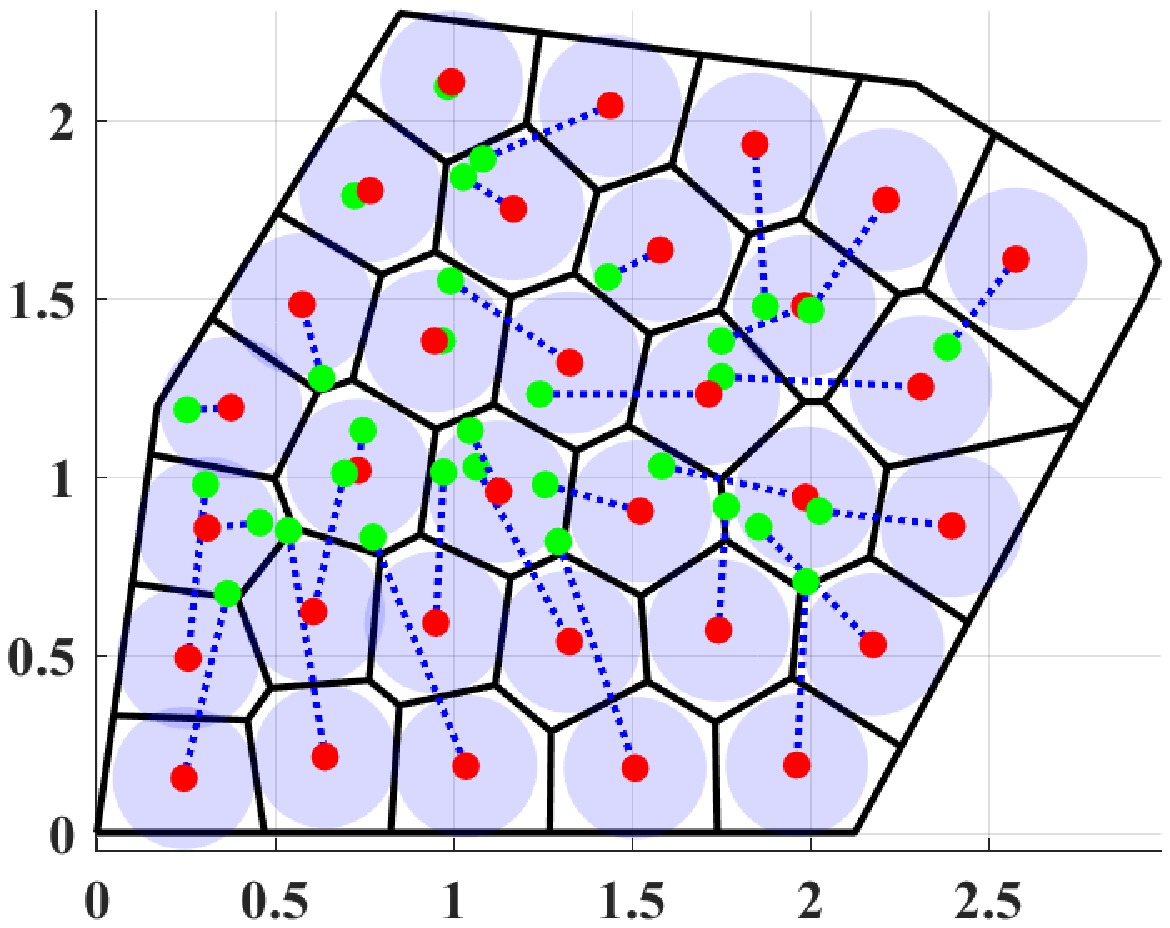}
\label{CCMLDeploymentWSN1}}
\hfil
\subfloat[]{\includegraphics[width=2.1in]{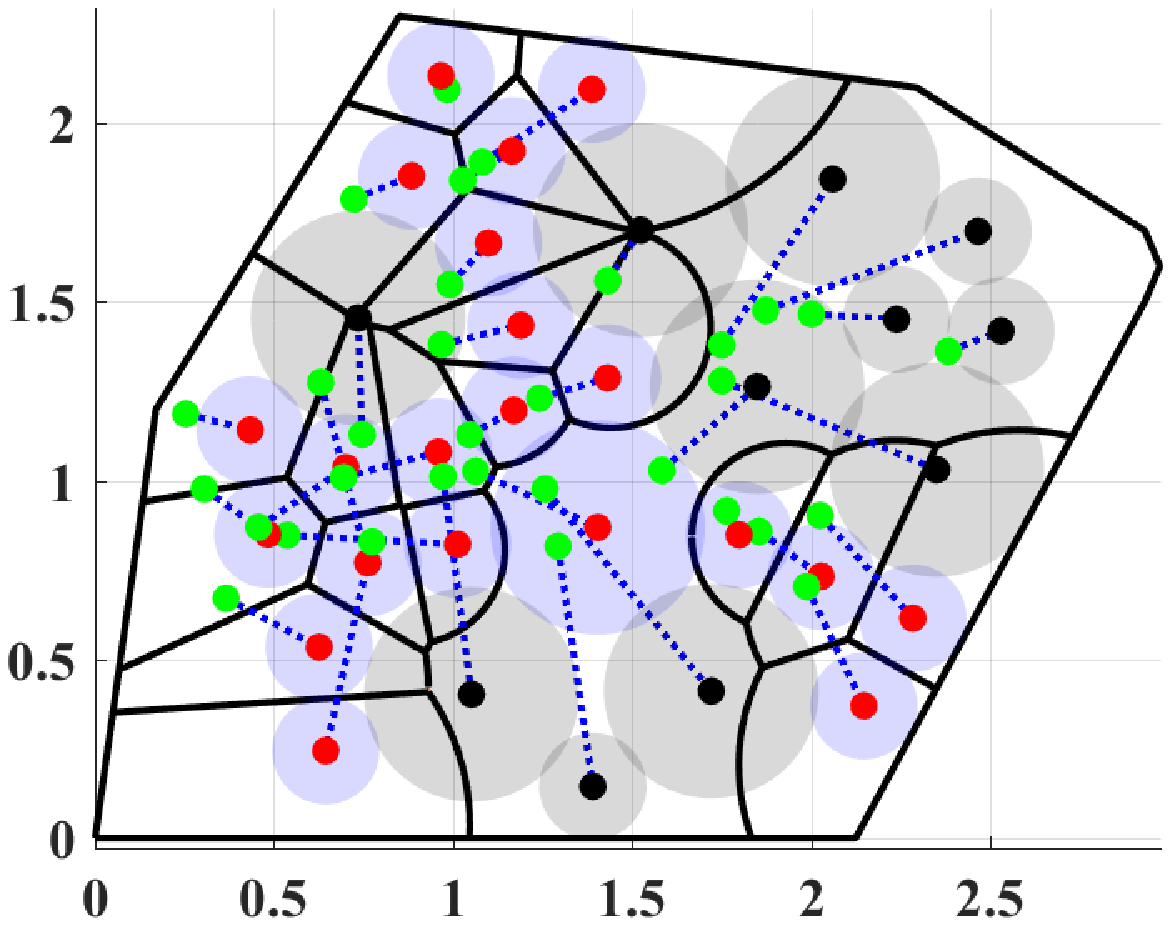}
\label{CVFADeploymentWSN2}}
\hfil
\subfloat[]{\includegraphics[width=2.1in]{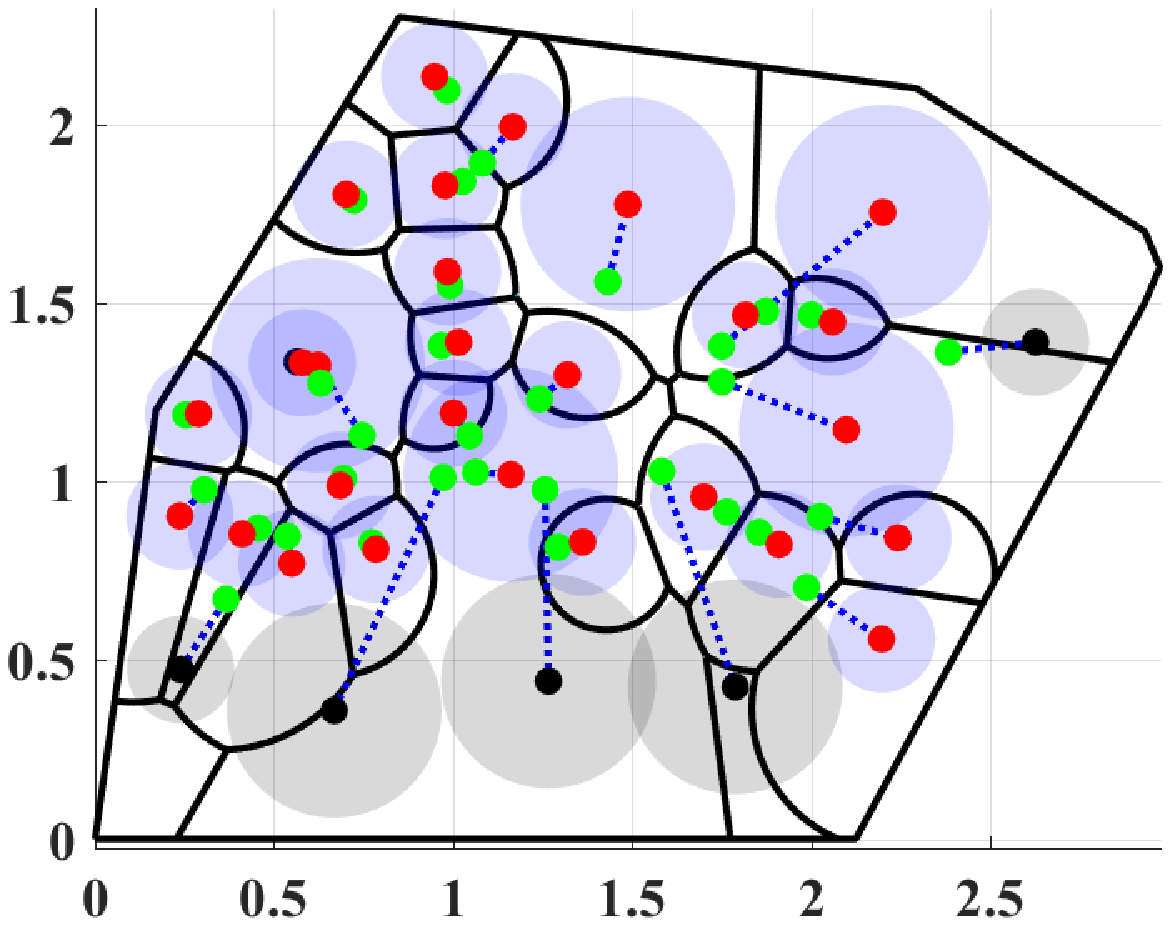}
\label{CLloydAlphaDeploymentWSN2}}
\hfil
\subfloat[]{\includegraphics[width=2.1in]{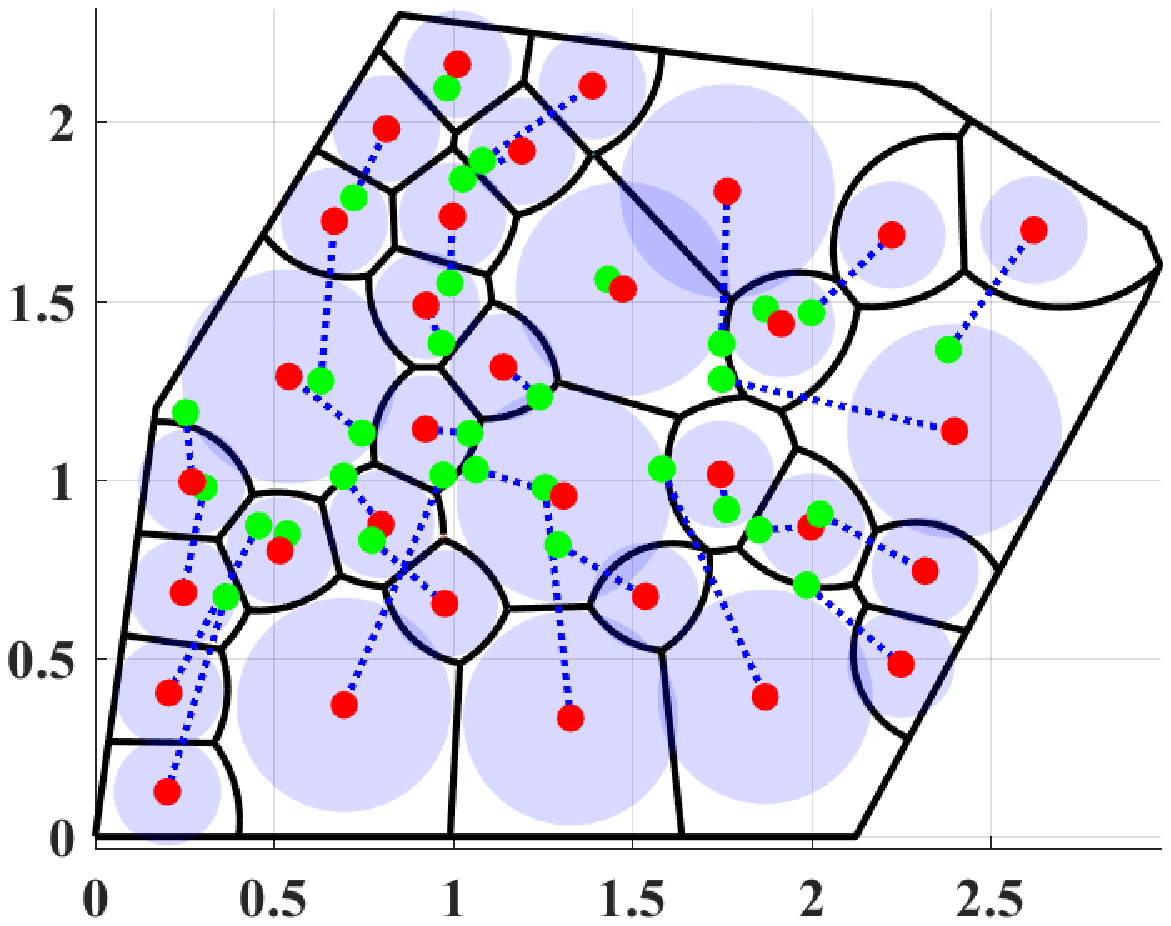}
\label{CCMLDeploymentWSN2}}
\hfil
\subfloat[]{\includegraphics[width=2.1in]{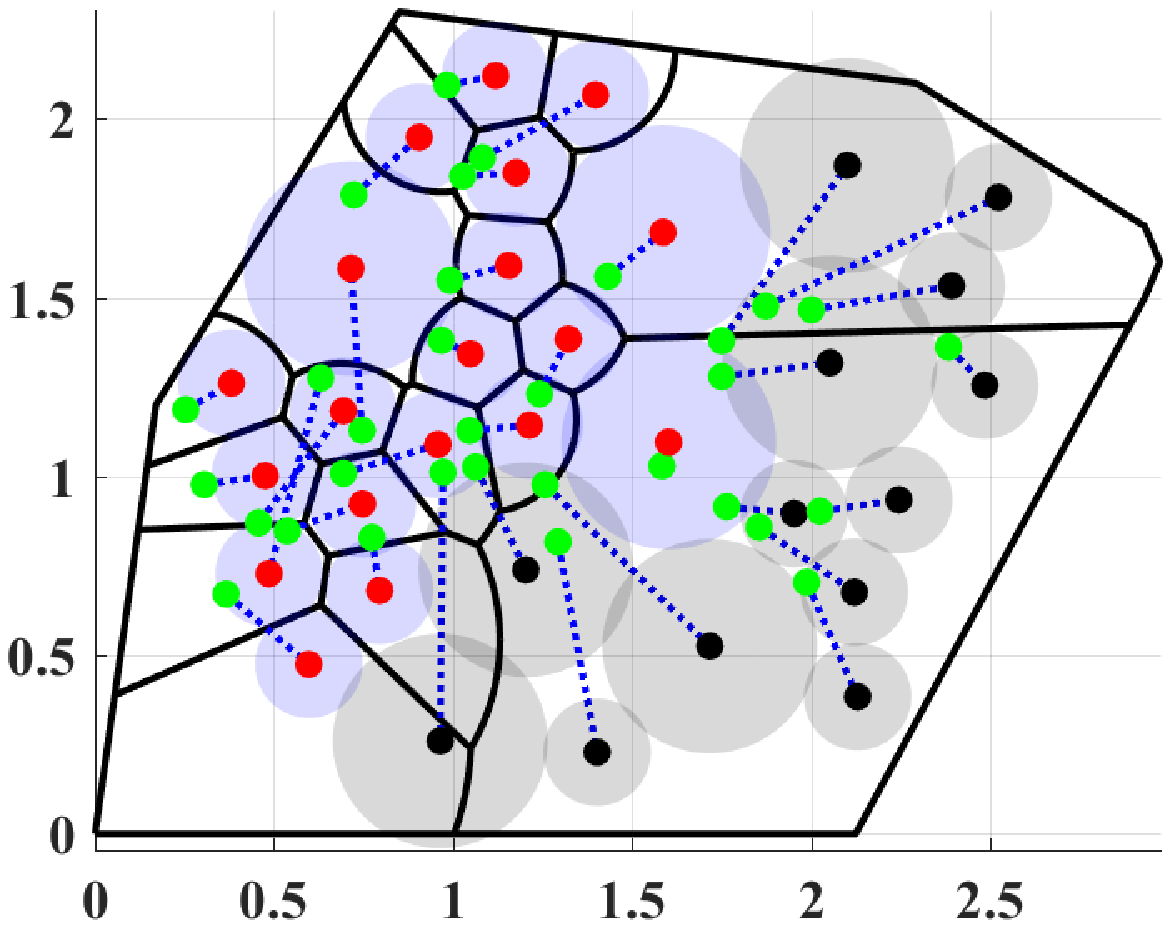}
\label{CVFADeploymentWSN3}}
\hfil
\subfloat[]{\includegraphics[width=2.1in]{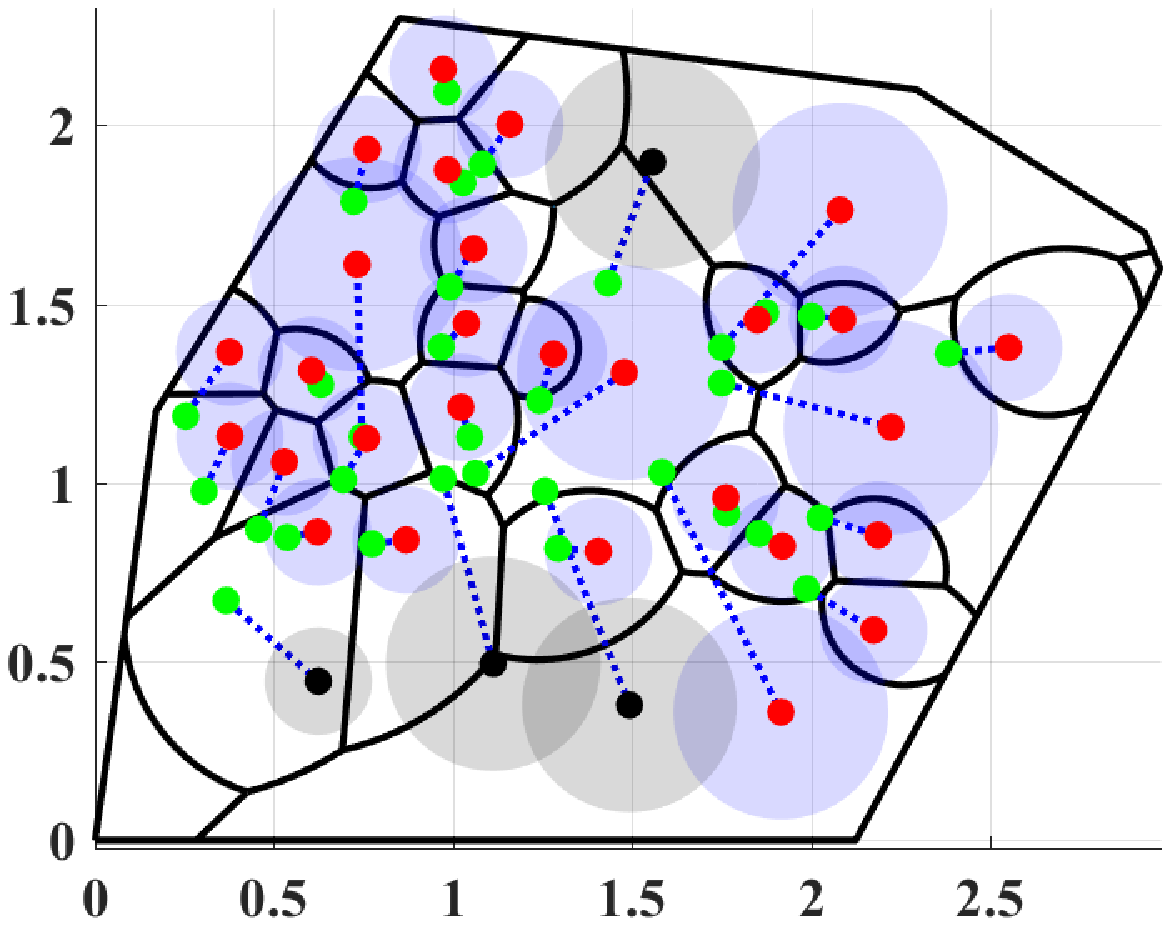}
\label{CLloydAlphaDeploymentWSN3}}
\hfil
\subfloat[]{\includegraphics[width=2.1in]{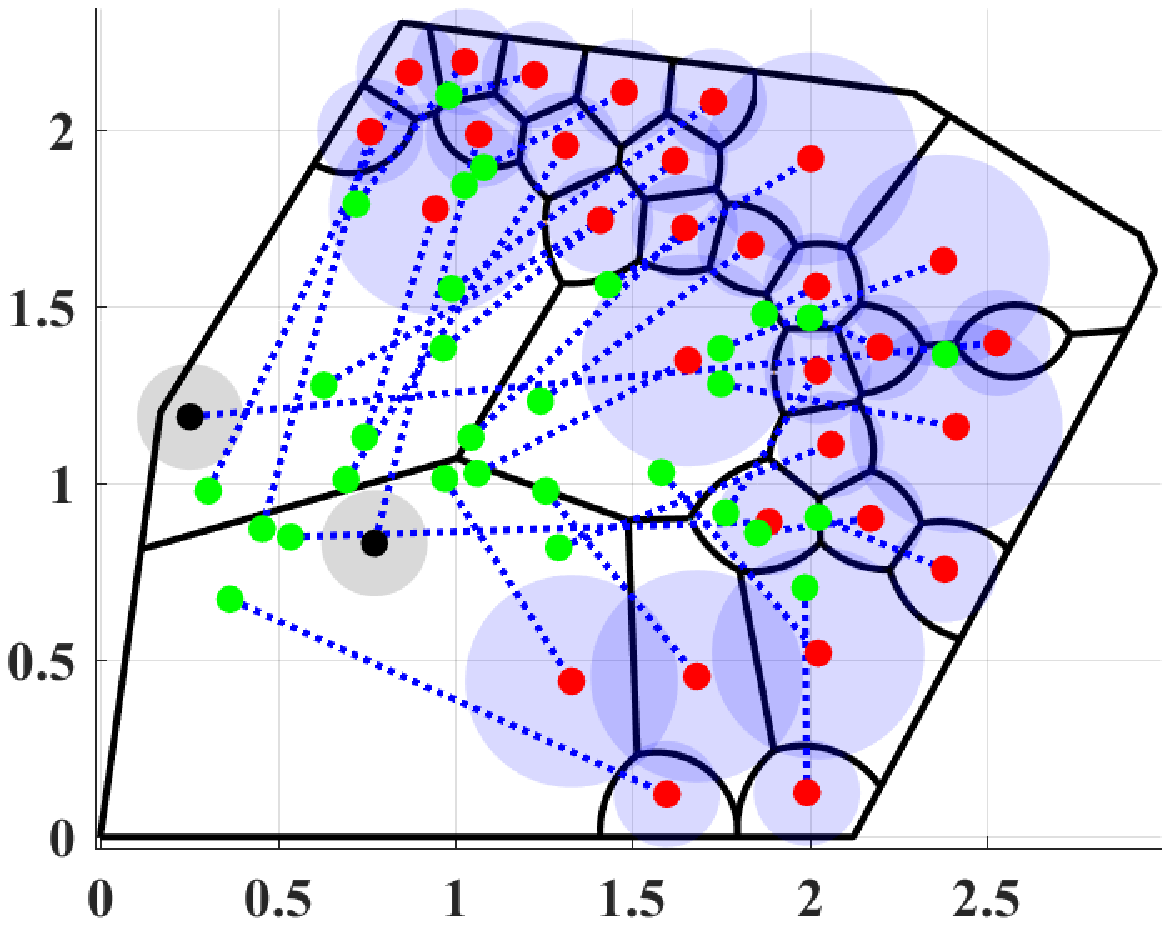}
\label{CCMLDeploymentWSN3}}
\hfil
\captionsetup{justification=justified}
\caption{Centralized sensor deployments: (a) VFA in MWSN1; (b) Lloyd-$\alpha$  in MWSN1; (c) CCML in MWSN1; (d) VFA in MWSN2; (e) Lloyd-$\alpha$ in MWSN2; (f) BCCML in MWSN3; (g) VFA in MWSN3; (h) Lloyd-$\alpha$  in MWSN3; (i)  BCCML in MWSN3. The initial sensor locations are denoted by green dots. The final locations of active and inactive sensors are denoted by red and black dots. The sensing regions of active and inactive sensors are denoted by blue and black. The movement paths are denoted by blue lines.}
\label{CentralizedDeployment}
\end{figure}

\setlength{\floatsep} {0pt plus 2pt minus 6pt}
\begin{figure}[!t]
\centering
\subfloat[]{\includegraphics[width=2.1in]{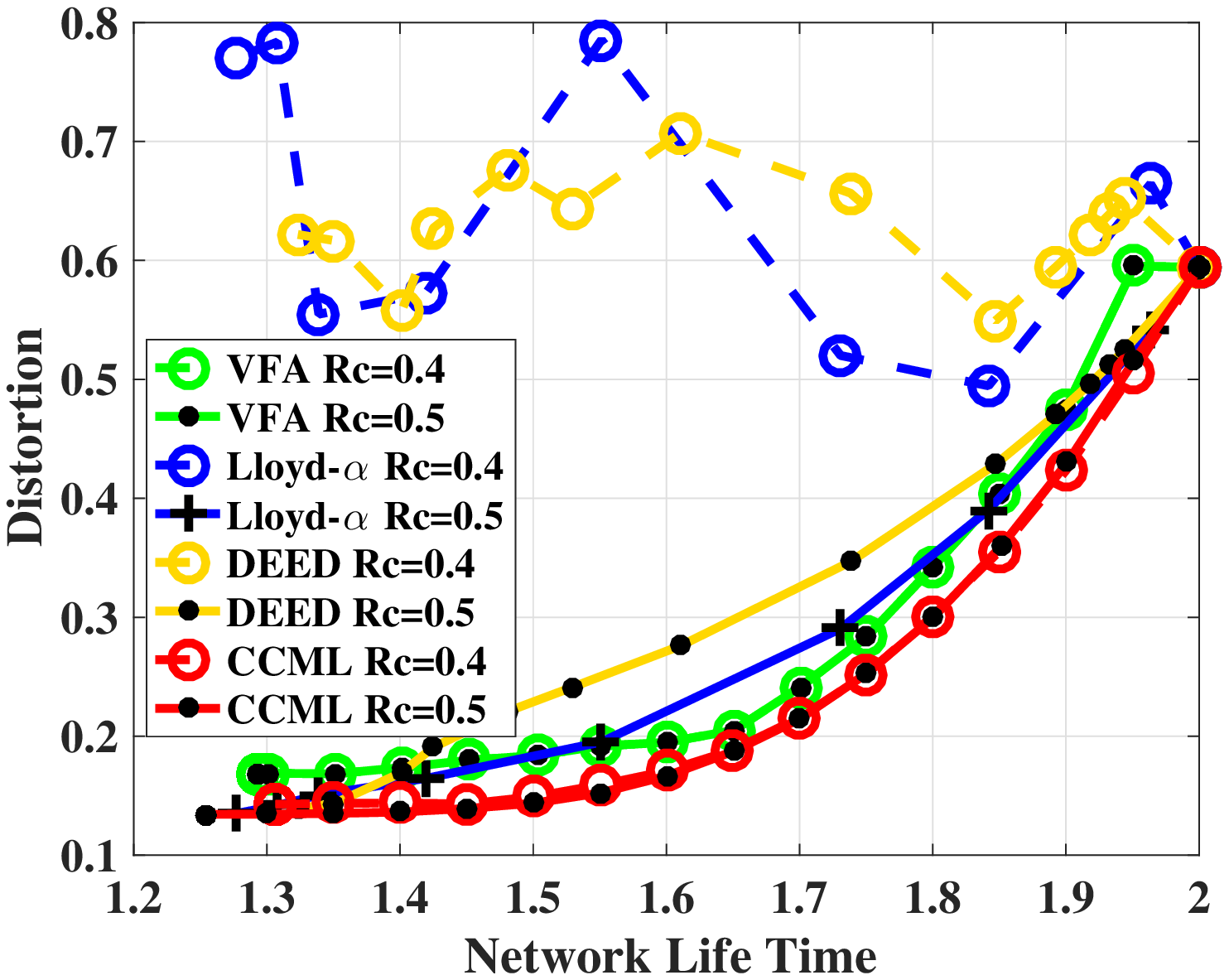}
\label{CentralizedPerformance1a}}
\hfil
\subfloat[]{\includegraphics[width=2.1in]{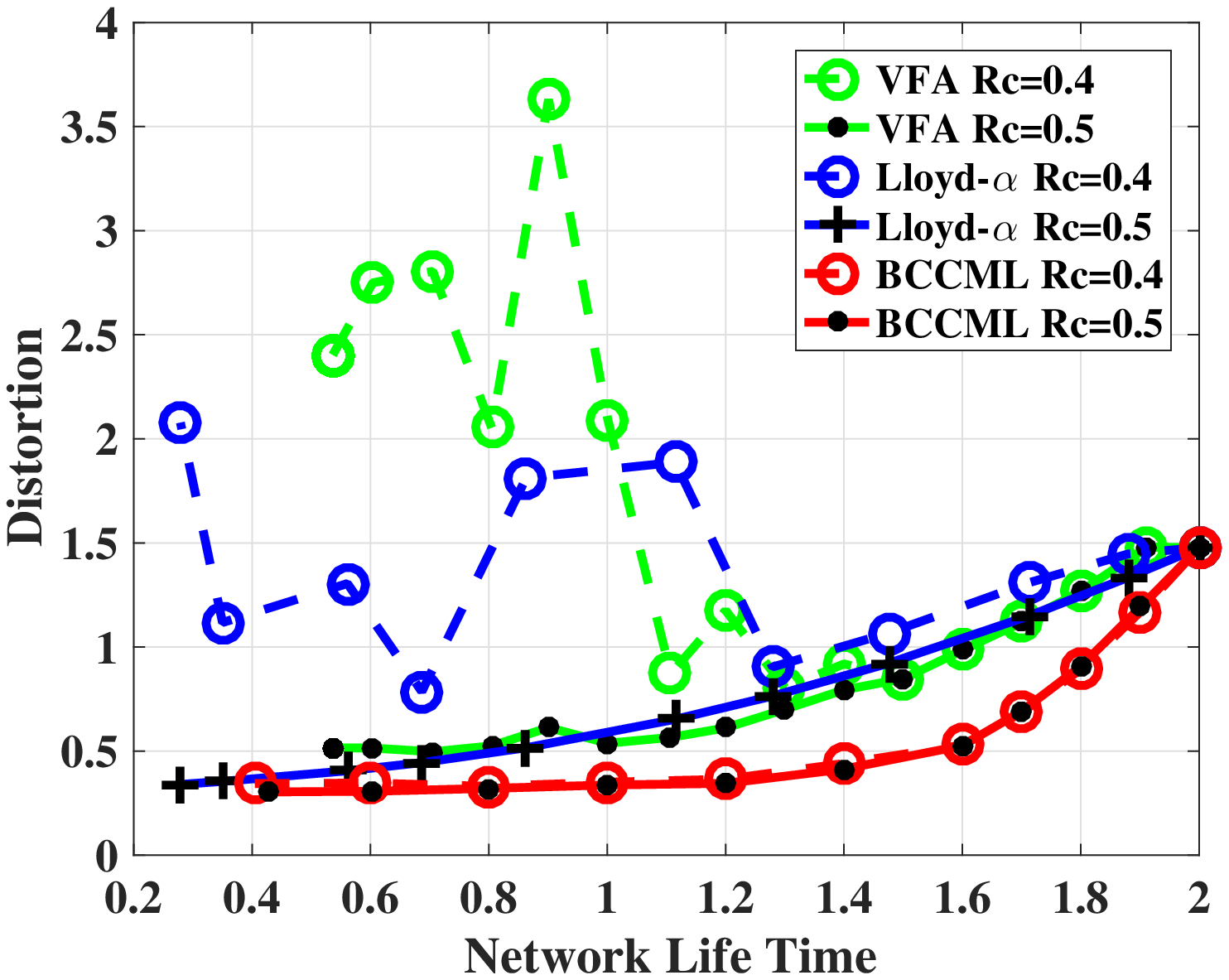}
\label{CentralizedPerformance2a}}
\hfil
\subfloat[]{\includegraphics[width=2.1in]{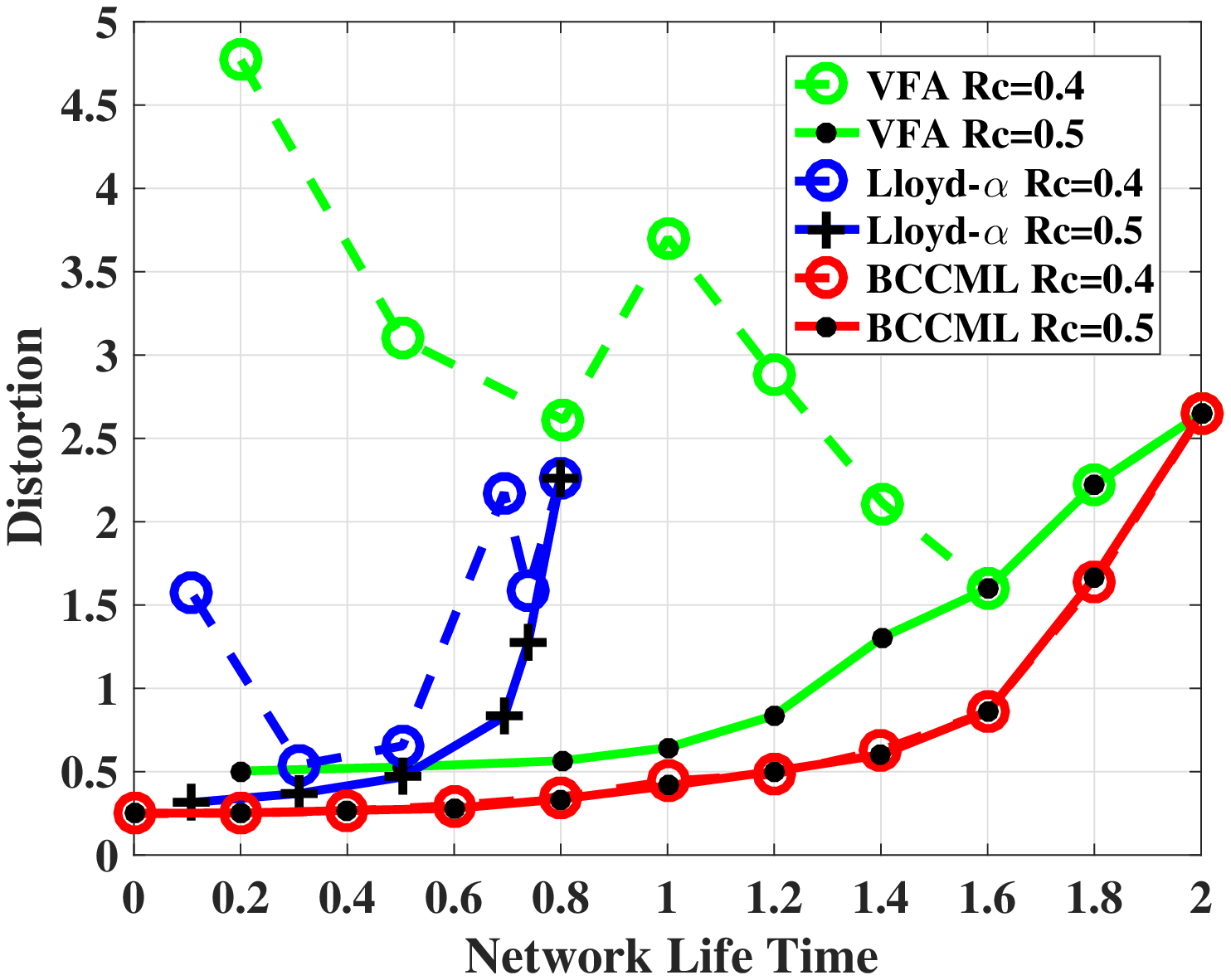}
\label{CentralizedPerformance3a}}
\caption{Distortion comparison for centralized sensor deployment. (a) MWSN1; (b) MWSN2; (c) MWSN3.}
\label{CentralizedDistortion}
\end{figure}

\begin{figure}[!t]
\centering
\subfloat[]{\includegraphics[width=2.1in]{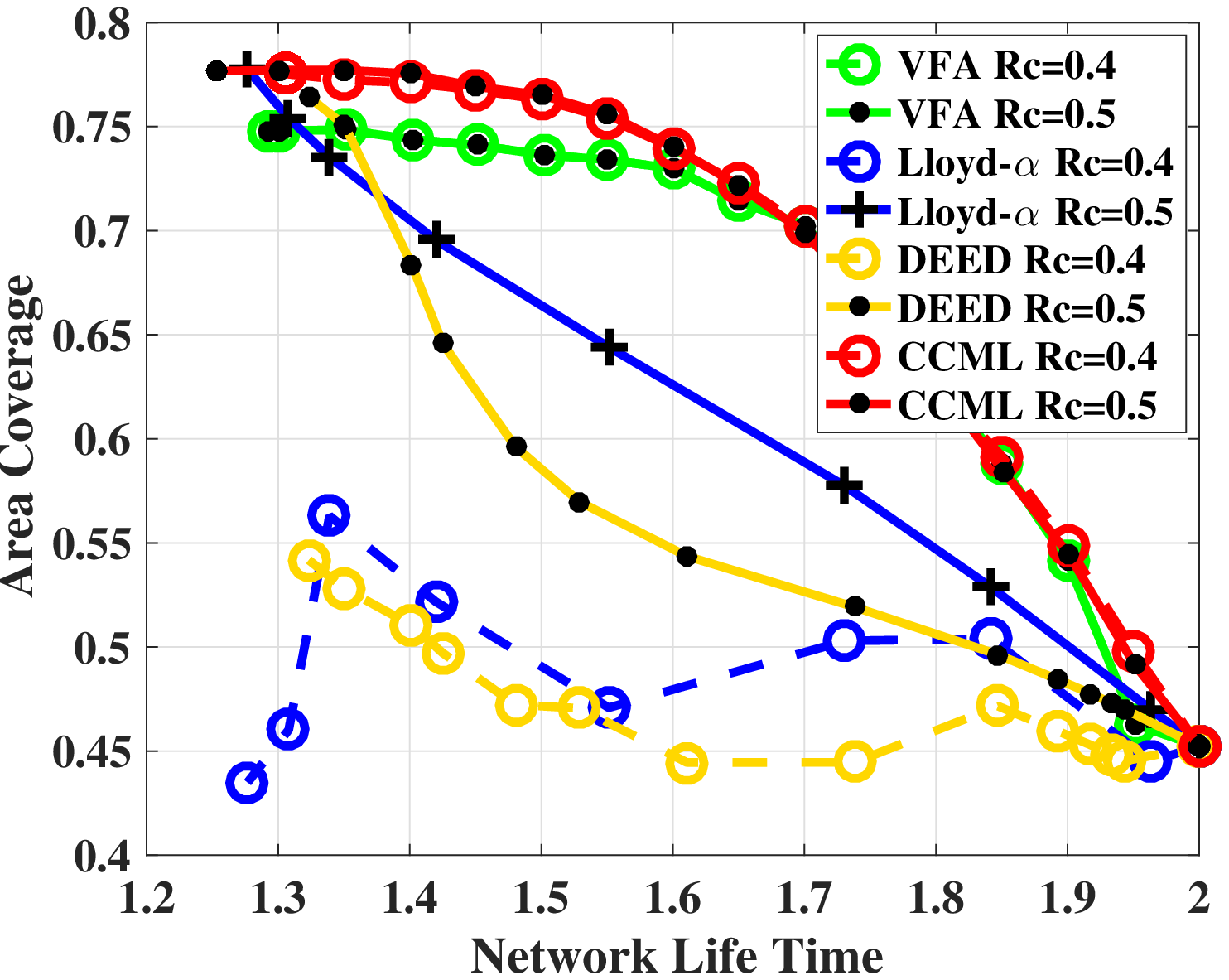}
\label{CentralizedPerformance1b}}
\hfil
\subfloat[]{\includegraphics[width=2.1in]{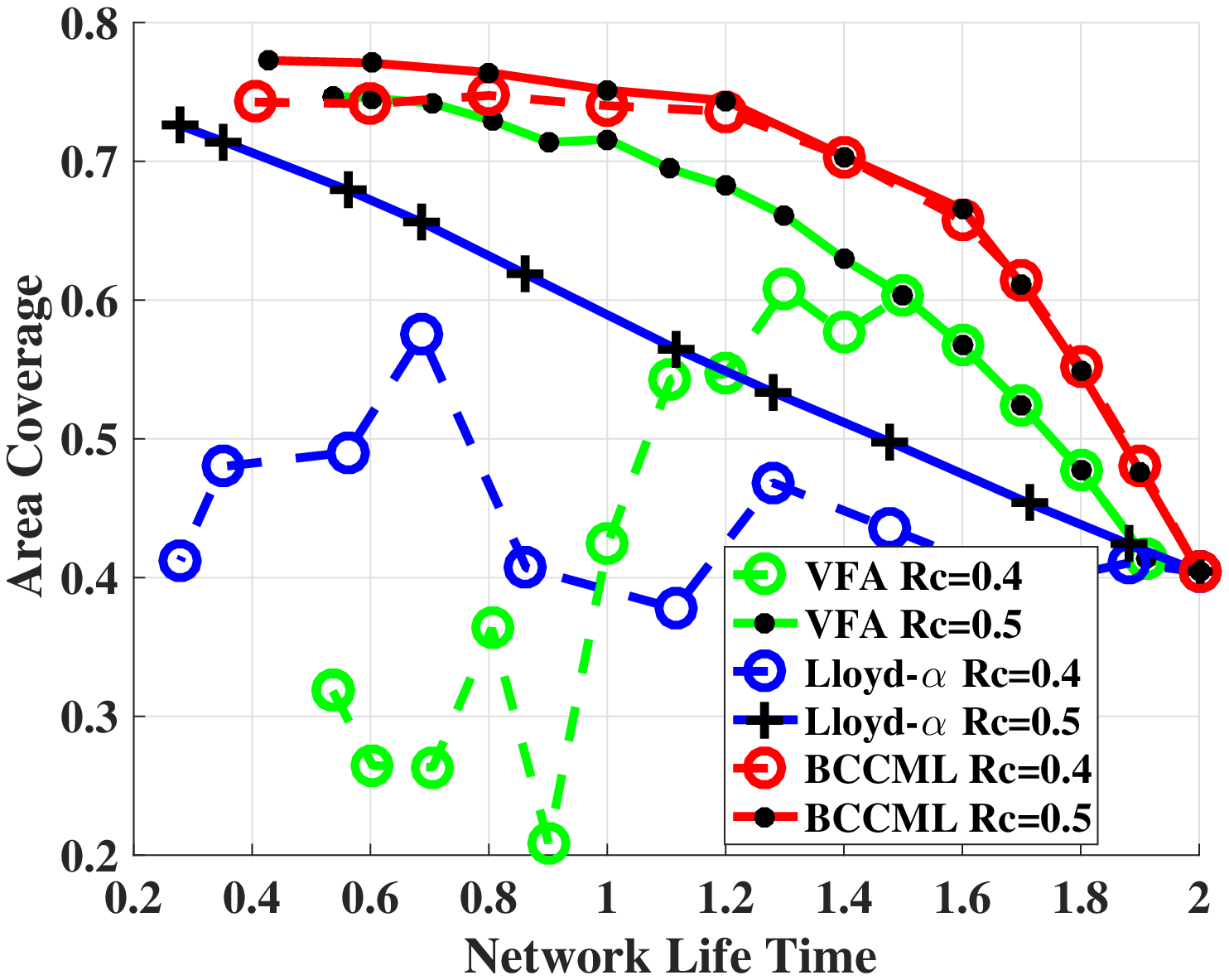}
\label{CentralizedPerformance2b}}
\hfil
\subfloat[]{\includegraphics[width=2.1in]{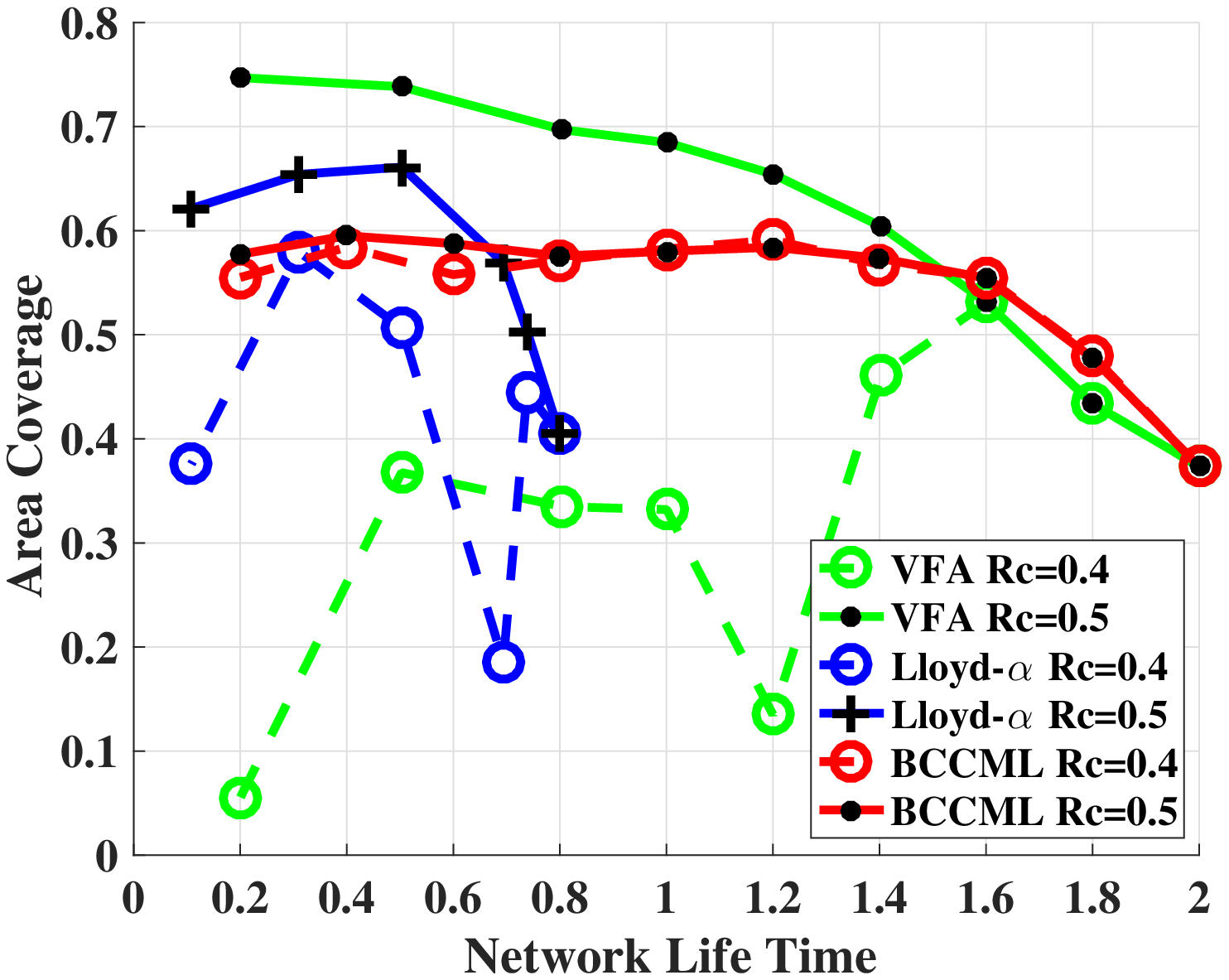}
\label{CentralizedPerformance3b}}
\caption{Area coverage comparison for centralized sensor deployment. (a) MWSN1; (b) MWSN2; (c) MWSN3.}
\label{CentralizedCoverage}
\end{figure}

Next, we analyze the time complexity of the above mentioned algorithms.
The computation of Vonoroi diagrams $V_n(\bP)$ and their centroids $c_n(\bP)$ dominate the running time.
In addition, Vonoroi partitioning is always followed by centroid computation in Lloyd-like algorithms.
Therefore, we concentrate on the number of computations of Voronoi partition.

Let $N$ be the number of sensors and $M$ be the number of iterations.
Since VFA does not rely on Vonoroi partitioning, its time complexity is just $O(1)$.
Note that Vonoroi partitioning is required after a sensor location update in Lloyd-like algorithms.
In Lloyd-$\alpha$, DEED, and DCML algorithms, sensor locations are updated simultaneously, and then Vonoroi partitioning is done once in each iteration.
Thus, the time complexity of Lloyd-$\alpha$, DEED, and DCML is $O(M)$.
However, in CCML Algorithm, sensor locations are updated sequentially, resulting in $N$ Vonoroi partitioning steps in each iteration.
Therefore, the time complexity of CCML Algorithm is $O(MN)$.

\begin{figure}[!t]
\centering
\subfloat[]{\includegraphics[width=2.1in]{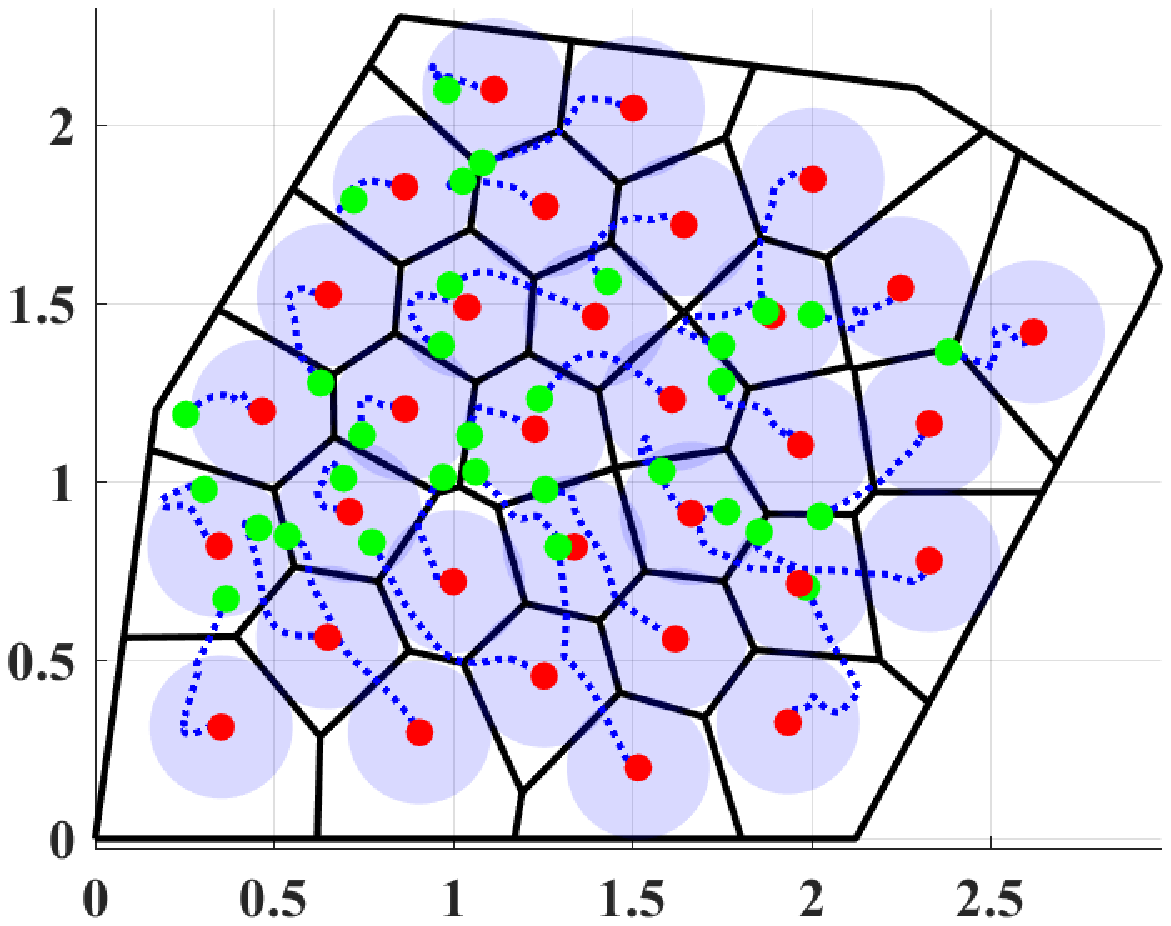}
\label{DVFADeploymentWSN1}}
\hfil
\subfloat[]{\includegraphics[width=2.1in]{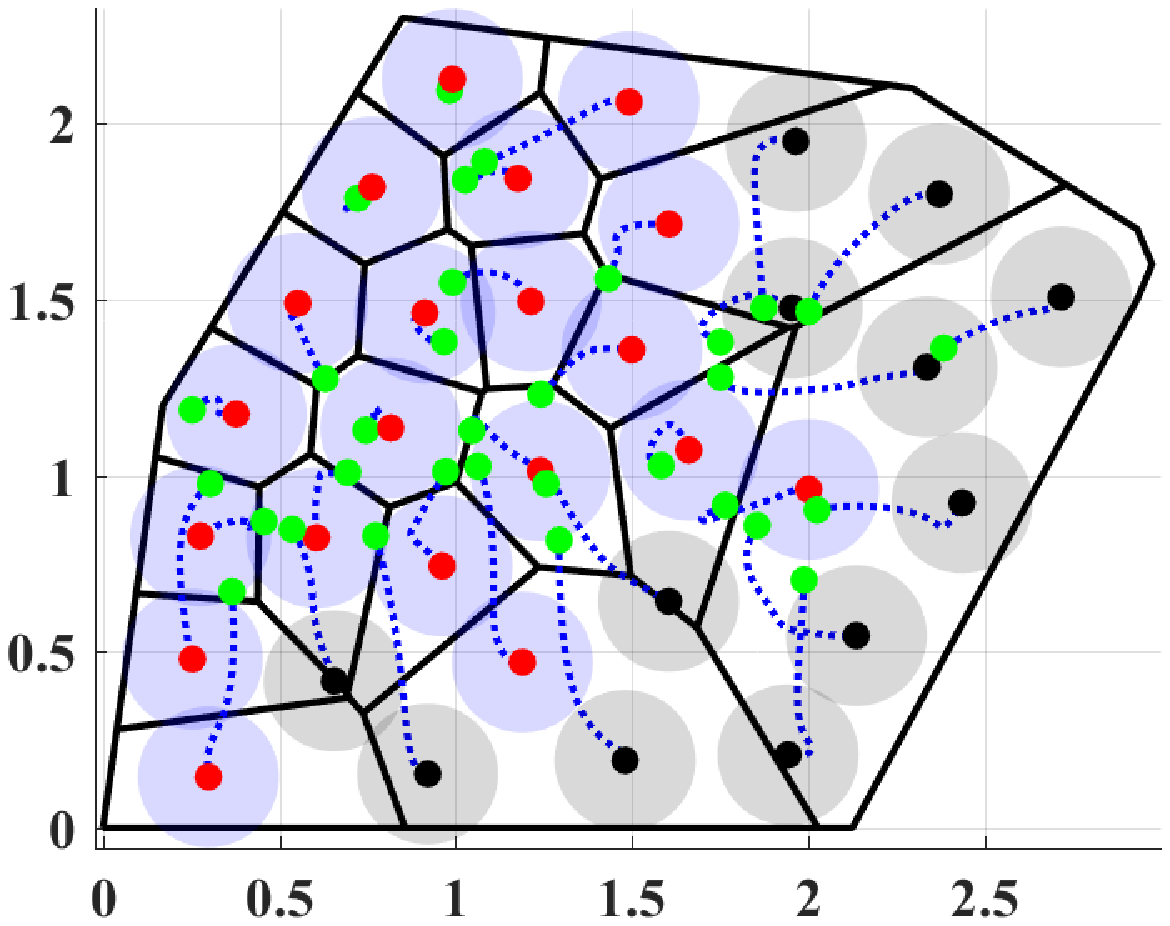}
\label{DLloydAlphaDeploymentWSN1}}
\hfil
\subfloat[]{\includegraphics[width=2.1in]{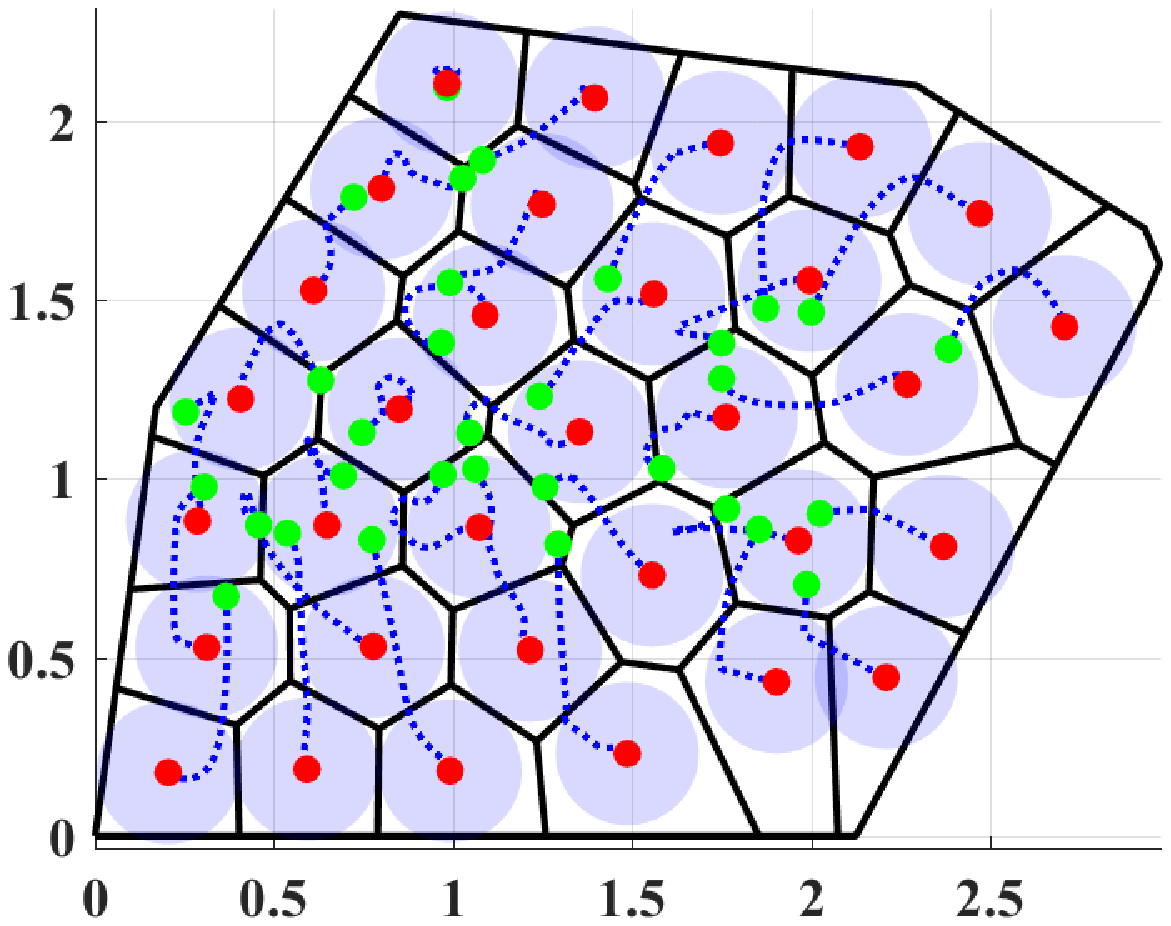}
\label{DCMLDeploymentWSN1}}
\hfil
\subfloat[]{\includegraphics[width=2.1in]{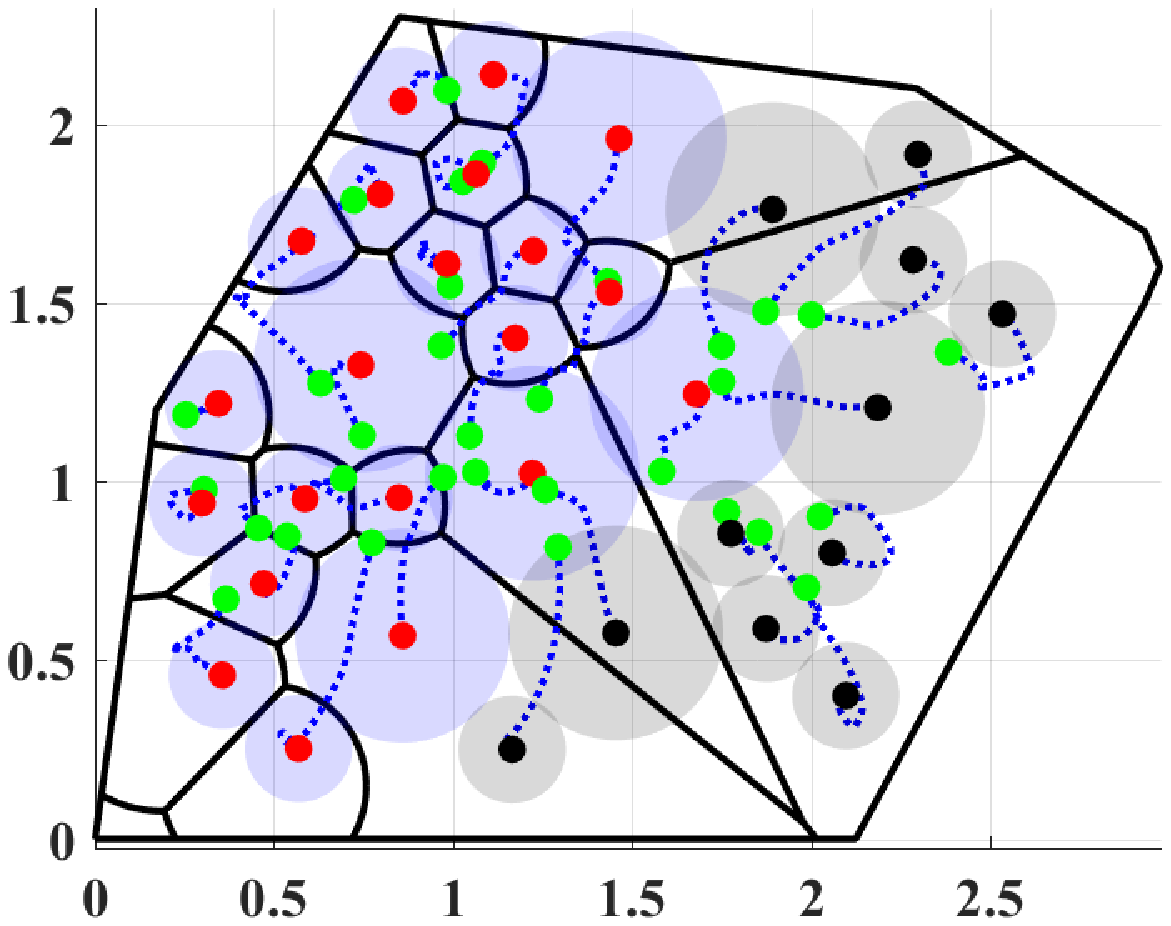}
\label{DVFADeploymentWSN2}}
\hfil
\subfloat[]{\includegraphics[width=2.1in]{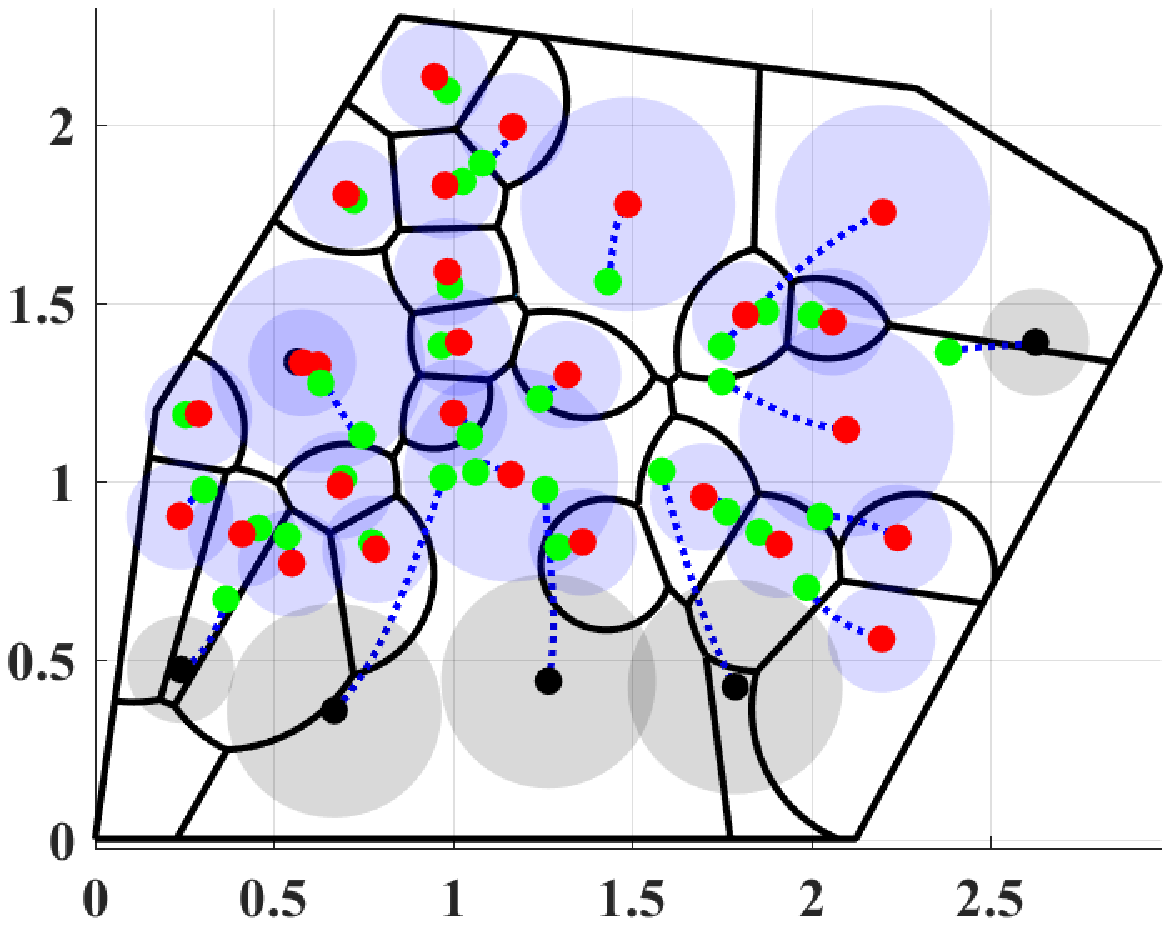}
\label{DLloydAlphaDeploymentWSN2}}
\hfil
\subfloat[]{\includegraphics[width=2.1in]{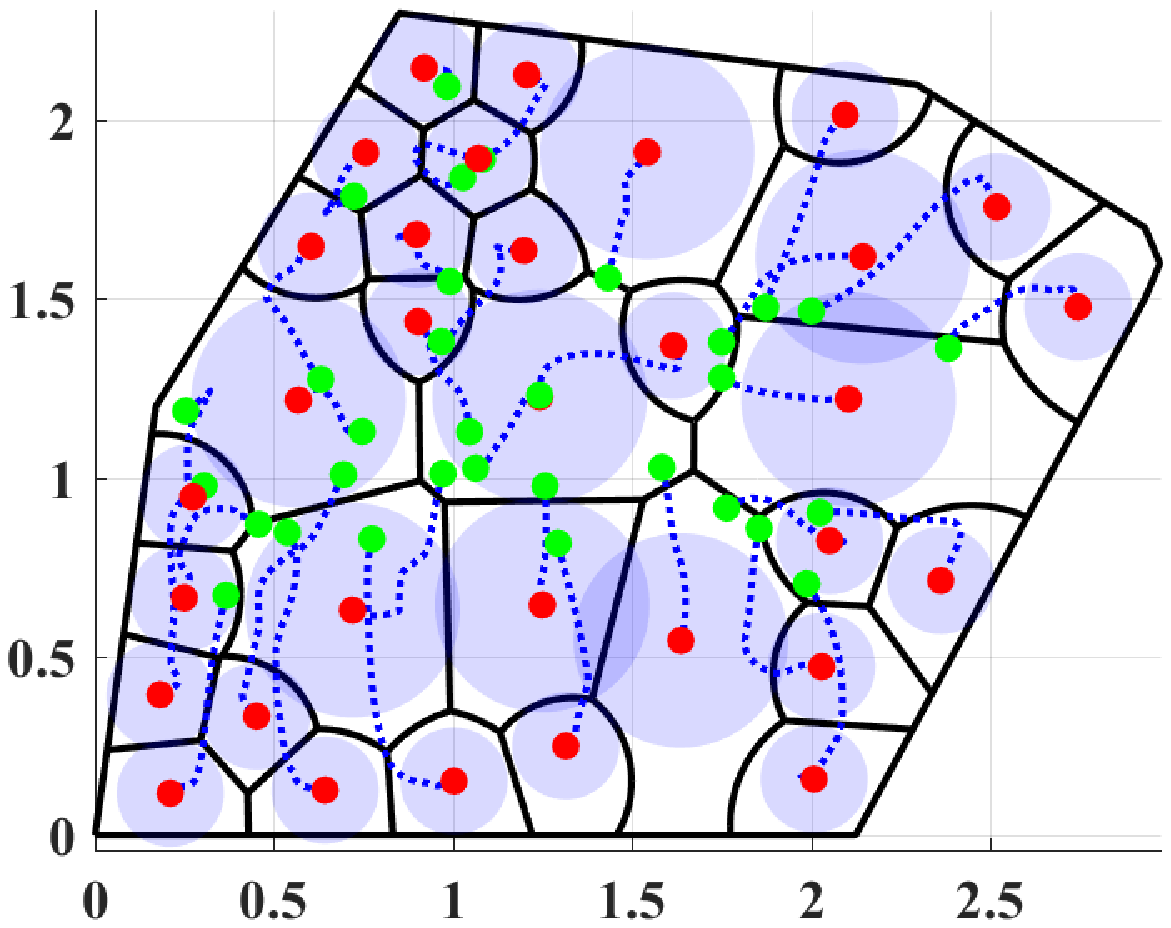}
\label{DCMLDeploymentWSN2}}
\hfil
\subfloat[]{\includegraphics[width=2.1in]{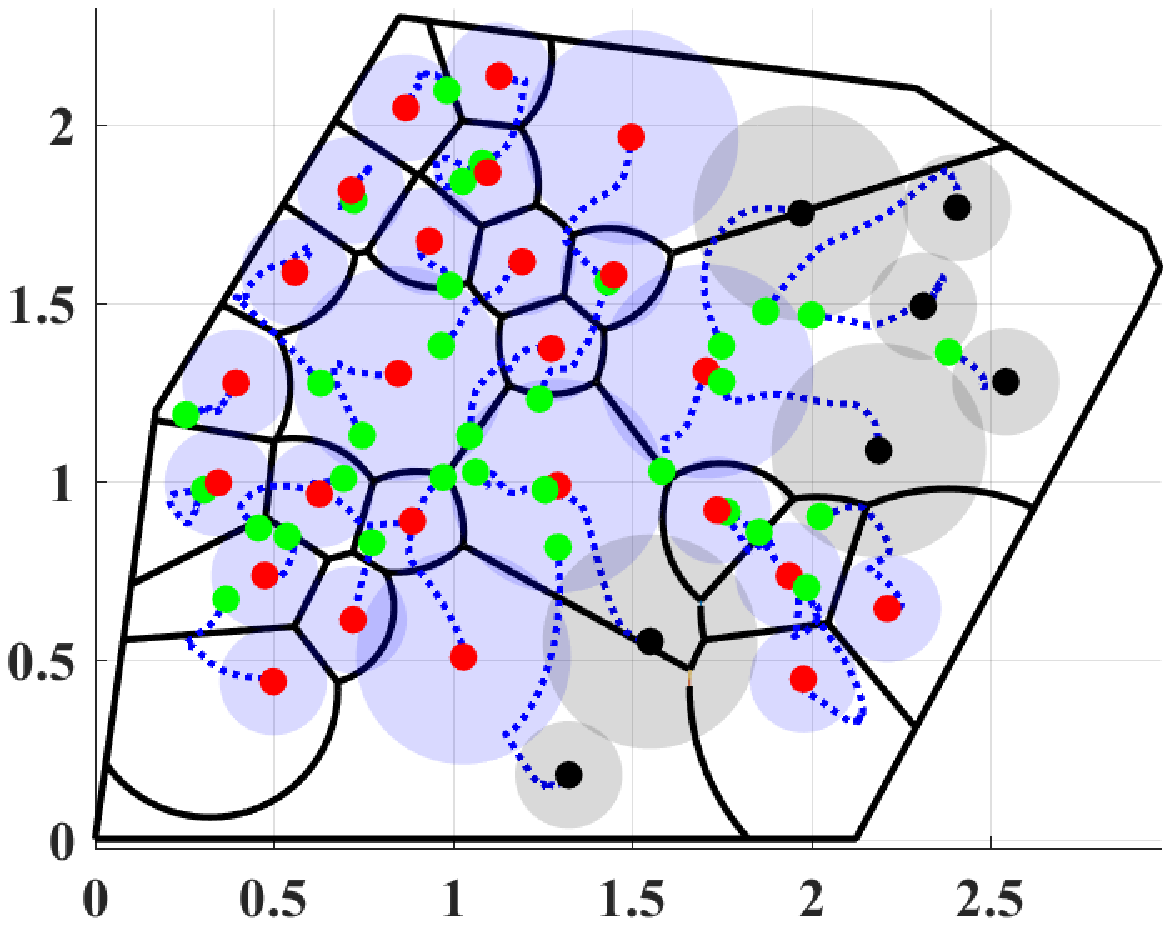}
\label{DVFADeploymentWSN3}}
\hfil
\subfloat[]{\includegraphics[width=2.1in]{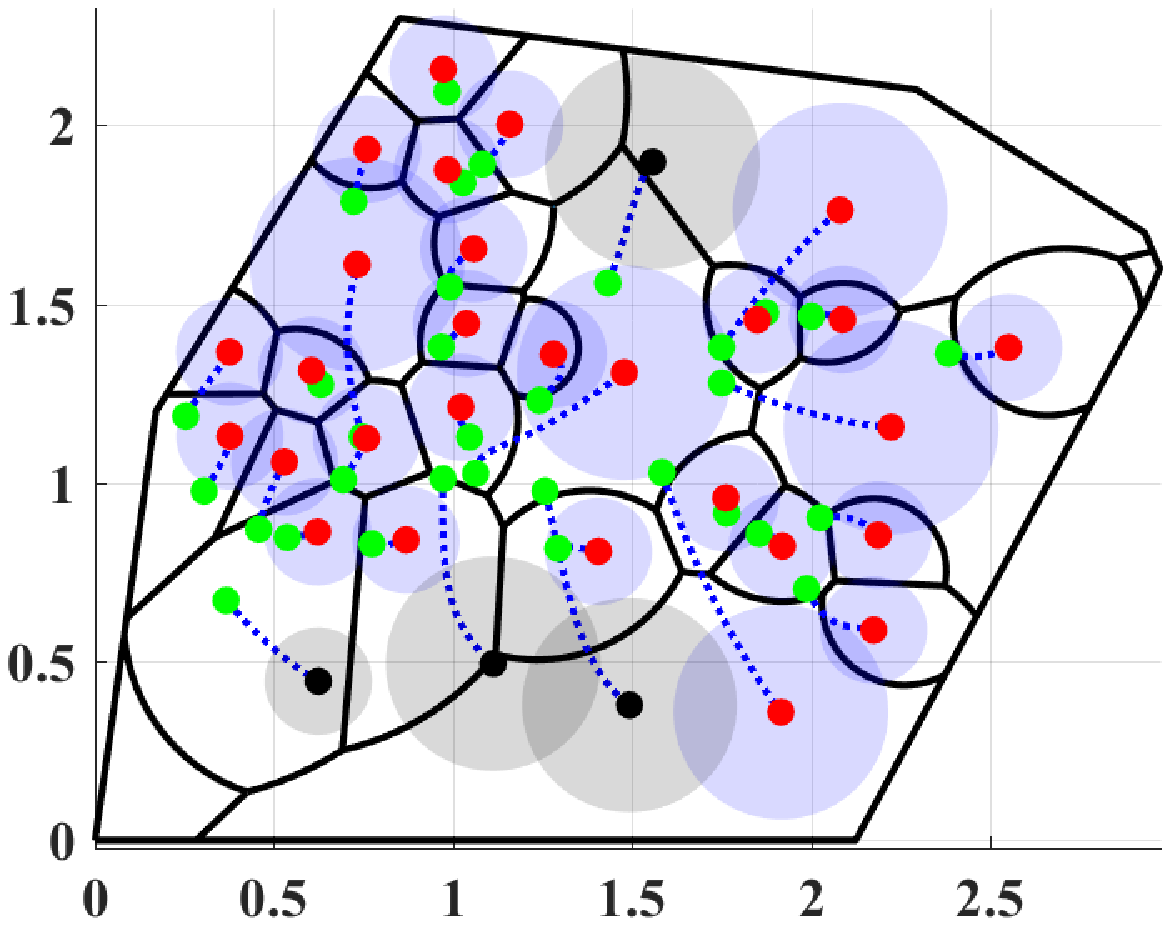}
\label{DLloydAlphaDeploymentWSN3}}
\hfil
\subfloat[]{\includegraphics[width=2.1in]{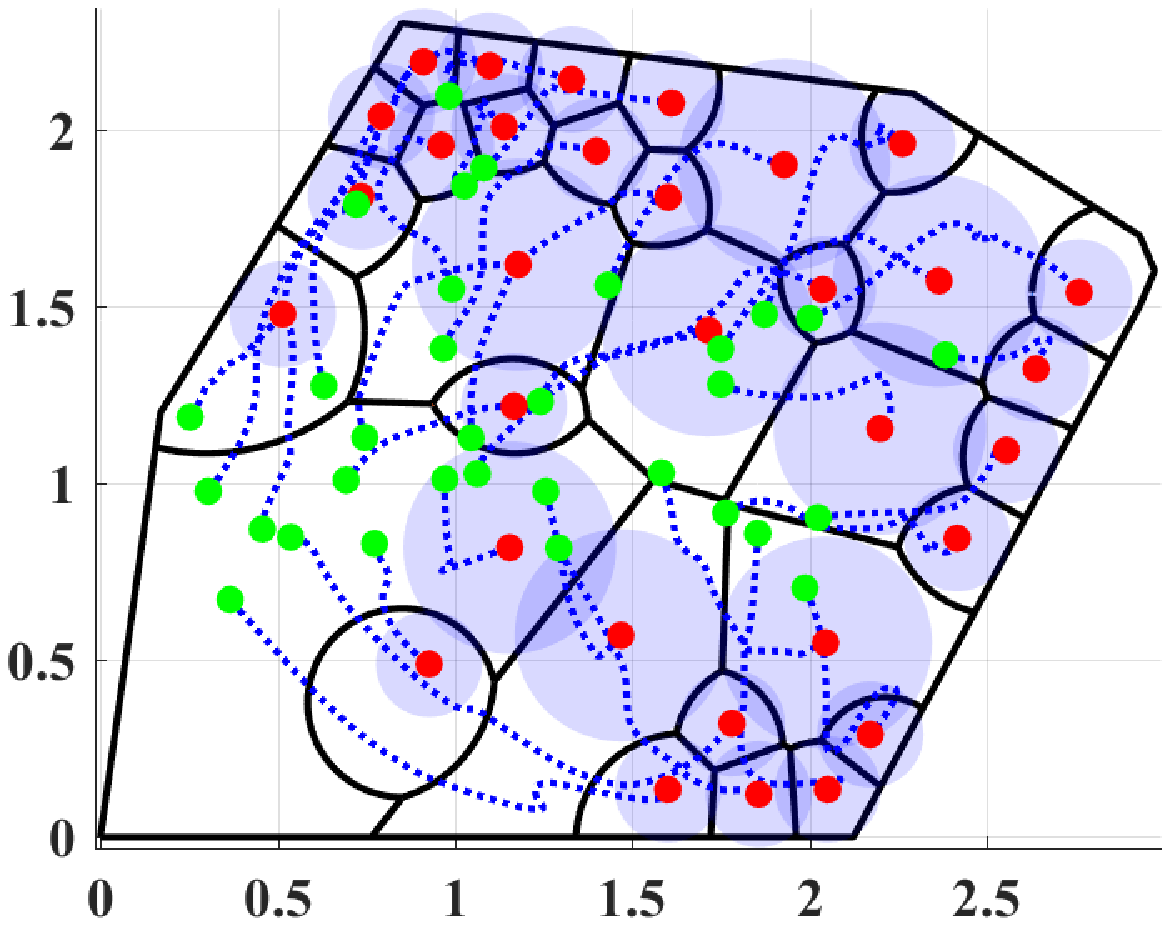}
\label{DCMLDeploymentWSN3}}
\hfil
\captionsetup{justification=justified}
\caption{Distributed sensor deployments: (a) VFA in MWSN1; (b) Lloyd-$\alpha$  in MWSN1; (c) CCML in MWSN1; (d) VFA in MWSN2; (e) Lloyd-$\alpha$ in MWSN2; (f) CCML in MWSN3; (g) VFA in MWSN3; (h) Lloyd-$\alpha$  in MWSN3; (i)  CCML in MWSN3. The initial sensor locations are denoted by green dots. The final locations of active and inactive sensors are denoted by red and black dots. The sensing regions of active and inactive sensors are denoted by blue and black. The movement paths are denoted by blue lines.}
\label{CCMLD}
\end{figure}

Simulation results for the centralized scheme are provided in Figs. \ref{CentralizedDeployment}, \ref{CentralizedDistortion}, and \ref{CentralizedCoverage}.
From Figs. \ref{CVFADeploymentWSN1} and \ref{CCMLDeploymentWSN1}, we observe that both VFA and CCML algorithms generate fully connected final deployments for the required network-lifetime, $T=1.3$.
By setting $\alpha$ to 0.2, Lloyd-$\alpha$ achieves a similar network-lifetime $T=1.31$.
However, Lloyd-$0.2$ generates a disconnected network where 12 sensors are placed out of the backbone network.
The corresponding distortions for VFA, Lloyd-$0.2$, and CCML are, respectively, 0.17, 0.78, and 0.14.

The centralized sensor relocations in heterogeneous MWSNs (MWSN2 and MWSN3) are illustrated in Figs. \ref{CVFADeploymentWSN2}, \ref{CLloydAlphaDeploymentWSN2}, \ref{CCMLDeploymentWSN2}, \ref{CVFADeploymentWSN3}, \ref{CLloydAlphaDeploymentWSN3}, and \ref{CCMLDeploymentWSN3}.
In MWSN2, BCCML keeps full-connectivity, but VFA and Lloyd-$\alpha$ generate dis-connected networks because of the limited communication range, which is 0.4.
The corresponding distortions for the deployments in Figs. \ref{CVFADeploymentWSN2}, \ref{CLloydAlphaDeploymentWSN2}, and \ref{CCMLDeploymentWSN2} are 2.75, 2.2, and 0.35, respectively.
In MWSN3, BCCML activates 30 sensors to sense the target region while the other two sensors are deactivated because of their low battery energy.
VFA and Lloyd-$\alpha$ attempt to use all sensors to finish the sensing task, but AP can only collect information from the active sensors shown by red dots.
Like the comparisons in MWSN1 and MWSN2, BCCML's distortion, $0.29$, is much smaller than that of VFA and Lloyd-$\alpha$, which are 4.16 and 0.66, respectively.

More detailed performance comparisons are provided in Figs. \ref{CentralizedPerformance1a}, \ref{CentralizedPerformance2a}, and \ref{CentralizedPerformance3a}.
In Fig. \ref{CentralizedPerformance1a}, for any given network lifetime, CCML Algorithm provides a lower distortion compare to  other algorithms in the homogeneous MWSN1.
In other words, CCML Algorithm outperforms other algorithms in homogeneous MWSNs.
Similarly,  BCCML Algorithm outperforms other algorithms in heterogeneous MWSNs shown in Figs. \ref{CentralizedPerformance2a} and \ref{CentralizedPerformance3a}.

We observe that all algorithms provide non-decreasing Distortion-Lifetime functions for the cases of $R_c=0.5$.
In fact, according to our simulation results, VFA, Lloyd-$\alpha$, and DEED keep full-connectivity for the three considered MWSNs when the communication range is large, e.g., $R_c=0.5$.
However, our proposed algorithms still outperform the other four algorithms even when $R_c=0.5$,
and the communication range is not a decisive parameter.
When $R_c=0.4$, VFA, Lloyd-$\alpha$, and DEED result in considerably large distortions because of the loss of full-connectivity, but BCCML Algorithm still provides small distortions.
For VFA, Lloyd-$\alpha$, and DEED algorithms, the distortions are significantly increased when $R_c$ is decreased from $0.5$ to $0.4$.
However, BCCML provides a similar performance in both cases.

\begin{figure}[!t]
\centering
\subfloat[]{\includegraphics[width=2.1in]{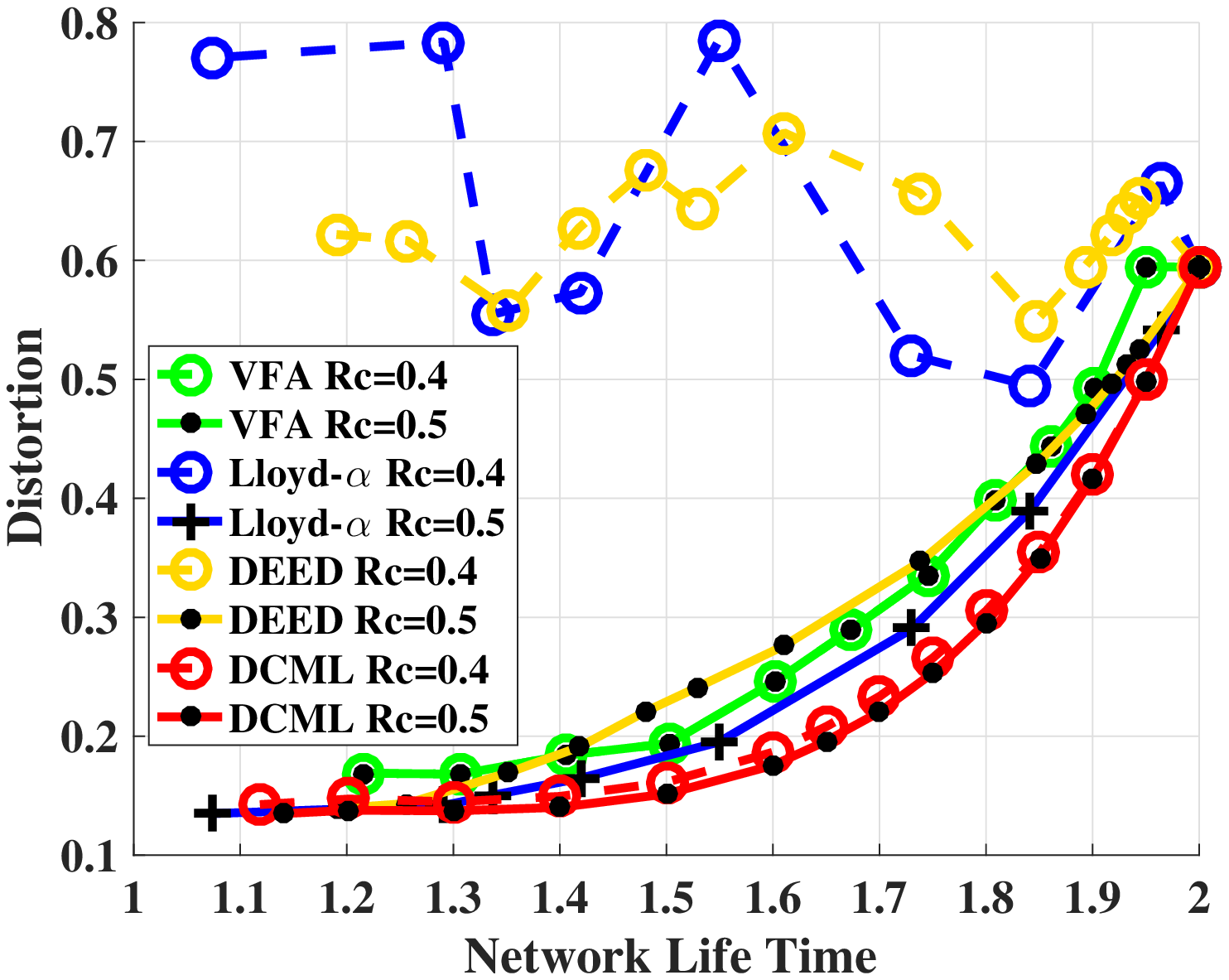}
\label{DistributedPerformance1a}}
\hfil
\subfloat[]{\includegraphics[width=2.1in]{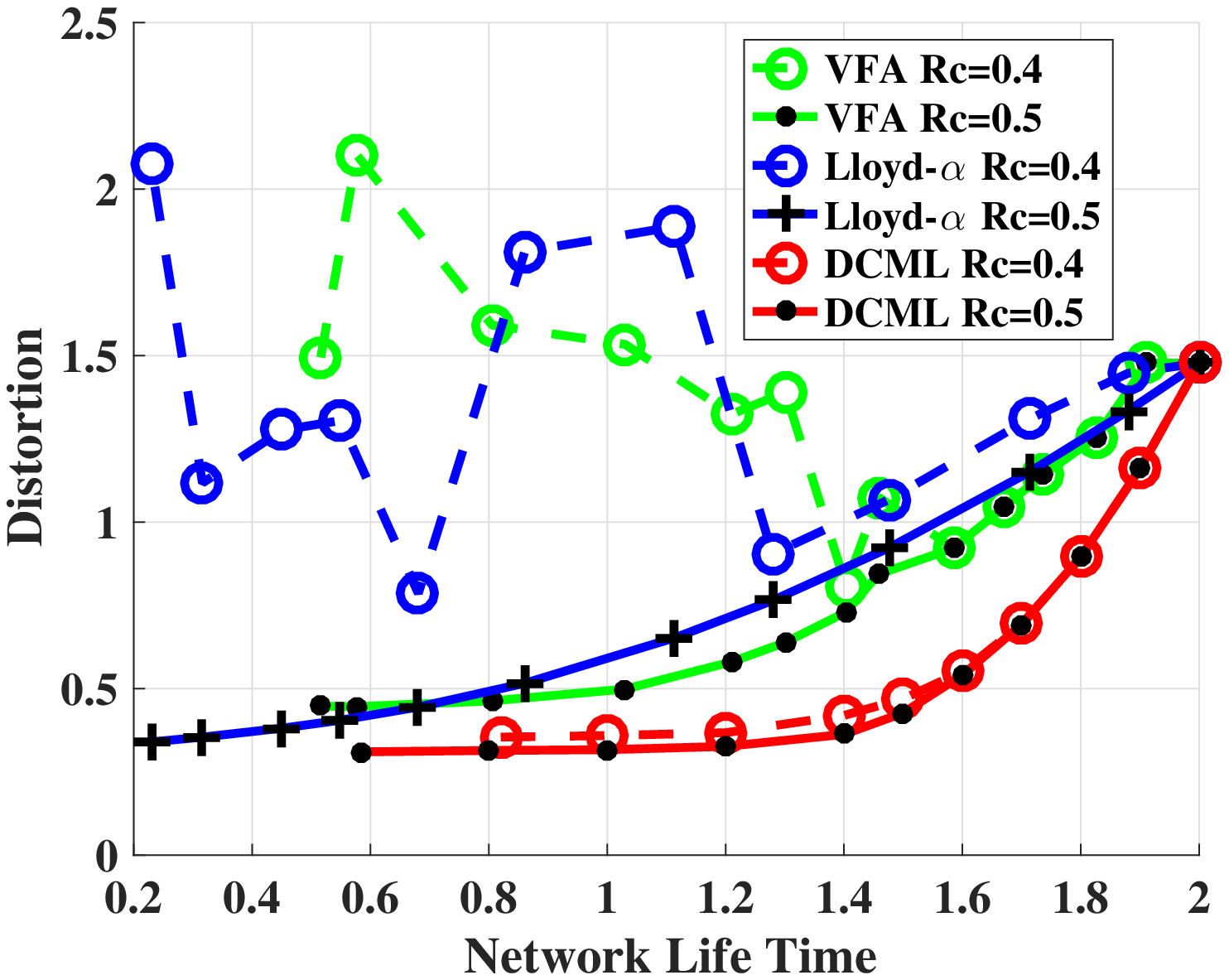}
\label{DistributedPerformance2a}}
\hfil
\subfloat[]{\includegraphics[width=2.1in]{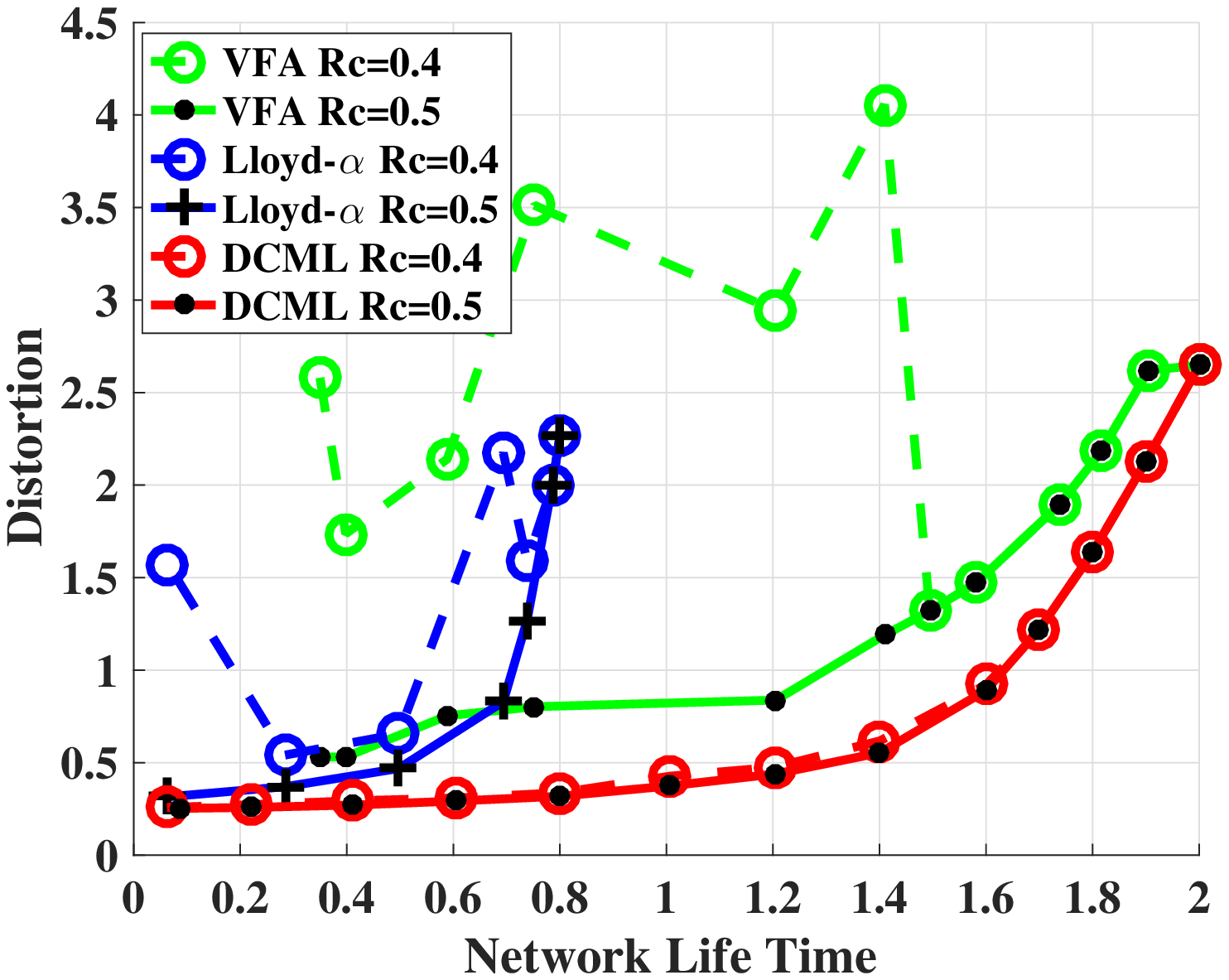}
\label{DistributedPerformance3a}}
\caption{The performance comparison for distributed sensor deployment in MWSN1.}
\label{DistributedDistortion}
\end{figure}

\begin{figure}[!t]
\centering
\subfloat[]{\includegraphics[width=2.1in]{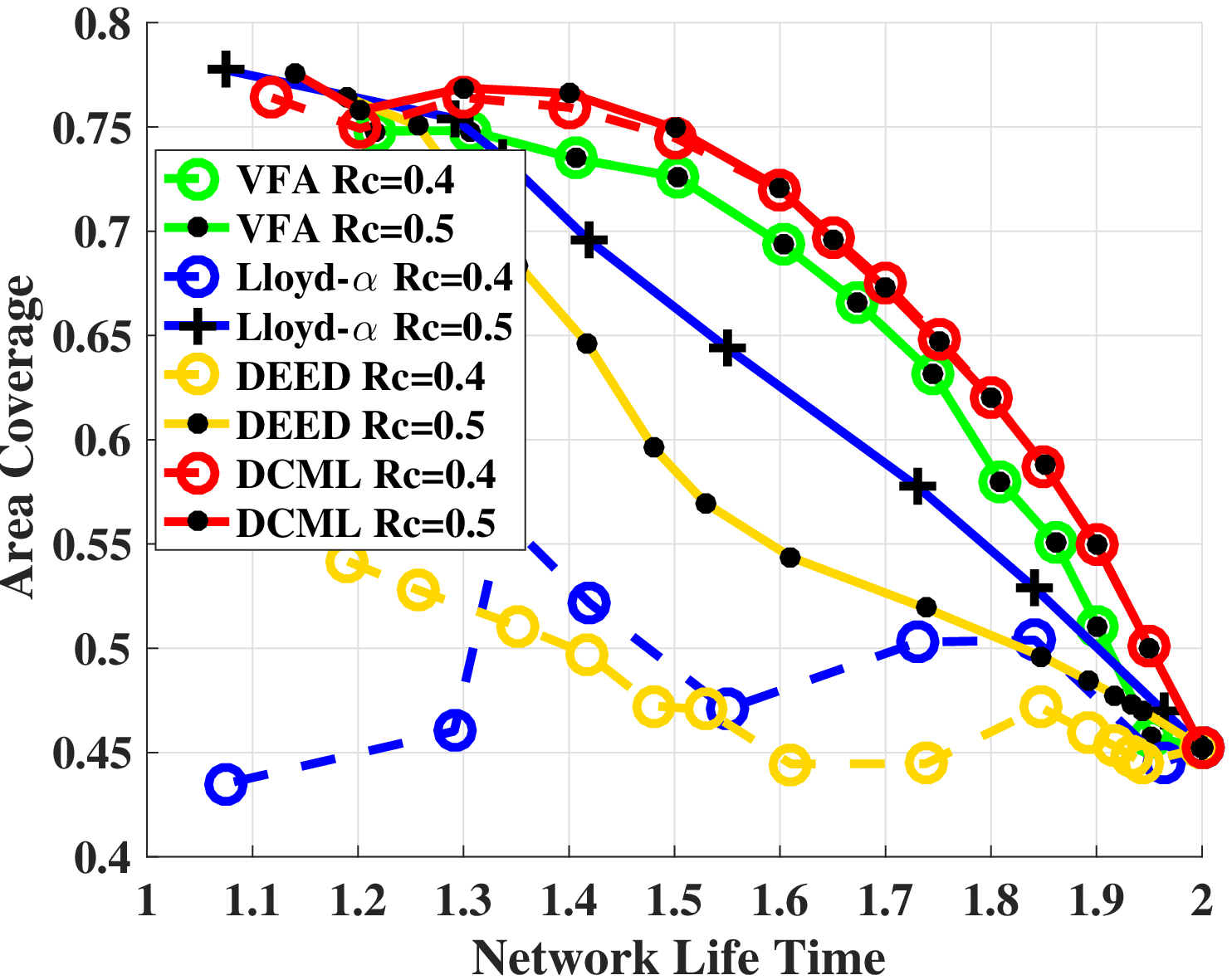}
\label{DistributedPerformance1b}}
\hfil
\subfloat[]{\includegraphics[width=2.1in]{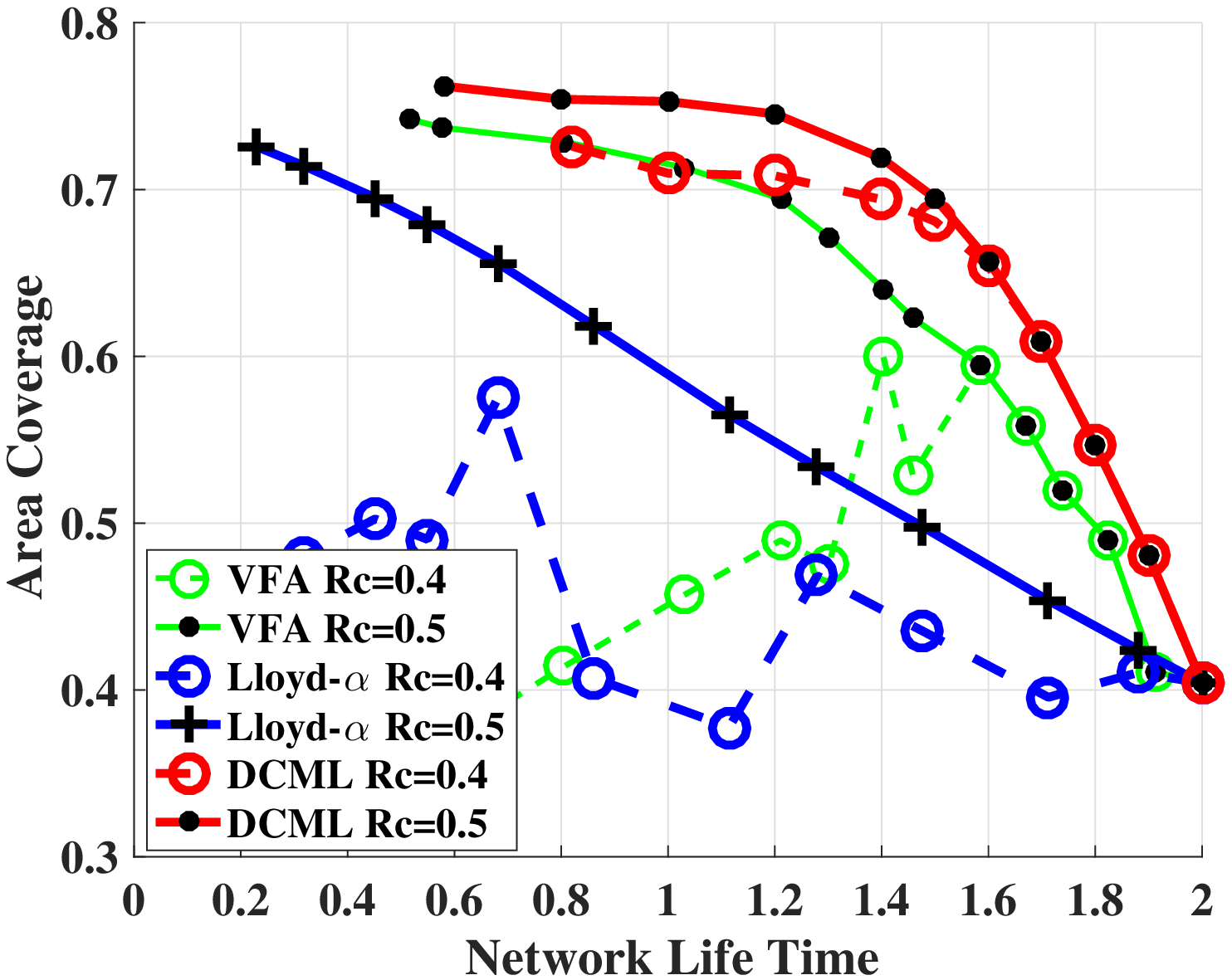}
\label{DistributedPerformance2b}}
\hfil
\subfloat[]{\includegraphics[width=2.1in]{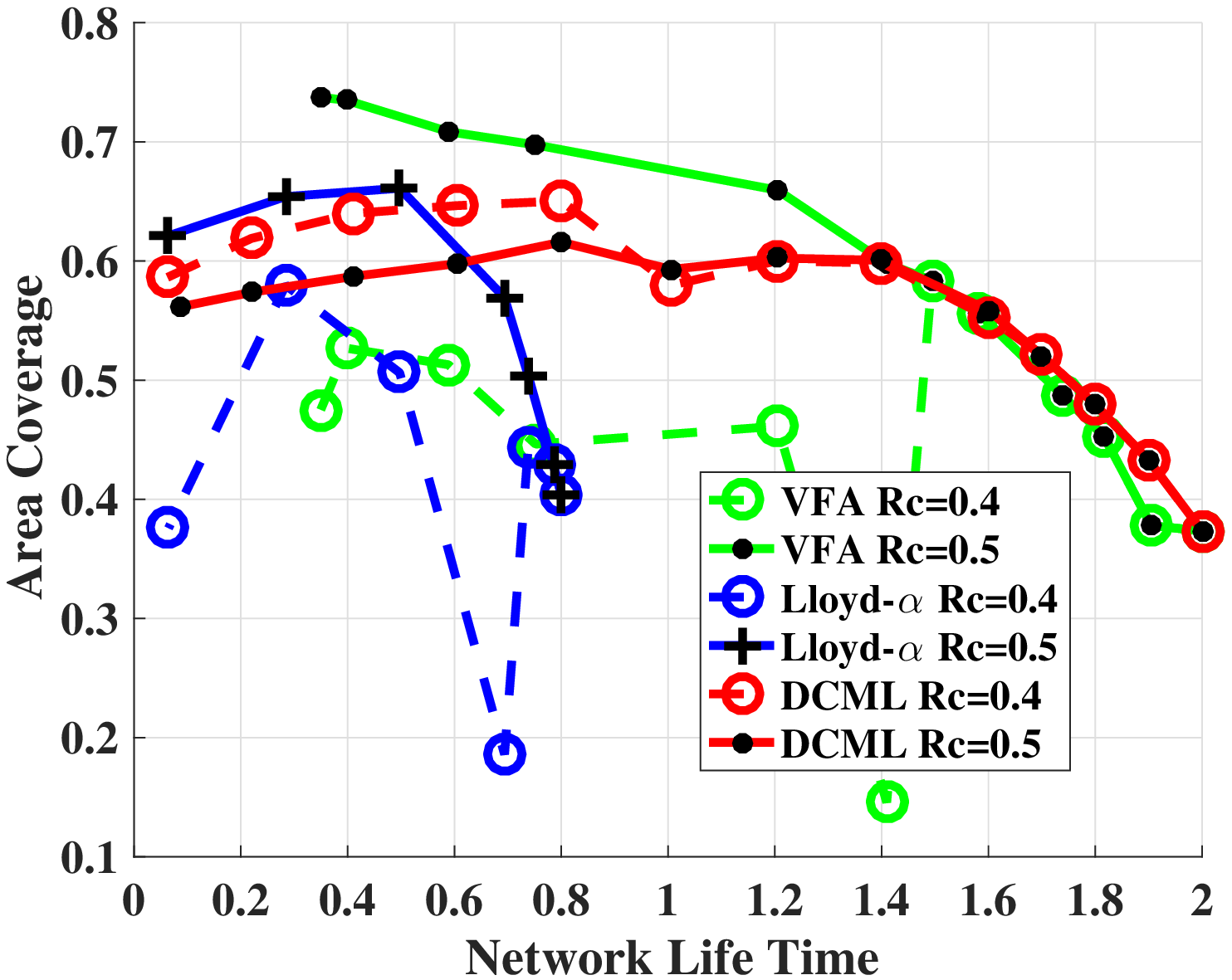}
\label{DistributedPerformance3b}}
\caption{The performance comparison for distributed sensor deployment in MWSN2.}
\label{DistributedCoverage}
\end{figure}

\begin{figure}[!t]
\centering
\subfloat[]{\includegraphics[width=2.1in]{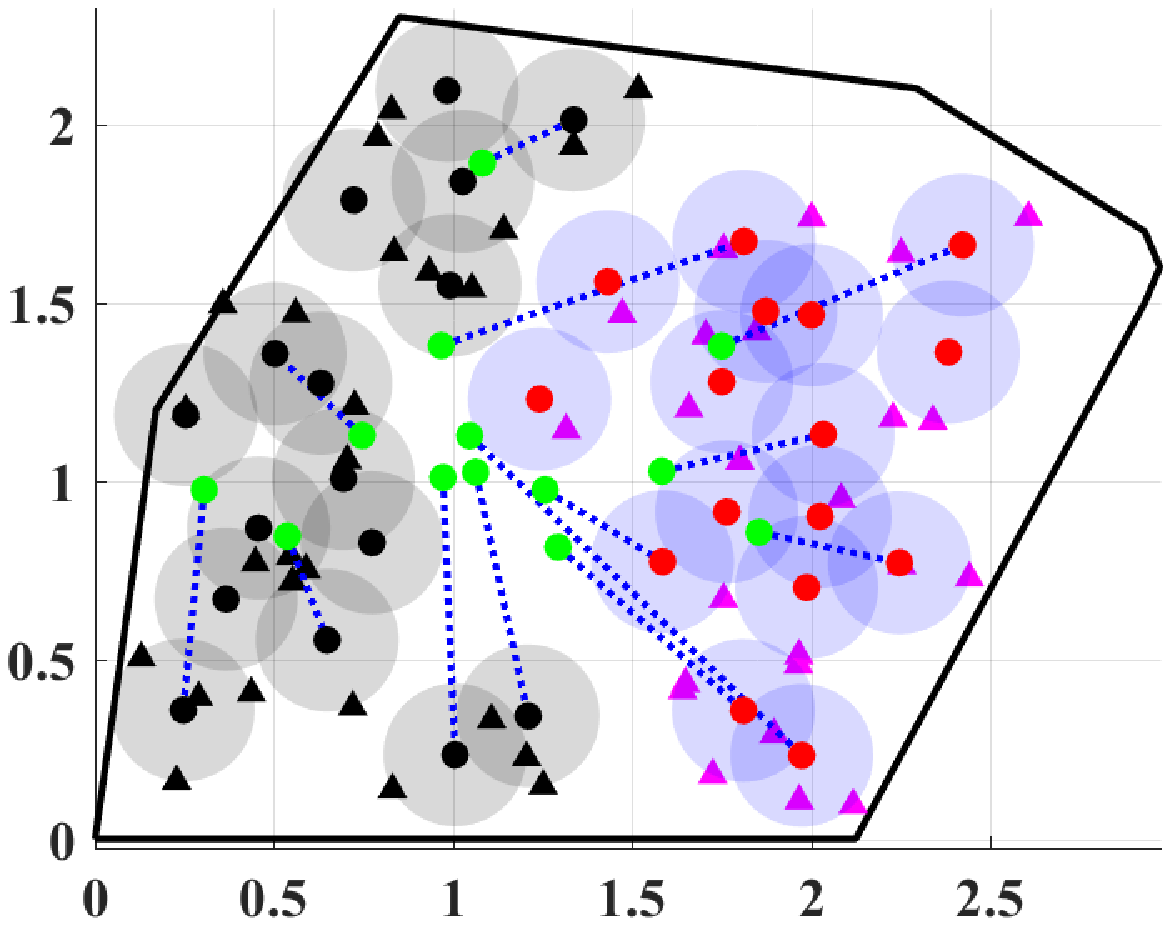}
\label{BasicECSTH}}
\hfil
\subfloat[]{\includegraphics[width=2.1in]{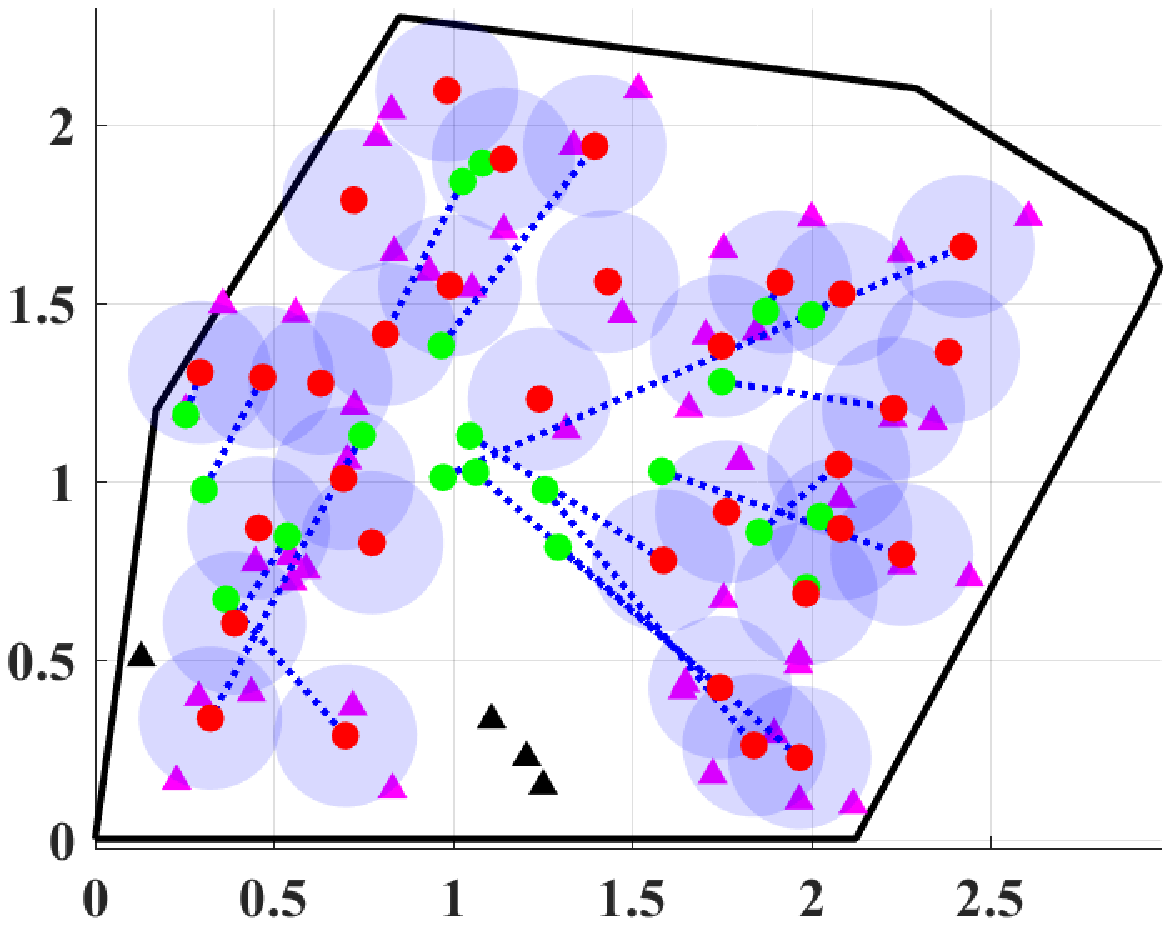}
\label{TVGreedyECSTH}}
\hfil
\subfloat[]{\includegraphics[width=2.1in]{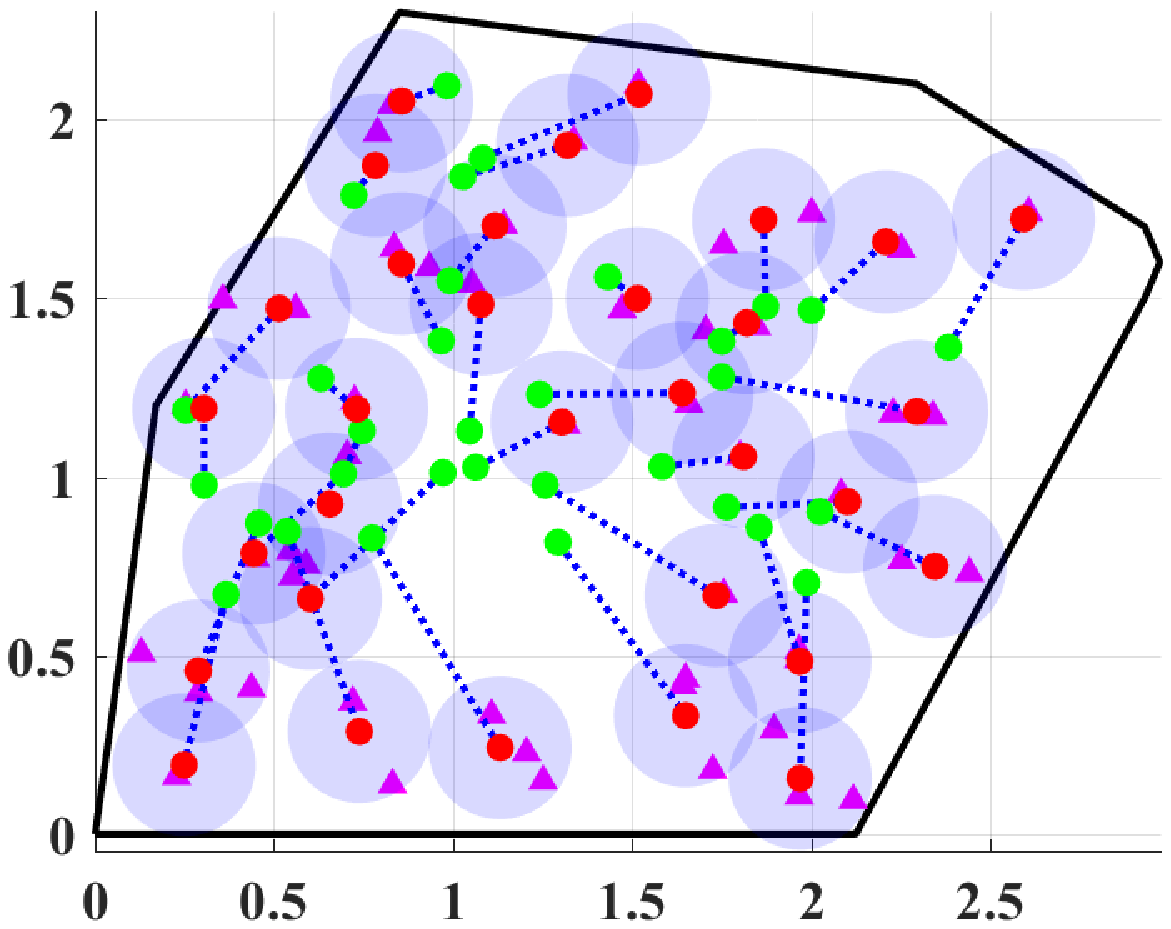}
\label{targetCoverageDeployment}}
\caption{The target coverage in MWSN1. (a) Basic+ECST-H; (b) TV-Greedy+ECST-H; (c) CCML. The covered targets and uncovered targets are denoted by magenta triangles and black triangles, respectively.}
\label{TargetCoverageDeployment}
\end{figure}

\begin{figure}[!t]
\centering
{\includegraphics[width=3in]{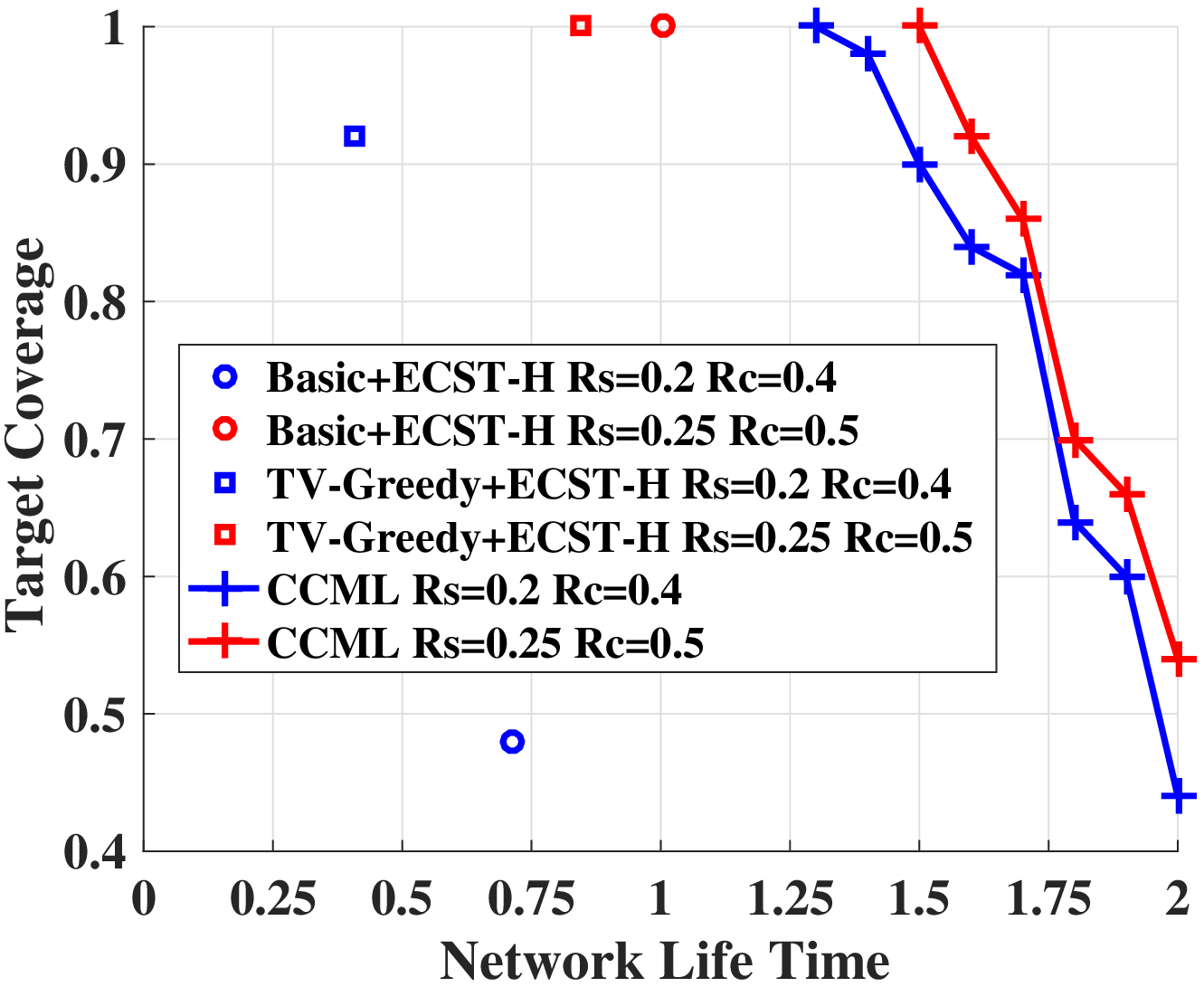}}
\captionsetup{justification=centering}
\caption{Target coverage comparison.}
\label{TargetCoverageComparison}
\end{figure}
Another advantage of BCCML Algorithm over Lloyd-$\alpha$ and DEED is its tractability.
BCCML Algorithm directly controls the energy consumption for relocation while Lloyd-$\alpha$ and DEED indirectly change energy consumption by tuning the hyperparameters $\alpha$ and $\delta$, respectively.
There is no explicit relationship between $\alpha$ (or $\delta$) and network lifetime.
Accordingly, one has to attempt different values of $\alpha$ (or $\delta$) in Lloyd-$\alpha$ (or DEED) to reach the required network lifetime.
Moreover, DEED needs both gradient and Hessian matrix of the objective function (\ref{distortion}).
To our best knowledge, the theoretical computation of Hessian matrix in heterogeneous MWSNs is still an open problem.
Although one can approximate the second-order derivatives by numerical methods, the corresponding extreme time complexity $O(N^2M)$ prevents DEED from being a feasible solution.
Therefore, DEED cannot be extended to heterogeneous MWSNs.
Different from DEED, Lloyd-$\alpha$, which only needs gradient, can be extended to heterogeneous MWSNs as the calculation of the gradient in heterogenous MWSNs already proposed in our previous work \cite{GJ}.
Unfortunately, when sensors are equipped with variant battery energies, Lloyd-$\alpha$ can only achieve a short network lifetime.
In MWSN3, where 4 sensors are equipped with a low battery energy, 0.8, Lloyd-$\alpha$ still uses all sensors to sense the target region.
Note that the network will die after the first node runs out of its battery energy.
As a result, Lloyd-$\alpha$ cannot achieve a network lifetime larger than 0.8 in MWSN3 (see Fig. \ref{CentralizedPerformance3a}).
However, the proposed BCCML Algorithm appropriately selects a subset of sensors to finish the sensing task.
In particular, to achieve a network lifetime larger than 0.8 in MWSN3, BCCML merely uses 28 sensors whose battery energy is larger than 0.8.

It is also noteworthy that low distortion is accompanied with high area coverage in MWSN1 and MWSN2 where the density function is uniform.
Unfortunately, such a relationship does not hold for MWSN3 where the density function is non-uniform.
Consequently, one can approximately optimize the area coverage by CCML and BCCML with uniform density function.

To confirm that CCML Algorithm can be extended to solve target coverage issues, we provide the sensor relocations of three different algorithms (Basic+ECST-H \cite{ZJSJG}, TV-Greedy+ECST-H \cite{ZJSJG}, and CCML) in Fig \ref{TargetCoverageDeployment}.
Both Basic+ECST-H and TV-Greedy+ECST-H consist of two stages: (i) A subset of sensors are placed to cover all targets; and (ii) other sensors are placed to guarantee connectivity.
According to our experiments, Basic+ECST-H and TV-Greedy+ECST-H require more sensors than CCML to achieve both full-coverage and full-connectivity.
In Fig. \ref{BasicECSTH}, all sensors are scheduled to cover targets in Stage (i) of Basic+ECST-H, and then no sensor is available
in Stage (ii).
As a results, Basic+ECST-H with $R_c=0.4$ is terminated with four disconnected subgraphs in which the largest sub-graph merely covers 48\% of targets.
In Fig. \ref{TVGreedyECSTH}, 32 sensors are not enough to achieve full-coverage in Stage (i) of TV-Greedy+ECST-H, and only 92\% of targets are covered.
However, with the same number of sensors, our proposed CCML Algorithm covers all targets and ensures full-connectivity.
The detailed comparison of Basic+ECST-H, TV-Greedy+ECST-H, and CCML are provided in Fig. \ref{TargetCoverageComparison}.
Compared to Basic+ECST-H and TV-Greedy+ECST-H, CCML Algorithm uses less energy to achieve full-coverage, and provides larger network lifetimes.
Moreover, different from Basic+ECST-H and TV-Greedy+ECST-H which attempt to obtain full-coverage with the minimum energy consumption, CCML Algorithm provides a more flexible trade-off between target coverage and energy consumption.

\vspace{-5pt}
\section{Conclusions and Discussion}\label{sec:conclusion}
\vspace{-7pt}
The trade-off between sensing quality and energy consumption, which is dominated by movement, is discussed in this paper.
We studied the optimal sensor deployment to minimize sensing uncertainty with a network life-time constraint in both homogeneous and heterogeneous mobile wireless sensor networks.
To make the model more practical, we take connectivity, which has a crucial influence on sensing performance, into consideration.
According to our analysis, full-connectivity is necessary to minimize the sensing uncertainty in homogeneous MWSNs.
The necessary condition for an optimal deployment implies that sensors should move towards the centroid within their own feasible regions, determined by both the battery energies and the communication range.
With the help of these necessary conditions, two centralized sensor relocation algorithms, Centralized Constrained Movement Lloyd Algorithm and Backwards-stepwise Centralized Constrained Movement Lloyd Algorithm, are designed for homogeneous and heterogeneous MWSNs, respectively.
Moreover, a distributed realization, Distributed Constrained Movement Lloyd Algorithm, whose performance is similar to the centralized scheme, is also provided in this paper.
Furthermore, by manually changing the density function, we extend the proposed sensor relocation algorithms to target coverage.
Our simulation results show that the proposed algorithms outperform the existing algorithms in the literature (VFA, Lloyd-$\alpha$, DEED) when a minimum network life-time is given in homogeneous and heterogeneous MWSNs.
Compared with the existing target coverage algorithms, such as Basic+ECST-H and TV-Greedy+ECST-H, CCML Algorithm provides a more flexible trade-off between target coverage and network life-time.
\appendices
\vspace{-20pt}
\section{Proof of Lemma 1}\label{appendixL1}
\vspace{-5pt}
Lemma \ref{connectivity} focuses on homogeneous MWSNs where $\eta_n=\eta$, $\xi_n=\xi$, and $\gamma_n=\gamma, \forall n\in\mathcal{I}_{\Omega}$.
Let $\bP^0=(p^0_1,\dots,p^0_N)$ and $\bP^*=(p^*_1,\dots,p^*_N)$ be the initial and the optimal sensor deployment in a MWSN with performance function (\ref{distortion}) and constraints (\ref{individualConstraint}), respectively.
For convenience, let $\mathcal{I}_{\Omega}$ be the set of all sensors, $\mathcal{S}(\bP)$ be the set of sensors that can communicate with the AP when the sensor deployment is $\bP$, and $\mathcal{N}_n(\bP)$ be Sensor $n$'s neighbors given deployment $\bP$.
$\mathcal{S}(\bP)$ is also referred to as the backbone network in Section \ref{sec:model}.
In Lemma \ref{connectivity}, we assumed that $\bP^0$ provides a fully connected network, i.e., $S(\bP^0)=\mathcal{I}_{\Omega}$.
Now, we assume that the optimal deployment is associated with a disconnected network, i.e., $\mathcal{S}(\bP^*)\ne\mathcal{I}_{\Omega}$.
In this case, we can find a sensor, $n\in\mathcal{S}(\bP^*)$, such that one of its neighbors in the initial deployment, $m\in\mathcal{N}_n(\bP^0)$, is not in the final backbone network, $m\notin\mathcal{S}(\bP^*)$.
An alternative point is defined as
\vspace{-5pt}
\begin{equation}
\vspace{-5pt}
p'_m=p^0_m+\min\left(0,\|p^*_n-p^0_m\|-R_c\right)\frac{p^*_n-p^0_m}{\|p^*_n-p^0_m\|}.
\label{alternative}
\vspace{-5pt}
\end{equation}
Replacing $p^*_m$ by $p'_m$, we get an alternative deployment $\bP'=\left(p^*_1,\dots,p'_m,\dots,p^*_N\right)$.
Next, we check if $\bP'$ satisfies the network lifetime constraints (\ref{individualConstraint}).
First, since sensors $m$ and $n$ are neighbors, we have
$\|p^0_m-p^0_n\|\le R_c$.
Second, following the constraints (\ref{individualConstraint}), we have
$\|p^*_n-p^0_n\|\le\frac{\gamma}{\xi}$.
Third, using the triangular inequality, we have
$\|p^0_n-p^0_m\|+\|p^*_n-p^0_n\|>\|p^0_m-p^*_n\|$.
Combining the above three inequalities, we obtain
$\|p^*_n-p^0_m\|<R_c+\frac{\gamma}{\xi}$.
According to (\ref{alternative}), $p'_m$ is placed between $p^*_n$ and $p^0_m$, and the moving distance is
\vspace{-8pt}
\begin{equation}
\vspace{-5pt}
\|p'_m-p^0_m\|=\min\left(0,\|p^*_n-p^0_m\|-R_c\right) < \frac{\gamma}{\xi}.
\vspace{-5pt}
\end{equation}
Thus, the deployment $\bP'$ satisfies the network lifetime constraints.\\
In what follows, we verify that $\bP'$ provides a smaller distortion compared to $\bP^*$.
The distance between $p'_m$ and $p^*_n$ can be calculated as
\vspace{-2pt}
\begin{equation}
\vspace{-8pt}
\|p^*_n-p'_m\| = \|p^*_n-p^0_m\| - \|p'_m-p^0_m\| =
\begin{cases}
    \|p^*_n-p^0_m\|, &\mbox{if $\|p^*_n-p^0_m\|\le R_c$}\\
    R_c, &\mbox{otherwise}
    \end{cases},
\vspace{-5pt}
\end{equation}
which means $\|p^*_n-p'_m\|\le R_c$.
In other words, when the deployment is $\bP'$, Sensor $m$ connects with $n$ and then should be taken into the calculation of distortion (\ref{distortion}).
In our model, sensors are initially deployed in the target region $\Omega$, i.e., $p^0_n\in\Omega$, where $\Omega$ is a convex region \cite{GJ}.
Moveover, it is self-evident that the optimal sensor locations are also in the target region, i.e., $p^*_n\in\Omega, \forall n\in\mathcal{I}_{\Omega}$.
By properties of a convex region, any point between $p^*_n$ and $p^0_m$ should be in the target region, e.g., $p'_m\in\Omega$.
Therefore, Sensor $m$ is associated with a non-empty MWVD $V_m(\mathcal{H}(\bP'))=\{\omega|\omega\in\Omega, \|\omega-p'_m\|<\|\omega-p^*_i\|, \forall i\in\mathcal{S}(\bP') i\ne m\}$.
The difference between the distortions at $\bP'$ and $\bP^*$ lays on $V_m(\mathcal{H}(\bP'))$ and can be calculated as
\begin{equation}
\vspace{-8pt}
\begin{aligned}
D(\bP')\!-\! D(\bP^*) {=}& \!\int_{V_m(\mathcal{H}(\bP'))}\!\!\!\!\!\eta\|\omega\!-\!p'_m\|^2f(\omega)d\omega
{-} \!\!\!\! \sum_{n\in\mathcal{S}(\mathcal{H}(\bP^*))}\!\!\!\!\int_{V_m(\mathcal{H}(\bP'))\bigcap V_n(\mathcal{H}(\bP^*))}\!\!\!\!\!\!\!\!\!\!\!\!\!\!\eta\|\omega\!-\!p'_m\|^2f(\omega)d\omega\\
{=}& \int_{V_m(\mathcal{H}(\bP'))}\eta\left(\|\omega-p'_m\|^2-\min_{n\in\mathcal{S}(\bP^*)}\|\omega-p^*_n\|^2\right)f(\omega)d\omega<0
\end{aligned}
\vspace{-2pt}
\end{equation}
Consequently, $\bP'$ is a better solution than $\bP^*$, which contradicts our assumption\footnote{Remark: In this proof, we ignore two special cases:
(i) $p'_m$ is placed on top of another sensor, i.e., $p'_m=p^*_i, i\ne m$.
In this case, $p'_m$ should be moved towards $p^0_m$ a little bit to avoid overlap without breaking the constraints, and then we will arrive at the same contradiction.
(ii) After replacing $p^*_m$ by $p'_m$, more than one sensor joins the backbone network.
In this case, the distortion at $\bP'$ will be further reduced, and therefore we will have the same contradiction.}.

\section{Proof of Theorem 1}\label{appendixP1}
\vspace{-5pt}
Let $\bR^*=(R^*_1,\dots,R^*_N)$ be the optimal partition and $\mathcal{I}^*$ be the optimal backbone network.
For simplicity, let $c^*_n=\frac{\int_{R^*_n}\omega f(\omega)d\omega}{\int_{R^*_n}f(\omega)d\omega}$ and $v^*_n=\int_{R^*_n}f(\omega)d\omega$ be, respectively, the geometric centroid and the Lebesgue measure (volume) of $R^*_n$, $\forall n\in\mathcal{I}_{\Omega}$.
Let $m\in\mathcal{I}^*$ be a specific sensor in the backbone network, $\mathcal{\bP^*}$.
First, we assume that other sensor locations, $\{p^*_n\}_{n\in\left(\mathcal{I}_{\Omega}-\{m\}\right)}$, are known, and then derive the constraint for $p^*_{m}$.
Since $\mathcal{I}^*$ is the backbone network, we have\footnote{The definitions of $\mathbb{D}_m\left(\bP,\mathcal{I}\right)$ and $\mathcal{W}\left(\bP,\mathcal{I}\right)$ and their relationships to backbone network are already provided in Section \ref{sec:ProblemB}}. $p^*_m\in\mathbb{D}_m\left(\bP^*,\mathcal{I}^*\right)\bigcap\mathcal{W}^{c}\left(\bP^*,\mathcal{I}^*\right)$.
Moreover, since Sensor $m$ should satisfy the energy constraints (\ref{individualConstraint}), we have $p^*_m\in \mathbb{B}(p^0_m,\frac{\gamma_m}{\xi_m})$.
In summary, Sensor $m$'s location is restrained by
\vspace{-2pt}
\begin{equation}
\vspace{-5pt}
\mathbb{D}_m\left(\bP^*,\mathcal{I}^*\right)\bigcap\mathcal{W}^c\left(\bP^*,\mathcal{I}^*\right)\bigcap \mathbb{B}\left(p^0_m, R_c\right)=\mathbb{F}_m\left(\bP^*,\mathcal{I}^*\right)\bigcap\mathcal{W}^c\left(\bP^*,\mathcal{I}^*\right).
\label{pmconstraint}
\vspace{-5pt}
\end{equation}
Second, the minimum distortion can be rewritten as
\vspace{-2pt}
\begin{align}
\vspace{-5pt}
D(\bP^*) &{=} \sum_{n\in\mathcal{I}^*}\int_{\bR^*_n}\eta_n\|p^*_n-\omega\|^2f(\omega)d\omega\\
&{=}\int_{\bR^*_m}\eta_m\|p^*_m-\omega\|^2f(\omega)d\omega+\!\!\sum_{n\in\left(\mathcal{I}^*-\{m\}\right)}\!\!\int_{\bR^*_n}\eta_n\|p^*_n-\omega\|^2f(\omega)d\omega\\
&{=}\eta_m\|c^*_m\!-\!p^*_m\|^2v^*_m\!+\!\int_{\bR^*_m}\!\!\eta_m\|c^*_m\!-\!\omega\|^2f(\omega)d\omega+\!\!\! \sum_{n\in\left(\mathcal{I}^*-\{m\}\!\right)}\!\!\int_{\bR^*_n}\!\!\eta_n\|p^*_n-\omega\|^2f(\omega)d\omega,
\label{optD}
\vspace{-5pt}
\end{align}
where the third equation follows from the parallel axis theorem.
Given the optimal partition $\bR^*$ and optimal locations $\{p^*_n\}_{n\in\left(\mathcal{I}_{\Omega}-\{m\}\right)}$, the second and third terms in (\ref{optD}) are constants.
Therefore,
$p^*_m$ should be a minimizer of the first term with the constraint (\ref{pmconstraint}), i.e.,
\begin{equation}
\vspace{-5pt}
\begin{aligned}
p^*_m
&{=}\arg\min_{p_m:p_m\in\mathbb{F}\left(\bP^*,\mathcal{I}^*\right)\bigcap\mathcal{W}^c\left(\bP^*,\mathcal{I}^*\right)}\eta_m\|p_m-c^*_m\|^2v^*_m.
\label{Pstar1}
\end{aligned}
\vspace{-5pt}
\end{equation}
Equation (\ref{Pstar1}) implies that the optimal solution should minimize the distance to $c^*_n$ within $\mathbb{F}(\bP^*,\mathcal{I}^*)\bigcap\mathcal{W}^c\left(\bP^*,\mathcal{I}^*\right)$.
In what follows, we discuss 3 different cases of $c^*_m$.

(a) If $c^*_m\in\left[\mathbb{F}(\bP^*,\mathcal{I}^*)\bigcap\mathcal{W}\left(\bP^*,\mathcal{I}^*\right)\right]$, we have $p^*_m\ne c^*_m$ because $p^*\in\left[\mathbb{F}\left(\bP^*,\mathcal{I}^*\right)\bigcap\mathcal{W}^c\right]$.
However, replacing $p^*_m$ by $c^*_m$, one can get a better solution $\bP'=\left(p^*_1,\dots,p^*_{m-1},c^*_m,p^*_{m+1},\dots,p^*_N\right)$ which not only follows the energy constraints but also provides a smaller distortion.
As a result, $c^*_m\in\left[\mathbb{F}(\bP^*,\mathcal{I}^*)\bigcap\mathcal{W}\left(\bP^*,\mathcal{I}^*\right)\right]$ is an impossible case for the optimal deployment.
In other words, we have the condition
\begin{equation}
\vspace{-5pt}
c^*_m\notin\left[\mathbb{F}(\bP^*,\mathcal{I}^*)\bigcap\mathcal{W}\left(\bP^*,\mathcal{I}^*\right)\right]
\label{impossible}
\vspace{-5pt}
\end{equation}
Since MWVD is the optimal partition, we have $\bR^*=V(\bP^*)$
and then
\begin{equation}
\vspace{-5pt}
c^*_m=\frac{\int_{V_m(\bP^*)}\omega f(\omega)d\omega}{\int_{V_n(\bP^*)}f(\omega)d\omega}=c_m(\bP^*), \forall n\in I_{\Omega}.
\label{c}
\vspace{-5pt}
\end{equation}
Moreover, the backbone network $\mathcal{I}^*$ is a function of deployment $\bP^*$ and can be represented by
\vspace{-5pt}
\begin{equation}
\vspace{-5pt}
\mathcal{I}^*=\mathcal{S}(\bP^*).
\label{I}
\vspace{-5pt}
\end{equation}
Substituting (\ref{c}) and (\ref{I}) to (\ref{impossible}), we get condition (i).

(b) If $c^*_m\in\left[\mathbb{F}(\bP^*,\mathcal{I}^*)\bigcap\mathcal{W}^c\left(\bP^*,\mathcal{I}^*\right)\right]$, Sensor $m$ should be placed at $c^*_m$, i.e.,
$p^*_m=c^*_m$.
Replacing $c^*_m$ by (\ref{c}) , we get the first case in condition (ii).

(c) If $c^*_m\notin\mathbb{F}(\bP^*,\mathcal{I}^*)$, a simple geometric argument reveals that the optimal solution should be on the boundary of $\mathbb{F}(\bP^*,\mathcal{I}^*)$.
Therefore, the optimal location can be represented as $p^*_m=\arg\min\limits_{q\in\partial{\left[\mathbb{F}_m(\bP^*,\mathcal{I}^*)\bigcap\mathcal{W}^c\left(\bP^*,\mathcal{I}^*\right)\right]}}\|q\!-\!c^*_m\|$.
Replacing $c^*_m$ and $\mathcal{I}^*$ by (\ref{c}) and (\ref{I}), respectively, we get the second case in condition (ii).

\vspace{-5pt}
\vspace{-10pt}
\section{Proof of Theorem 3}\label{appendixP3}
Let $\bP^{k}=\left(p^k_1,\dots,p^k_N\right)$ and $\bP^{k}=\left(p^{k+1}_1,\dots,p^{k+1}_N\right)$ be the current and next sensor deployment.
Let $m$ and $n$ be two sensors such that they are MST neighbors at $\bP^{k}$.
Then, the distance between $p^k_m$ and $p^k_n$ is no larger than the communication range $R_c$, i.e., $\|p^k_m-p^k_n\|\le R_c$.
According to the definition of semi-desired region (\ref{SDR}), we have $\mathbb{D}^s_m(\bP)\subset \mathbb{B}\left(\frac{p^k_m+p^k_n}{2},\frac{R_c}{2}\right)$ and $\mathbb{D}^s_n(\bP)\subset \mathbb{B}\left(\frac{p^k_m+p^k_n}{2},\frac{R_c}{2}\right)$.
Moreover, $m$ and $n$ move within their semi-desired regions, i.e., $p^{k+1}_m\in\mathbb{D}^s_m\left(\bP^k\right)$ and $p^{k+1}_n\in\mathbb{D}^s_n\left(\bP^k\right)$.
Thus, both $p^{k+1}_m$ and $p^{k+1}_n$ lie in the circle $\mathbb{B}\left(\frac{p^k_m+p^k_n}{2},\frac{R_c}{2}\right)$.
Therefore, the distance between $p^{k+1}_m$ and $p^{k+1}_n$ is no larger than the communication range, i.e., $\|p^{k+1}_m-p^{k+1}_n\|\le R_c$.
In other words, Sensors $m$ and $n$ are still connected with each other after the relocation.
Consequently, the network at $\bP^{k+1}$, $\mathcal{G}\left(\bP^{k+1}\right)$, retains all edges in the MST at $\bP^{k}$, indicating that the network, $\mathcal{G}\left(\bP^{k+1}\right)$, is fully connected.

\end{document}